# Suppression of Surfaces States at Cubic Perovskite (001) Surfaces by CO2 Adsorption

*Kostiantyn V. Sopiha[1], Oleksandr I. Malyi[2]\*, Clas Persson[2], Ping Wu[1]\**

1 – Entropic Interface Group, Engineering Product Development, Singapore University of Technology and Design, 8 Somapah Road, 487372 Singapore, Singapore

2 – Centre for Materials Science and Nanotechnology, Department of Physics, University of Oslo, P. O. Box 1048 Blindern, NO-0316 Oslo, Norway

E-mails: oleksandrmalyi@gmail.com (O.I.M), wuping@sutd.edu.sg (W.P)

**Abstract**

By using first-principles approach, the interaction of $CO_2$ with (001) surfaces of six cubic $ABO_3$ perovskites (A = Ba, Sr and B = Ti, Zr, Hf) is studied in detail. We show that $CO_2$ adsorption results in the formation of highly stable $CO_3$-like complexes with similar geometries for all investigated compounds. This reaction leads to the suppression of the surfaces states, opening the band gaps of the slab systems up to the corresponding bulk energy limits. For most AO-terminated $ABO_3(001)$ perovskite surfaces, a $CO_2$ coverage of 0.25 was found to be sufficient to fully suppress the surface states, whereas the same effect can only be achieved at 0.50 $CO_2$ coverage for the $BO_2$ terminations. The largest band gap modulation among the AO-terminated surfaces was found for $SrHfO_3(001)$ and $BaHfO_3(001)$, whereas the most profound effect among the $BO_2$ terminations was identified for $SrTiO_3(001)$ and $BaTiO_3(001)$. Based on these results and considering practical difficulties associated with measuring conductivity of highly resistive materials, $TiO_2$-terminated $SrTiO_3(001)$ and $BaTiO_3(001)$ were identified as the most prospective candidates for chemiresistive $CO_2$ sensing applications.

**Keywords:** perovskites, $CO_2$ adsorption, sensing mechanism, surface states, band gap modulation

## Introduction

An early discovery of gas sensing features in metal-oxides[1] has triggered an intensive search for functional materials having superior gas sensing characteristics.[2-4] Despite this apparent success, in general, chemiresistive sensing of chemically inert gases remains practically difficult. Specifically, measuring $CO_2$ concentration is currently performed using optical gas detectors, which are not well suited for mass-scale applications due to their high production cost imposed by the structural complexity.[5] From the chemical perspective, a majority of classical sensing oxides are practically incapable of chemiresistive $CO_2$ detection due to high stability of the $CO_2$ molecule.[3, 4, 6] In an attempt to overcome this obstacle, research focus has gradually shifted to alternative sensing materials, including various inorganic perovskites possessing prospective $CO_2$ sensing characteristics.[7-17]

The first efficient perovskite-based $CO_2$ sensor was developed by Meyer *et al.*, who discovered profound chemiresistive features of $BaTiO_3$-containing composites.[7, 8] Since then, the optimized composition consisting of $BaTiO_3$, $CuO$, and $La_2O_3$ has long been the best-performing $CO_2$ sensing material reported in the literature[4, 8] until the recent development of $BaSnO_3$-based chemiresistor.[9] To understand the sensing phenomenon, $CO_2$ sensing properties of the $BaTiO_3$-$CuO$ composites were examined by different authors.[18-26] The studies revealed the formation of surface carbonate groups and proposed a $CO_2$ sensing mechanism based on modulation of the electron energy levels at $BaTiO_3/CuO$ heterointerfaces.[21-24] Although this mechanism can explain chemiresistive response for the composites, it fails to justify experimental findings revealing $CO_2$ sensing features of pristine[9-11] and doped[12-17] perovskites. Moreover, despite considerable progress achieved in the field, so far, reported values of the $CO_2$ sensing response remain too low for practical use, highlighting a need for further research and functionalization of the perovskite-based sensing materials. The main obstacle for the further development, however, lays in a poor understanding of the adsorption chemistry of perovskite surfaces and its relation to the electrical conductivity.

To unveil the background surface chemistry of sensing, $CO_2$ adsorption on stable (001) perovskite surfaces has been intensively investigated using first-principles methods. It was found that the $CO_2$ adsorption results in the formation of highly stable $CO_3$-like complexes for all considered materials.[27-29] Despite this knowledge, the exact mechanism on how the adsorption affects the surface conductivity has long remained uncertain. Recently, it has been demonstrated that molecular chemisorption on $K_{1-y}Na_yTa_{1-x}Nb_xO_3(001)$[29, 30] and $SrTiO_3(001)$[31] results in the suppression of surface states emergent at their clean (001) facets. Since the existence of surface states can increase the surface conductivity[32-34] and even contribute to the formation of 2-dimensional electron gas (2DEG),[35] this suppression might be a long-missing link between the $CO_2$ adsorption and conductivity change during the sensing. However, since up to date the band gap modulation has only been explored for two representative perovskite materials, a wider family of the compounds should be considered to provide a better understanding and allow further generalization of the mechanism. Motivated by this, we investigated $CO_2$ interaction with (001) surfaces of different cubic $ABO_3$ perovskites (A = Ba, Sr and B = Ti, Zr, Hf) focusing on the adsorption-induced changes in the surface electronic properties. The material selection for this study was dictated by similar geometry and chemistry of the compounds, which can allow to reveal common adsorption trends for cubic perovskites by analyzing the roles played by different parameters (such as lattice constant, band gap, surface termination, *etc.*). The obtained results are expected to guide further optimization of the perovskite-based sensing materials for $CO_2$ sensing applications, catalysis, photocatalysis, and beyond.

**Methods**

The first-principles calculations were performed using Vienna *ab initio* simulation package (VASP)[36-38] utilizing Perdew-Burke-Ernzerhof (PBE) functional[39] and projected augmented wave (PAW) pseudopotentials.[40, 41] Barium $5s^25p^66s^2$, strontium $4s^24p^65s^2$, hafnium $5s^25p^65d^26s^2$, zirconium $4s^24p^64d^25s^2$, titanium $3s^23p^63d^24s^2$, oxygen $2s^22p^4$, and carbon $2s^22p^2$ were treated as valence electrons. The Brillouin-zone integrations were performed on $\Gamma$-centered 5×5×1 Monkhorst-Pack grids[42] for ionic relaxations and the 10×10×1 grids for electronic structure calculations. The cutoff energies of 400 eV and atomic force threshold of 0.01 eV/Å were employed in all calculations. Since PBE is known to underestimate the band gap energies,[43] electronic structures of the bulk and clean perovskite surfaces were also computed using hybrid Heyd-Scuseria-Ernzerhof (HSE06) functional with the default mixing coefficient of 0.25 for the exact exchange.[44] The obtained results were analyzed using Vesta[45] and pymatgen.[46]

Bulk unit cells of cubic perovskites were relaxed imposing $Pm\overline{3}m$ symmetry. These phases has been found experimentally in all perovskite systems considered herein, although distorted phases are often the ground states.[47] The optimized lattice parameters were used to construct model slabs consisting of 13 atomic layers (6.5 unit cells) of the perovskites and about 20 Å (5.5 unit cells) of vacuum. Since previous first-principles[48, 49] and experimental[50] observations revealed that (001) facets are the most stable for the perovskite systems, both the AO- and $BO_2$-terminated $ABO_3$(001) were investigated herein. Experimentally observed surface reconstructions were disregarded in this work due to their thermodynamic instability at ambient conditions.[51-57] All atoms from five bulk-like middle layers were kept fixed during the relaxations to maintain the cubic symmetry of the deep atomic layers. In order to avoid dipole-dipole interaction due to the cell periodicity,[58] $CO_2$ chemisorption on both sides of the symmetrical slabs was modeled. This model was validated for $SrTiO_3$(001) surfaces in our recent study[31] and compared to the alternative approaches[27, 28] therein. Stability of $CO_2$ adsorption conformations was quantified by the adsorption energy calculated as $E_{Ads} = \left( E(slab + n \cdot CO_2) - E(slab) - n \cdot E(CO_2) \right)/n$, where $E(slab + n \cdot CO_2)$, $E(slab)$, and $E(CO_2)$ are total energies of the slab containing $n$ adsorbed $CO_2$ molecules, clean slab relaxed after $CO_2$ desorption, and free $CO_2$ molecule, respectively. The $CO_2$ coverage ($\Theta$) was determined as the number of $CO_2$ molecules per unit cell of the surface. All local density of states (LDOS) for the slabs are presented in respect to the valence band maxima (VBM) of the corresponding bulk systems, where the alignment was performed based on average electrostatic potentials in cores of B-site cations in the bulk systems and three middle bulk-like atomic layers of the slabs. To demonstrate localization of the surface states, surface state densities were computed as charge densities for the states within energy limits from the conduction band minima (CBM) or VBM levels of the slabs to those of the corresponding bulk compounds.

### Results and discussion

Before investigating an impact of $CO_2$ adsorption on electronic properties of the perovskite surfaces, we first established benchmarks for the bulk compounds, as summarized in Table S1. It was found that the considered compounds have lattice parameters varying from 3.95 Å for $SrTiO_3$ to 4.26 Å for $BaZrO_3$, following the corresponding trends in the crystal ionic radii of the cations ($Sr^{2+}$: 1.32, $Ba^{2+}$: 1.49, $Ti^{4+}$: 0.745, $Zr^{4+}$: 0.86, and $Hf^{4+}$: 0.85 Å for the 6-coordinated ions).[59] These systems have band gap energies between 1.69 and 3.74 eV computed on PBE level for $BaTiO_3$ and $SrHfO_3$, respectively (see Fig. 1a). For both Sr- and Ba-containing perovskites, the energy gaps correlate with electronegativities of B-site cations (Ti: 1.54, Zr: 1.33, and Hf: 1.30 in Pauling scale).[60, 61] This trend reflects different localizations of *d*-like orbitals of B-site cations determining CBM positions in all considered compounds, as shown in Fig. S1.

As briefly mentioned in the methods section, the $Pm\overline{3}m$ crystal phases can undergo distortions at low temperatures.[47] Although this tendency is largely constrained at the first-principles level by the fixation of bulk-like atomic layers (see methods), akin relaxations may still occur at the surfaces. Indeed, our first-principles analysis revealed that the ideal surfaces obtained from the ionic relaxation of as-cleaved slabs are metastable with respect to structural perturbation. Depending on the compound, several distortion patterns were found, as summarized in Fig. S2. The largest energy changes due to the distortion were computed for SrO-terminated $SrZrO_3$(001) and $SrHfO_3$(001), which were stabilized in respect to the ideal surfaces by 30.7 and 19.7 meV/Å$^2$, respectively (see Fig. S3). This behavior can be understood from the analysis of the Goldschmidt tolerance factor,[62] which was developed to determine relative stability and distortion of perovskite structures.[47, 63] In this regard, the smallest tolerance factors were found for $SrZrO_3$ and $SrHfO_3$ (see Table S1), implying their low stability in cubic $Pm\overline{3}m$ phases near absolute zero. It should be noted that despite reaching the local minimum of energy, computed distortion patterns might be artificially stabilized by the imposed slab geometry constraints. However, accounting for the surface distortion is critical to separate adsorption-driven relaxations from those triggered by the broken symmetry. Therefore, the optimized slabs with the most stable distortion patterns were used as reference systems for the $CO_2$ adsorption energy calculations and addressed as clean surfaces henceforth.

By analyzing electronic properties of the stable (001) perovskite surfaces, we found that their band gaps are smaller compared to those of the corresponding bulk systems (see Fig. 1a), in agreement with previous reports.[31, 35, 64, 65] Although these energy gap reductions were observed for most investigated surfaces, the underlying mechanisms differ significantly among them. For the $BO_2$-terminated $ABO_3$(001), the effect is originated from the surface states emerging above the VBM of the bulk systems, as evident from the layer-resolved LDOS and projected surface state density shown in Fig. 1c,e on example of $HfO_2$-terminated $SrHfO_3$(001). These states are populated by *2p*-like electrons of O from the outermost $BO_2$ atomic layer. Importantly, we found that magnitude of the band gap reduction is not strongly dependent on the energy gap of the bulk system, falling within a range of 0.41-0.82 eV for all $BO_2$-terminated surfaces. In contrast, emerging surface states at the AO-terminated $ABO_3$(001) can have two different origins. In both cases, the states emerge below CBM of the bulk compounds but can be localized either at the subsurface $BO_2$ or above the outermost AO atomic layer. The first scenario applies to SrO-terminated $SrTiO_3(001)$, where vacant $3d$-like orbitals from the subsurface Ti atoms (see Fig. 1f) reduce the band gap energy by 0.24 eV. Since this change is much smaller compared to that at the corresponding $TiO_2$-terminated surface (0.78 eV), the band gap reduction effect has less practical importance here. Similar behavior was also found for BaO-terminated $BaTiO_3(001)$, where the surface state density at CBM implies formation of the surface states (see Fig. S5), but no reduction in the overall band gap energy of the slab system was observed, as reflected in Table S2. This result indicates that surface states at BaO-terminated $BaTiO_3(001)$, if present, are within the error bar for electronic structure calculations and cannot alter the surface conductivity significantly. The second scenario leads to the band gap reductions of 0.27-0.39 eV for the Zr-containing and 0.77-0.94 eV for the Hf-containing perovskites. These states are centered about 2 Å above the outermost AO atomic layer and dominated by the orbitals of both A-site cations and O from the outermost AO atomic layer, as shown in Figs. 1b,d for SrO-terminated $SrHfO_3(001)$. Regardless of the origin, these results indicate that the band gap reduction is indeed a common feature of the cubic perovskite family.

Although LDOS in known to provide a good insight into the spatial distribution of electronic properties,[31, 35, 66, 67] the computed quantity can be affected by the limitations of the projection scheme utilizing spherical approximation of the Wigner-Seitz cells.[68] In particular, the surface charge densities estimated from the layer-resolved LDOS (see Fig. 1b,c) can be affected. To evince that the surface states can accommodate enough charge carriers to alter the surface conductivities considerably, we calculated cumulative charges for the surface states by integrating the surface state densities (see methods, Figs. S4 and S5) over the slab volumes and represented them per unit surface area. The obtained quantity can be interpreted as maximum number of charge carriers that can be accommodated per unit area in the states at the surface of interest. As can be seen in Fig. 1g, the cumulative charges are generally larger for the states at the $BO_2$-terminated surfaces, with the exception of $SrHfO_3(001)$. Importantly, for all studied surfaces except BaO-terminated $BaTiO_3(001)$, the charges are greater than $5\times10^{-3}$ $e$/Å$^2$, which is comparable to the electron densities of 2DEG on $SrTiO_3$ surfaces and interfaces.[69-71] This result indicates that these states can indeed alter the surface conductivities of perovskites considerably and even contribute to the formation of the 2DEG, as it was previously discussed for $SrTiO_3(001)$.[35]

To identify the most stable $CO_2$ adsorption conformations, we analyzed all configurations obtained from the screening of over 300 $CO_2$ adsorption positions on SrO- and $TiO_2$-terminated $SrTiO_3(001)$ surfaces presented in our recent work.[31] Most of them were also found after relaxation on the other perovskite surfaces (see Figs. 2a,c and S6), as it could be anticipated from the similar geometries and chemistries of the cubic perovskites. For higher $CO_2$ coverage of 0.50, six and two different coverage modes were analyzed at the AO and $BO_2$ terminations, respectively (see Fig. S7). The most stable adsorption conformations and coverage modes were also investigated accounting for the reduced symmetry due to the surface distortion patterns, as well as by introducing random perturbations to the surface ions. We found that the perturbations can lead to minor stabilization of up to 5 meV/molecule, which signifies thermodynamic stability of the identified conformations and sets an error bar for the $CO_2$ adsorption energy calculations in this work.

The most stable adsorption conformations at $\Theta=0.25$ are nearly identical for all considered materials, with minor differences originated from the surface distortion patterns, as illustrated in Fig. S8. In particular, $CO_2$ interaction with the AO-terminated $ABO_3(001)$ results in the formation of highly stable $CO_3$-like complexes, where C and O atoms of the adsorbed molecules are bonded to the surface oxygen (denoted by $O_S$ here) and two separate A-site cations, respectively (see Fig. 2a). From the thermodynamic perspective, molecular $CO_2$ chemisorption on the AO-terminated $ABO_3(001)$ is highly favorable, as suggested by their low adsorption energies varying from -2.19 eV for $SrZrO_3$ to -1.79 eV for $BaHfO_3$ (see Fig. 2b). Interestingly, for both Sr- and Ba-containing perovskites, the strongest $CO_2$ adsorption was found for the compounds containing Zr. Considering that optimized geometries of the $CO_3$-like complexes are almost independent of the compound (see Table S5), higher stability of $CO_2$ adsorption here can be associated with lower stresses due to the largest lattice constants of the Zr-containing perovskites (see Table S1). At higher $CO_2$ coverage of 0.50, the chemisorption becomes weaker. Specifically, the molecule-molecule interaction increases the $CO_2$ adsorption energies by 0.24-0.45 eV and even results in stabilization of alternative adsorption modes for $BaTiO_3$ and $BaHfO_3$ (see Fig. S9). These results suggest that the AO-terminated $ABO_3(001)$ surfaces can firmly trap atmospheric $CO_2$ by forming a thin layer of $CO_3$-like complexes upon exposure of as-synthesized perovskites to atmospheric air.

CO$_2$ interaction with the BO$_2$-terminated surfaces also leads to the formation of CO$_3$-like complexes, where C and O atoms of the adsorbed molecules are bonded to the surface oxygen (O$_S$) and B-site atoms, respectively, as shown in Fig. 2c. The C-O$_S$ bonds and O-C-O angles of the formed CO$_3$-like complexes are larger here as compared to those at the AO terminations (see Table S5). The CO$_2$ adsorption energies for the BO$_2$-terminated surfaces at $\Theta=0.25$ are within a range from -1.75 eV for BaHfO$_3$(001) to -1.24 eV for SrTiO$_3$(001), implying less stable chemisorption compared to that on the AO terminations (see Fig. 2d). The further increase in CO$_2$ coverage to 0.50 increases adsorption energy by less than 0.1 eV, and therefore, the interaction between the chemisorbed molecules on the BO$_2$ terminations is weaker.

The CO$_2$ adsorption on the perovskite surfaces alters their electronic properties by suppressing the surface states. Specifically, energy gaps of the AO-terminated ABO$_3$(001) slabs increase considerably to the corresponding bulk values at $\Theta=0.25$ of CO$_2$ (see Fig. 3a). This change is due to the suppression of the surfaces states at the CBM levels of the clean surfaces, as illustrated in Figs. S11-S26. It should be noted that the energy gaps of some slabs containing CO$_2$ molecules exceed the corresponding bulk values, which can be attributed to the model geometry constraints and may be eliminated using thicker slabs. These electronic properties persist at $\Theta=0.50$ as well, as shown in Fig. 3a,c. For the BO$_2$-terminated surfaces, the band gap energies at $\Theta=0.25$ are 0.31-0.46 eV larger than those for the corresponding clean slabs but are 0.10-0.39 eV smaller than the bulk values, as shown in Fig. 3b. Instead, the surface states at CBM levels are fully suppressed at $\Theta=0.50$ (see Fig. 3d), suggesting a more gradual change of the surface electronic properties upon CO$_2$ chemisorption on the BO$_2$-terminated surfaces. However, more detailed analysis within a wider CO$_2$ coverage range is needed to predict the concentration-dependent sensing response.

Since sensing response is experimentally defined as a ratio of material resistances in different environments,[2, 3] a similar first-principles descriptor is needed to quantify the band gap modulation effect. Considering that CO$_2$ adsorption leads to the complete suppression of the surface states at $\Theta=0.50$, the absolute difference in the band gaps of the bulk and clean slab systems can be employed for such quantification. Although this parameter ignores possible changes in the carrier mobilities, it captures a dominant role played by the band gap changes and accounts for the exponential dependence of the charge carrier concentrations on the energy gap. It should be noted, however, that this approach can only be used to compare sensitivities of the surface layers, whereas electrical conductance of the bulk-like regions is implicitly neglected.

To examine the sensitivity of the proposed parameter to the computational approach, we also computed the band gap reductions for the cubic perovskite surfaces using the hybrid HSE06 functional. As can be seen in Fig. S10, despite the significant difference between the band gap energies of the bulk perovskites computed using PBE and HSE06, the magnitude of the band gap reduction for all investigated $BO_2$-terminated surfaces does not strongly depend on the employed formalism. Specifically, the maximum difference of 0.15 eV only was obtained for $HfO_2$-terminated $SrHfO_3(001)$. Furthermore, no drastic change in the surface electronic properties was found for AO-terminated $BaTiO_3(001)$ and $SrTiO_3(001)$. In contrast, surface states at all other AO-terminated surfaces reduce the band gap energies more drastically when analyzed using the HSE06 approach. Here, the computed differences in the band gap reductions were in 0.34-0.39 eV range for all four representative surfaces. Although the HSE06 approach changes the values of band gap reduction, it does not alter general trends deduced for the surface state, as evident from Fig. S10. Therefore, we can conclude that the PBE results can be employed for the analysis of chemiresistive $CO_2$ sensing response in the considered perovskite systems.

The computed band gap differences for the cubic perovskites are shown in Fig. 3e. The largest band gap modulations of 0.94 and 0.77 eV at AO-terminated $SrHfO_3(001)$ and $BaHfO_3(001)$ imply their strong potential for chemiresistive $CO_2$ detection. It should be noted, however, that the use of these perovskites can be limited by their relatively large band gap energies, which lead to high electrical resistance, thus making conductivity measurements practically difficult.[72] At the same time, $TiO_2$-terminated $BaTiO_3(001)$ and $SrTiO_3(001)$ have the strongest $CO_2$ detection potentials among the considered $BO_2$-terminated surfaces, which is due to the energy gap reductions of 0.82 and 0.78 eV, respectively. It worth mentioning that $BaTiO_3$-containing composites have already been employed as active materials for chemiresistive $CO_2$ detection with great success,[7, 8, 18-21, 23-26] while $SrTiO_3$ nanostructures are effectively used to sense various reactive gases.[73-76] Moreover, the use of Ti-containing perovskites is further promoted by their relatively small energy gaps, which can result in lower electrical resistance, and thus simplify electrical circuit needed for the future devices.[72]

Suppression of the surface states represents a new approach to chemiresistive $CO_2$ detection, which may challenge classical ionization mechanism adopted for detection of reactive gases.[2-4] However, to unleash its full capability, in practice, usage of nanoscale materials with high surface-to-bulk ratios is required due to the strong spatial confinement of the surface states. Apart from that, when choosing optimal perovskite material for $CO_2$ detection, low band gap perovskites might be favored as small energy gap generally leads to higher conductivity preferable for the conductivity measurements. It may also be valuable to hypothesize an impact the band gap modulation effect on the optical properties of nanostructured perovskites. Since band gap energy is among the main parameters determining light absorption characteristics of the material, the band gap reduction can also be utilized to develop optical $CO_2$ sensors. However, such detectors may require compound with the band gap energy in the visible spectral region, whereas more research on the technical aspects of the perovskite-based optical sensors is needed.

Although this work considers the band gap modulation effect from the gas sensing perspective, it can also have considerable impact in other related fields, including catalysis and photocatalysis. From one side, layer of $CO_3$-like complexes at the most stable (001) surfaces can form a barrier preventing further molecular chemisorption, while from the other side, the formation of such complexes might be a first step in the $CO_2$ reduction pathways.[6] Moreover, adsorption-induced change in the surface electronic properties may signify a need to reconsider both concentration and localization of photoelectrons as well as overpotentials between the band gap edges of the perovskite-based catalysts and target redox potentials of the catalytic reactions. Finally, since the perovskites are often used as substrate materials for epitaxial growth,[50] the presence of chemisorbed molecules and the corresponding changes in the surface electronic properties may affect the composition and band alignment at the formed interfaces. Because of this, care should be taken to control the concentration of $CO_2$ species adsorbed on the perovskite substrates during the epitaxial growth.

**Conclusions**

In this work, the effect of band gap modulation by $CO_2$ adsorption on (001) surfaces of six different cubic perovskites with the general formula of $ABO_3$ (A = Ba, Sr and B = Ti, Zr, Hf) was assessed using first-principles methods. The results reveal an appearance of surface states at most clean surfaces. For the $BO_2$-terminated $ABO_3$(001), the states emerge above the bulk VBM levels and are confined to the outermost $BO_2$ atomic layer, reducing the energy gaps of the slab systems by 0.41-0.82 eV. For the AO-terminated $ABO_3$(001), the surface states appear below the bulk CBM level and are localized either at the subsurface $BO_2$ layer or about 2 Å above the outermost AO atomic layer. The first scenario applies to $TiO_2$-terminated $SrTiO_3$(001) and results in the band gap reduction of 0.24 eV, while the second is observed for the Zr- and Hf-containing compounds and leads to the narrowing of the band gaps by 0.27-0.39 and 0.77-0.94 eV, respectively. Moreover, the surface electronics of the studied perovskites can change upon the $CO_2$ chemisorption, which is accompanied by the formation of stable $CO_3$-like complexes. For most AO-terminated surfaces, $\Theta=0.25$ coverage is sufficient to suppress the surface states completely, whereas for the $BO_2$-terminated surface, the same effect is only achieved at the higher $\Theta=0.50$ coverage. Based on these results, we can conclude that $TiO_2$-terminated $BaTiO_3(001)$ and $SrTiO_3(001)$ are the most prospective $CO_2$ sensing materials among all studied systems due to the high computed band gap reductions of 0.82 and 0.78 eV, respectively, and relatively low band gap energies of their bulk forms. The AO-terminated $SrHfO_3(001)$ and $BaHfO_3(001)$ are also promising candidates for chemiresistive $CO_2$ detection due to even larger band gap modulations of 0.94 and 0.77 eV, respectively. However, the application of these surfaces can be limited by practical difficulties associated with measuring low conductivities inherent to the compounds with large band gaps.

**Electronic supplementary information (ESI) available**

Additional information on the surface geometry, electronic properties, and optimized slab structures of the perovskite systems are provided.

**Conflicts of interest**

There are no conflicts to declare.

**Acknowledgments**

The authors acknowledge the support from the SUTD-ZJU (ZJURP1200101), MOE Tier2 (T2 MOE1201-Singapore), and Research Council of Norway (contracts: 221469, 250346, 251131). Most of this work was performed on the Abel cluster, owned by the University of Oslo and the Norwegian Metacenter for Computational Science (NOTUR), and operated by the Department for Research Computing at USIT, the University of Oslo IT-department. The authors also acknowledge PRACE for awarding access to resource MareNostrum based in Spain at BSC-CNS.

# Figures and Captions

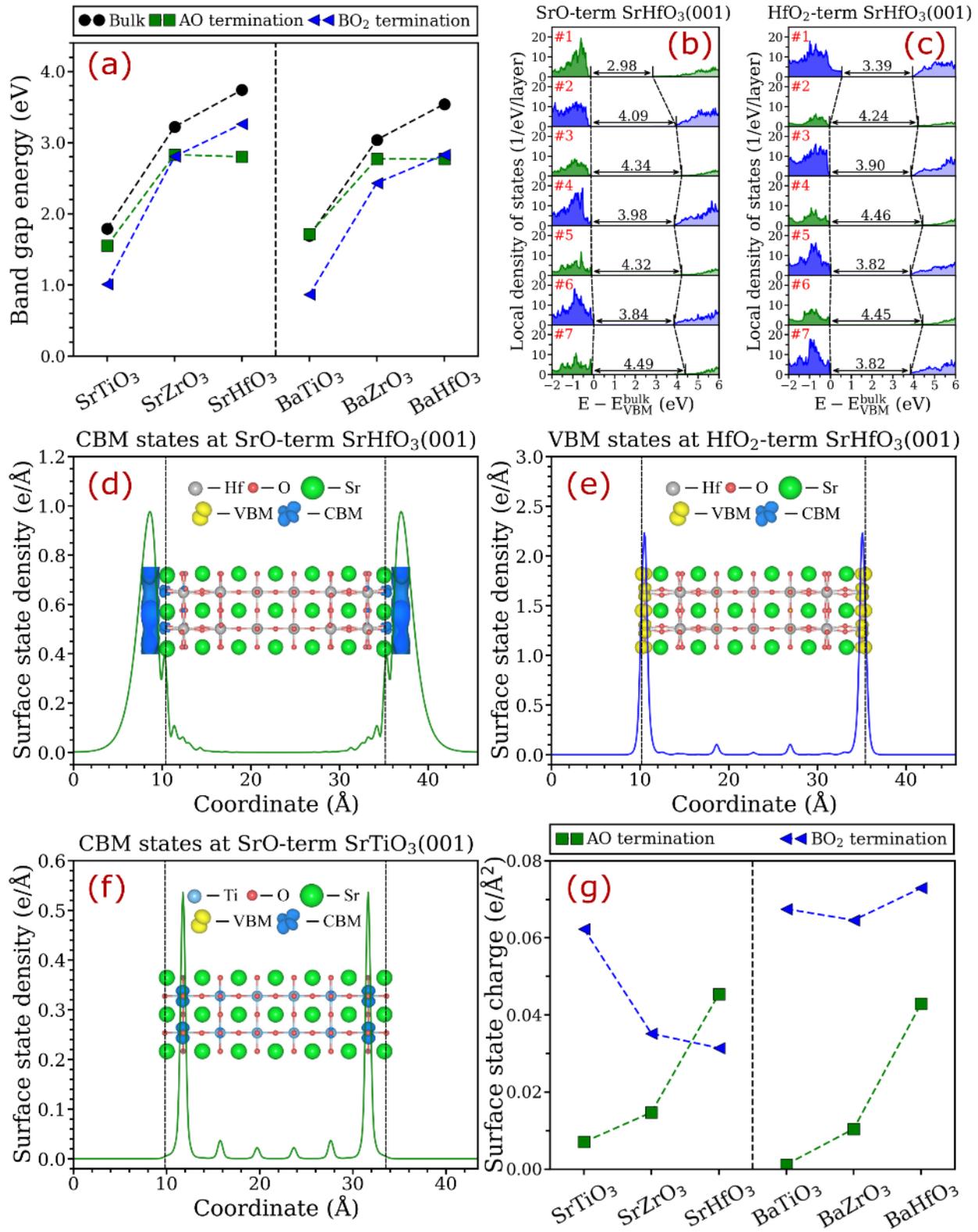

Figure 1. Electronic properties of (001) cubic perovskite surfaces. (a) Computed band gap energies of the bulk and slab $ABO_3$ systems. Typical layer-resolved local density of states (LDOS)

illustrated for clean (b) SrO- and (c) HfO$_2$-terminated SrHfO$_3$(001) surfaces (only one half of the slab is presented due to the symmetry; indexing is from the surface to the middle layers; numbers represent effective band gaps for the atomic layers computed form the layer-resolved LDOS neglecting the population densities below 0.3 1/eV/layer). More details on electronic properties of the other considered surfaces are given in Figs. S11-S26. In-plane averaged surface state densities for (d) SrO- and (e) HfO$_2$-terminated SrHfO$_3$(001), as well as (f) SrO-terminated SrTiO$_3$(001) surfaces (see Figs. S4 and S5 for the surface state localization at the other perovskite surfaces). (g) Cumulative charges for surface states at the ABO$_3$(001).

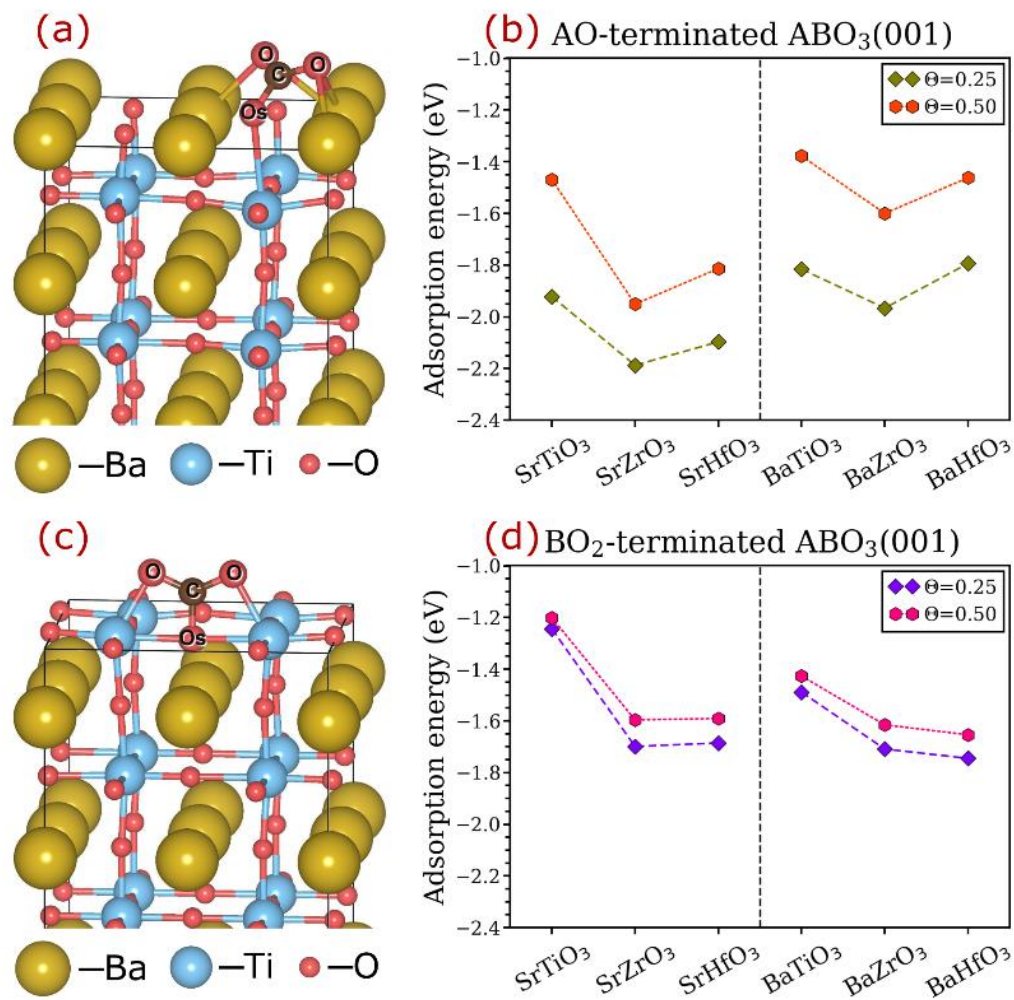

Figure 2. Most stable CO$_2$ adsorption conformations on (a) BaO- and (c) TiO$_2$-terminated BaTiO$_3$(001) surfaces corresponding to $\Theta$=0.25. More details on CO$_2$ adsorption conformations for the other perovskite surfaces, metastable configurations, and higher ($\Theta$=0.50) coverage can be found in Figs. S6-S9. Computed CO$_2$ adsorption energies for the (b) AO- and (d) BO$_2$-terminated ABO$_3$(001) surfaces. The adsorption energies are tabulated in Tables S3 and S4.

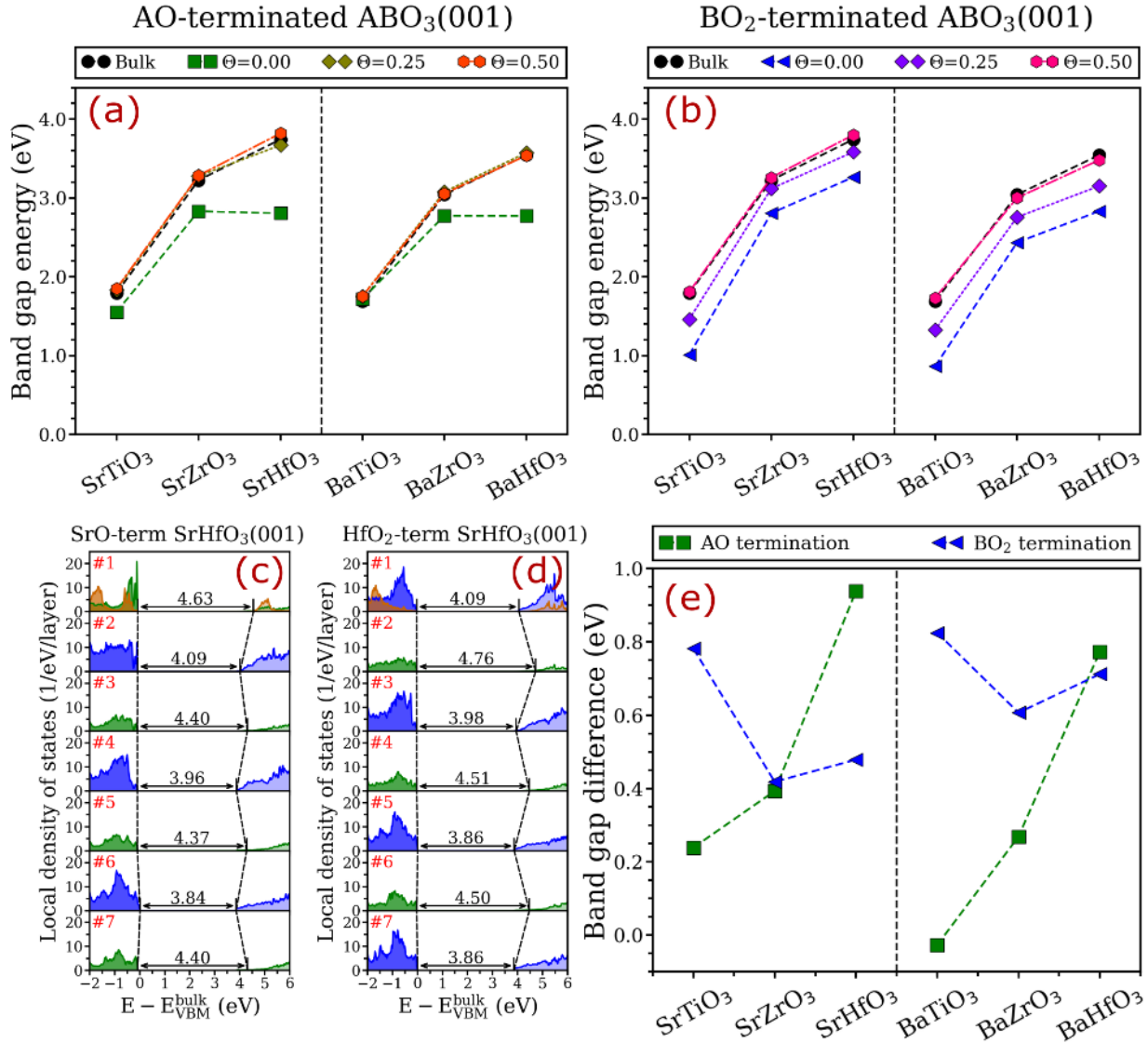

Figure 3. Computed band gap energies of the (a) AO- and (b) $BO_2$-terminated $ABO_3$(001) surfaces containing zero ($\Theta$=0.00), one ($\Theta$=0.25), and two ($\Theta$=0.50) adsorbed $CO_2$ molecules. Band gap energies of the corresponding bulk systems are given for comparison. Typical layer-resolved local density of states (LDOS) for (c) SrO- and (d) $HfO_2$-terminated $SrHfO_3$(001) containing $\Theta$=0.50 $CO_2$ coverage (only one half of the slab is presented due to the symmetry; indexing is from the surface to the middle layers; numbers represent effective band gaps for the atomic layers computed form the layer-resolved LDOS neglecting the population densities below 0.3 1/eV/layer). (e) Computed difference in the band gap energies of the bulk and clean slab perovskite systems. More details on electronic properties of the other perovskite systems containing chemisorbed $CO_2$ are given in Figs. S11-S26.

**Supplementary information:**

**Suppression of Surfaces States at Cubic Perovskite (001) Surfaces by CO₂ Adsorption**

*Kostiantyn V. Sopiha[1], Oleksandr I. Malyi[2*], Clas Persson[2], Ping Wu[1*]*

1 – Entropic Interface Group, Engineering Product Development, Singapore University of Technology and Design, 8 Somapah Road, 487372 Singapore, Singapore

2 – Centre for Materials Science and Nanotechnology, Department of Physics, University of Oslo, P. O. Box 1048 Blindern, NO-0316 Oslo, Norway

E-mails: oleksandrmalyi@gmail.com (O.I.M), wuping@sutd.edu.sg (W.P)

**List of tables**

**List of figures**

## List of structures

Table S1. Consolidated properties of the cubic perovskites. Goldschmidt tolerance factors were calculated using ionic radii of 12-fold coordinated A-site cations in +2 state, 6-fold coordinated B-site cation in +4 state, and 6-fold coordinated $O^{2-}$, as tabulated by Shannon.[59]

| Compound | Lattice constant (Å) | PBE band gap (eV) | Goldschmidt tolerance factor |
|---|---|---|---|
| $SrTiO_3$ | 3.95 | 1.79 | 1.002 |
| $SrZrO_3$ | 4.20 | 3.22 | 0.947 |
| $SrHfO_3$ | 4.14 | 3.74 | 0.952 |
| $BaTiO_3$ | 4.04 | 1.69 | 1.062 |
| $BaZrO_3$ | 4.26 | 3.04 | 1.004 |
| $BaHfO_3$ | 4.21 | 3.54 | 1.009 |

Table S2. Consolidated properties of the clean $ABO_3(001)$ surfaces. Stabilization energies refer to the differences in surface energies for undistorted relaxed surfaces and those with the most stable distortion patterns.

| Compound | Termination | PBE band gap (eV) | PBE band gap reduction (eV) | Stabilization energy (meV/Å²) |
|---|---|---|---|---|
| SrTiO$_3$ | SrO | 1.55 | 0.24 | 0.0 |
| | TiO$_2$ | 1.01 | 0.78 | 2.0 |
| SrZrO$_3$ | SrO | 2.83 | 0.39 | 30.7 |
| | ZrO$_2$ | 2.81 | 0.41 | 12.8 |
| SrHfO$_3$ | SrO | 2.80 | 0.94 | 19.7 |
| | HfO$_2$ | 3.26 | 0.48 | 6.8 |
| BaTiO$_3$ | BaO | 1.72 | -0.03* | 2.2 |
| | TiO$_2$ | 0.87 | 0.82 | 10.5 |
| BaZrO$_3$ | BaO | 2.77 | 0.27 | 0.9 |
| | ZrO$_2$ | 2.44 | 0.60 | 0.0 |
| BaHfO$_3$ | BaO | 2.77 | 0.77 | 0.0 |
| | HfO$_2$ | 2.83 | 0.71 | 0.0 |

*within the accuracy limit*

Table S3. Consolidated properties of the $ABO_3(001)$ surfaces at $\Theta=0.25$ $CO_2$ coverage.

| Compound | Termination | PBE band gap (eV) | $CO_2$ adsorption energy (eV) |
|---|---|---|---|
| SrTiO$_3$ | SrO | 1.83 | -1.92 |
| | TiO$_2$ | 1.46 | -1.24 |
| SrZrO$_3$ | SrO | 3.28 | -2.19 |
| | ZrO$_2$ | 3.12 | -1.70 |
| SrHfO$_3$ | SrO | 3.67 | -2.10 |
| | HfO$_2$ | 3.58 | -1.69 |
| BaTiO$_3$ | BaO | 1.76 | -1.82 |
| | TiO$_2$ | 1.33 | -1.49 |
| BaZrO$_3$ | BaO | 3.08 | -1.97 |
| | ZrO$_2$ | 2.76 | -1.71 |
| BaHfO$_3$ | BaO | 3.57 | -1.79 |
| | HfO$_2$ | 3.15 | -1.75 |

Table S4. Consolidated properties of the $ABO_3(001)$ surfaces at $\Theta=0.50$ $CO_2$ coverage.

| Compound | Termination | PBE band gap (eV) | $CO_2$ adsorption energy (eV) |
|---|---|---|---|
| SrTiO$_3$ | SrO | 1.85 | -1.47 |
| | TiO$_2$ | 1.81 | -1.20 |
| SrZrO$_3$ | SrO | 3.28 | -1.95 |
| | ZrO$_2$ | 3.26 | -1.60 |
| SrHfO$_3$ | SrO | 3.82 | -1.81 |
| | HfO$_2$ | 3.80 | -1.59 |
| BaTiO$_3$ | BaO | 1.75 | -1.38 |
| | TiO$_2$ | 1.73 | -1.43 |
| BaZrO$_3$ | BaO | 3.05 | -1.60 |
| | ZrO$_2$ | 3.00 | -1.62 |
| BaHfO$_3$ | BaO | 3.54 | -1.46 |
| | HfO$_2$ | 3.48 | -1.65 |

Table S5. Geometries of the $CO_3$-like complexes formed upon $CO_2$ chemisorption on the $ABO_3(001)$ surfaces of cubic perovskites at $\Theta=0.25$.

| Compound | Termination | Average C-O bond length (Å) | C-O$_S$ bond length (Å) | O-C-O angle (degree) |
|---|---|---|---|---|
| SrTiO$_3$ | SrO | 1.29 | 1.33 | 122.2 |
| | TiO$_2$ | 1.27 | 1.37 | 130.8 |
| SrZrO$_3$ | SrO | 1.28 | 1.35 | 122.2 |
| | ZrO$_2$ | 1.27 | 1.36 | 130.7 |
| SrHfO$_3$ | SrO | 1.28 | 1.35 | 122.7 |
| | HfO$_2$ | 1.27 | 1.37 | 131.3 |
| BaTiO$_3$ | BaO | 1.28 | 1.34 | 123.3 |
| | TiO$_2$ | 1.27 | 1.35 | 130.6 |
| BaZrO$_3$ | BaO | 1.28 | 1.35 | 122.9 |
| | ZrO$_2$ | 1.27 | 1.37 | 129.2 |
| BaHfO$_3$ | BaO | 1.28 | 1.36 | 123.7 |
| | HfO$_2$ | 1.27 | 1.37 | 129.9 |

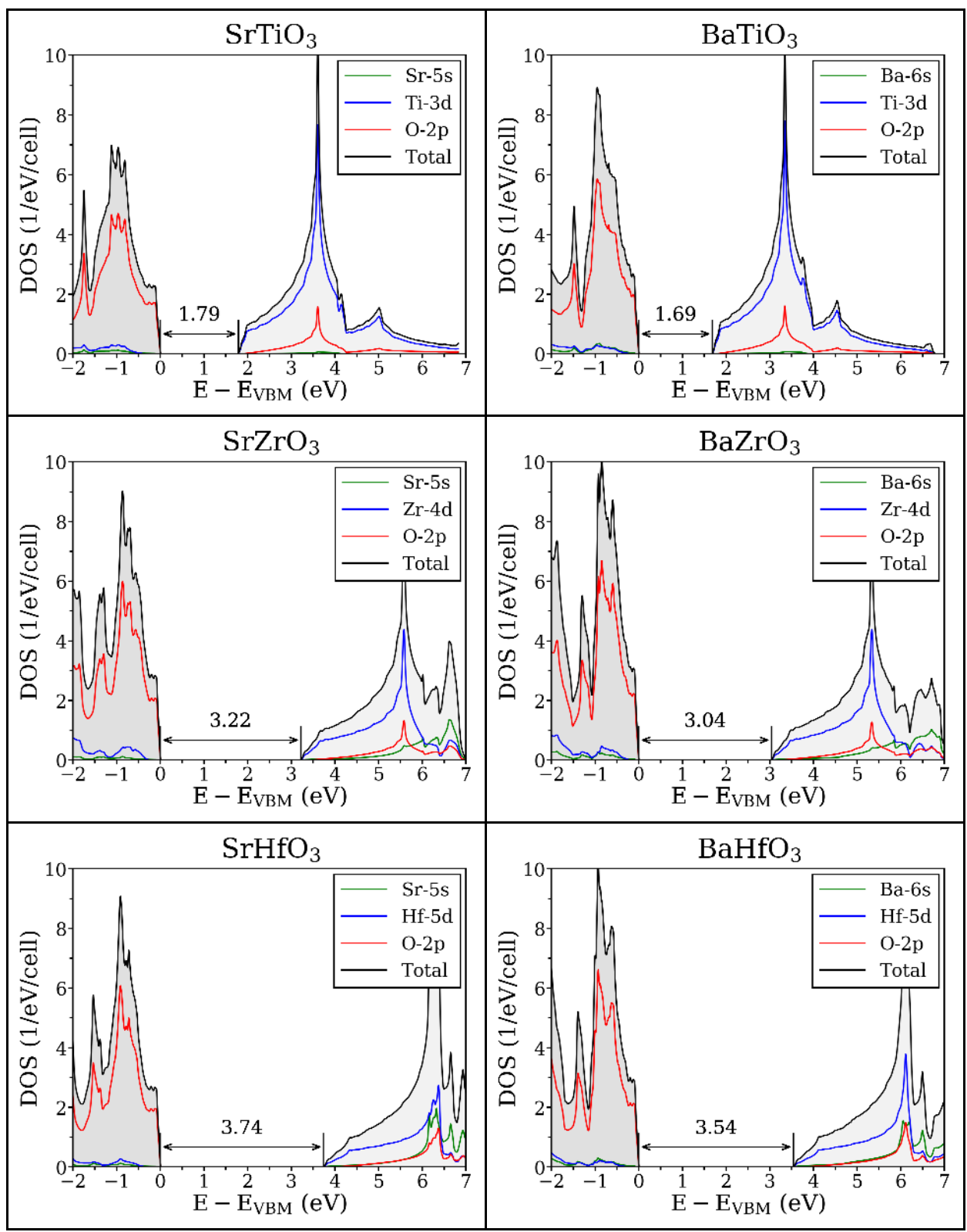

Figure S1. Projected density of states (DOS) for all considered cubic perovskite materials.

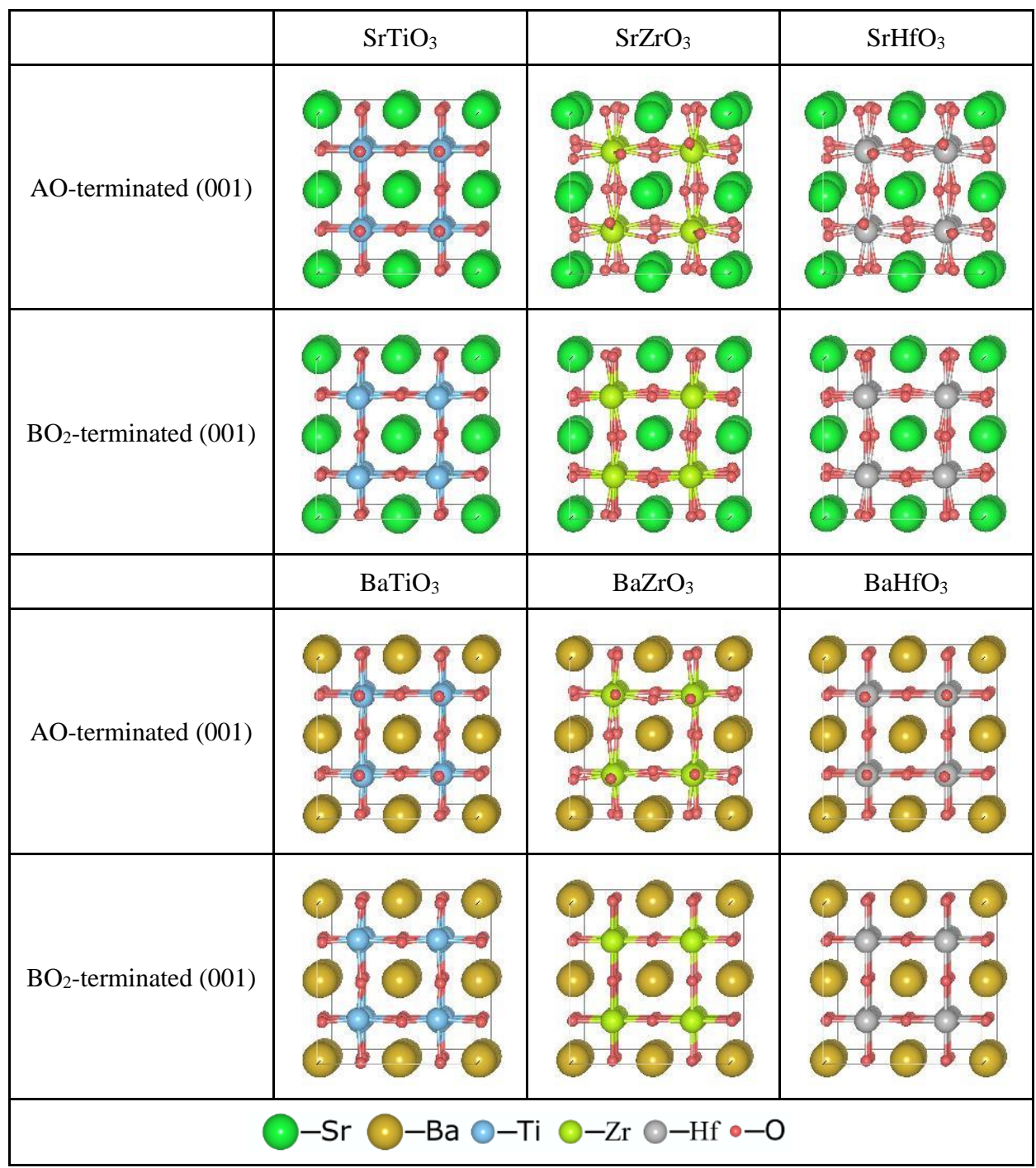

Figure S2. Most stable surface distortion for all considered cubic perovskite surfaces.

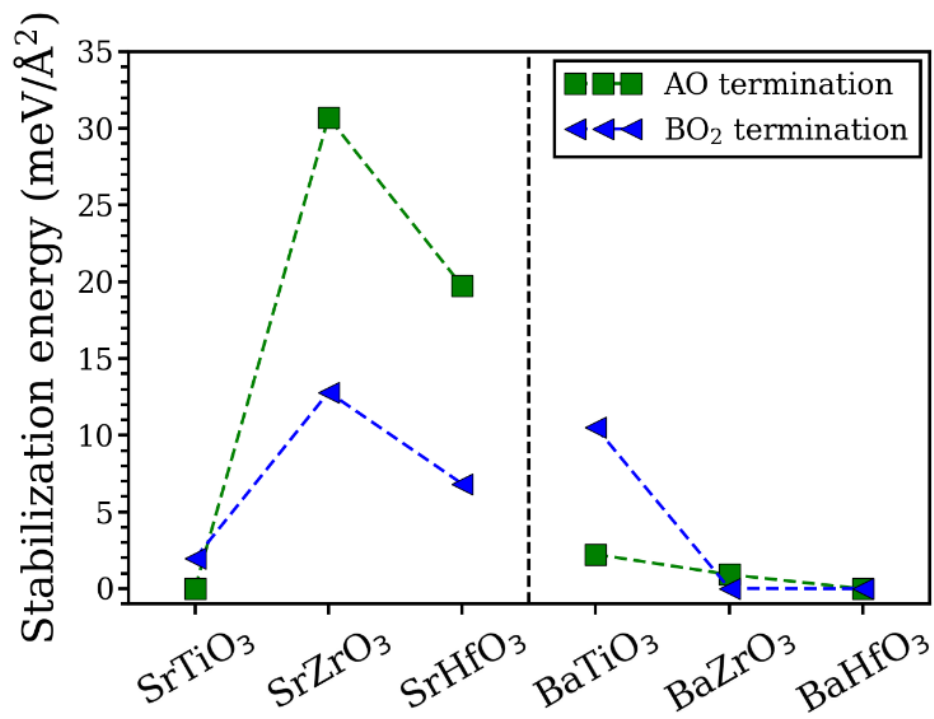

Figure S3. Stabilization energies for different perovskite ABO$_3$(001) surfaces. Stabilization energies were calculated as differences in surface energies for undistorted relaxed surfaces and those with the most stable distortion patterns. The values are also tabulated in Table S2.

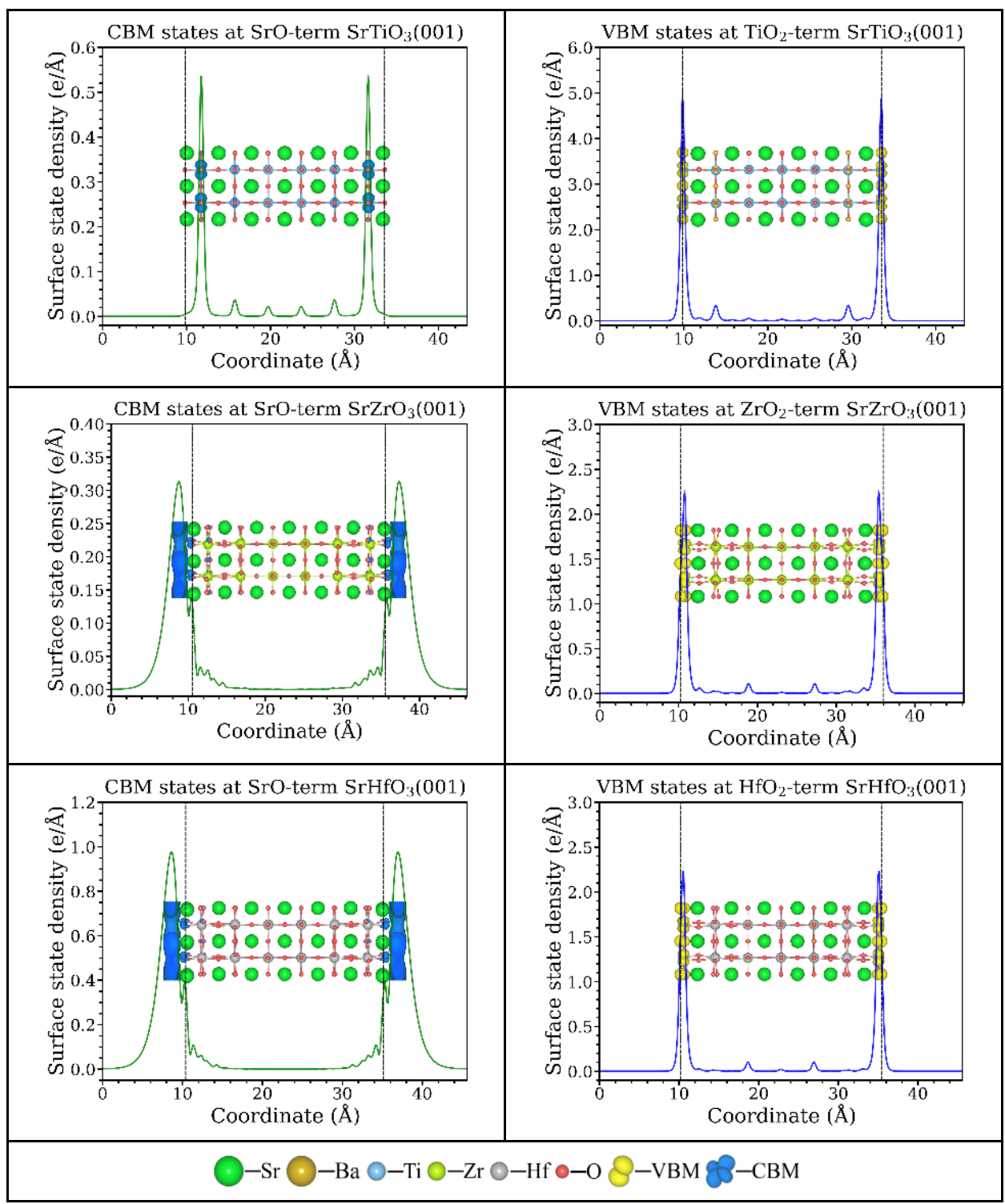

Figure S4. Localization of the surface states at the Sr-containing ABO₃(001). The surface state densities were computed as charge densities for the states within energy limits from the conduction band minima (CBM) or valence band maxima (VBM) levels of the slabs to those of the corresponding bulk compounds.

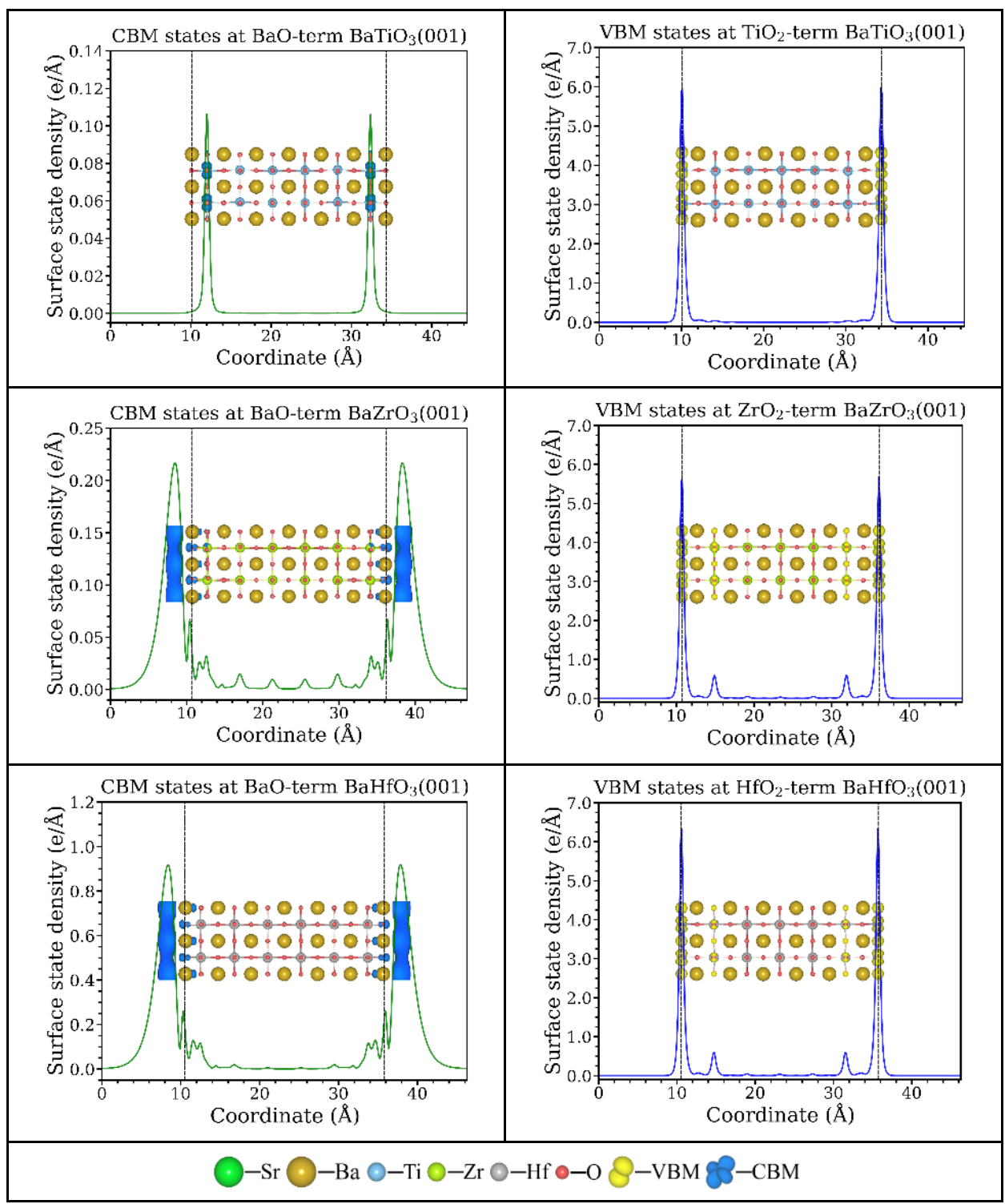

Figure S5. Localization of the surface states at the Ba-containing ABO₃(001). The surface state densities were computed as charge densities for the states within energy limits from the conduction band minima (CBM) or valence band maxima (VBM) levels of the slabs to those of the corresponding bulk compounds.

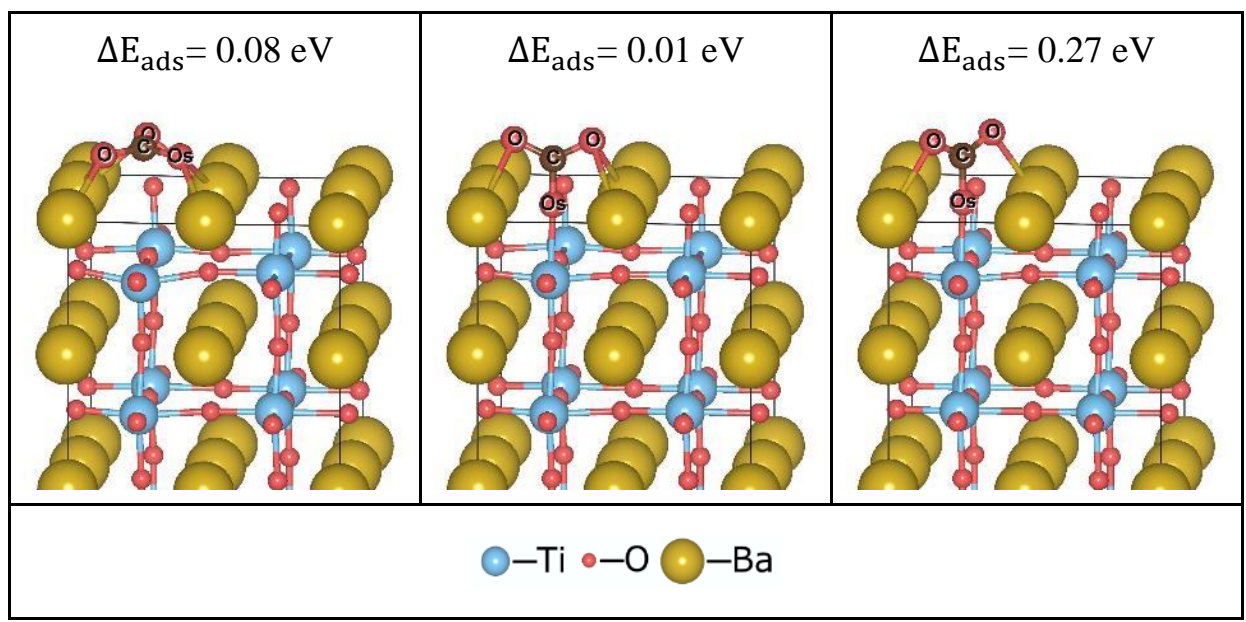

Figure S6. Metastable $CO_2$ adsorption conformations observed on AO-terminated BaTiO$_3$(001) surface. All systems correspond to $\Theta$=0.25 $CO_2$ coverage. $\Delta E_{ads}$ denotes energy difference between the most stable and specific metastable $CO_2$ adsorption configuration.

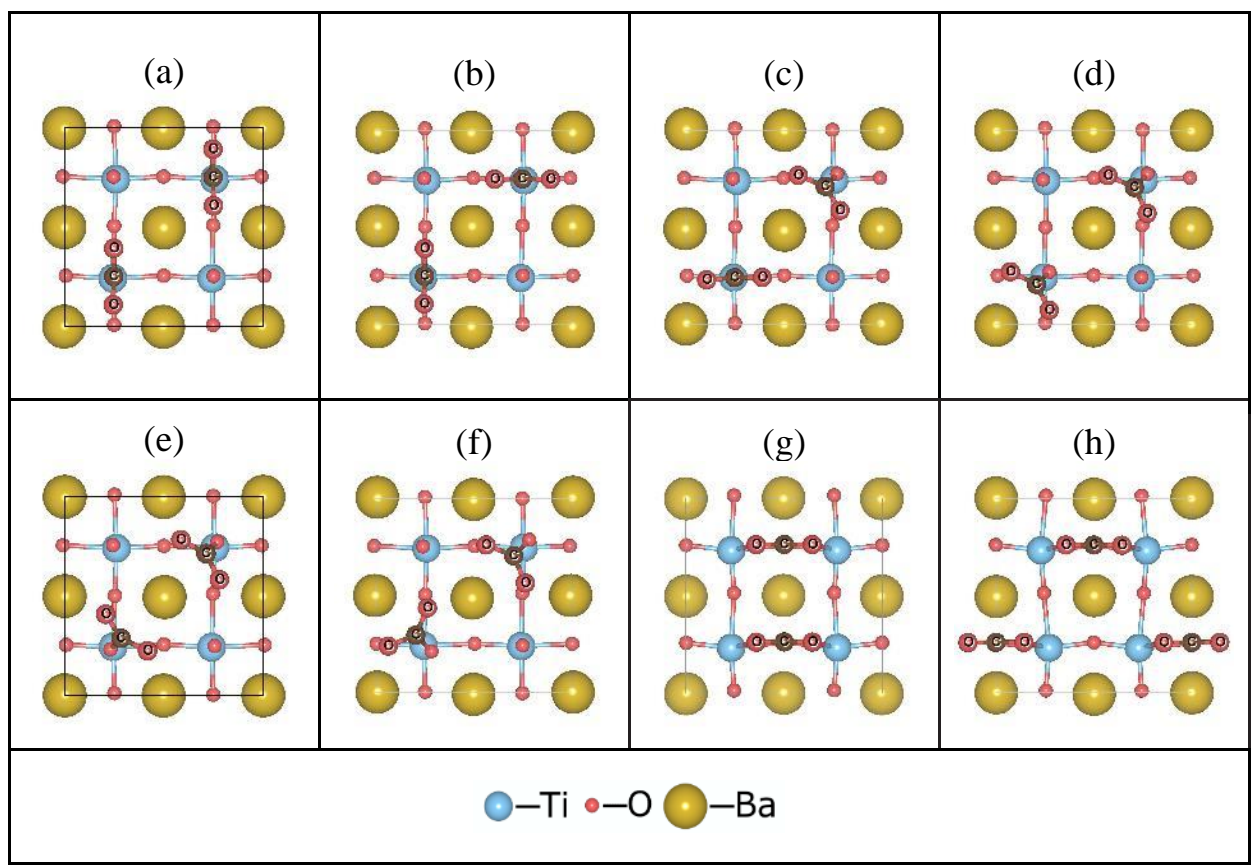

Figure S7. Considered modes for CO$_2$ adsorption demonstrated for (a-f) BaO- and (g-h) TiO$_2$-terminated BaTiO$_3$(001) surfaces. All systems correspond to $\Theta$=0.50 CO$_2$ coverage.

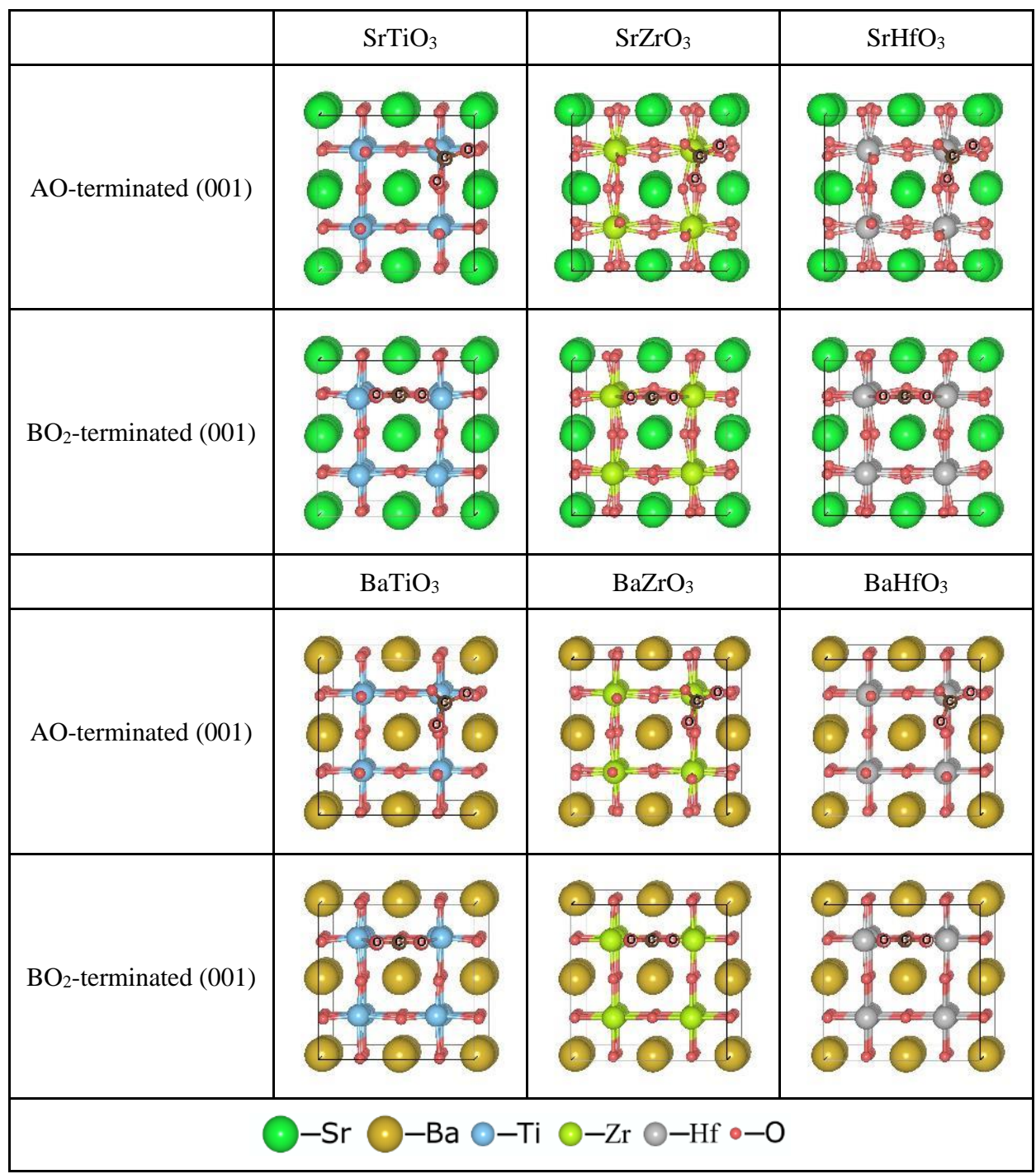

Figure S8. Surface relaxations of all considered perovskite surfaces upon $CO_2$ adsorption at $\Theta$=0.25 coverage.

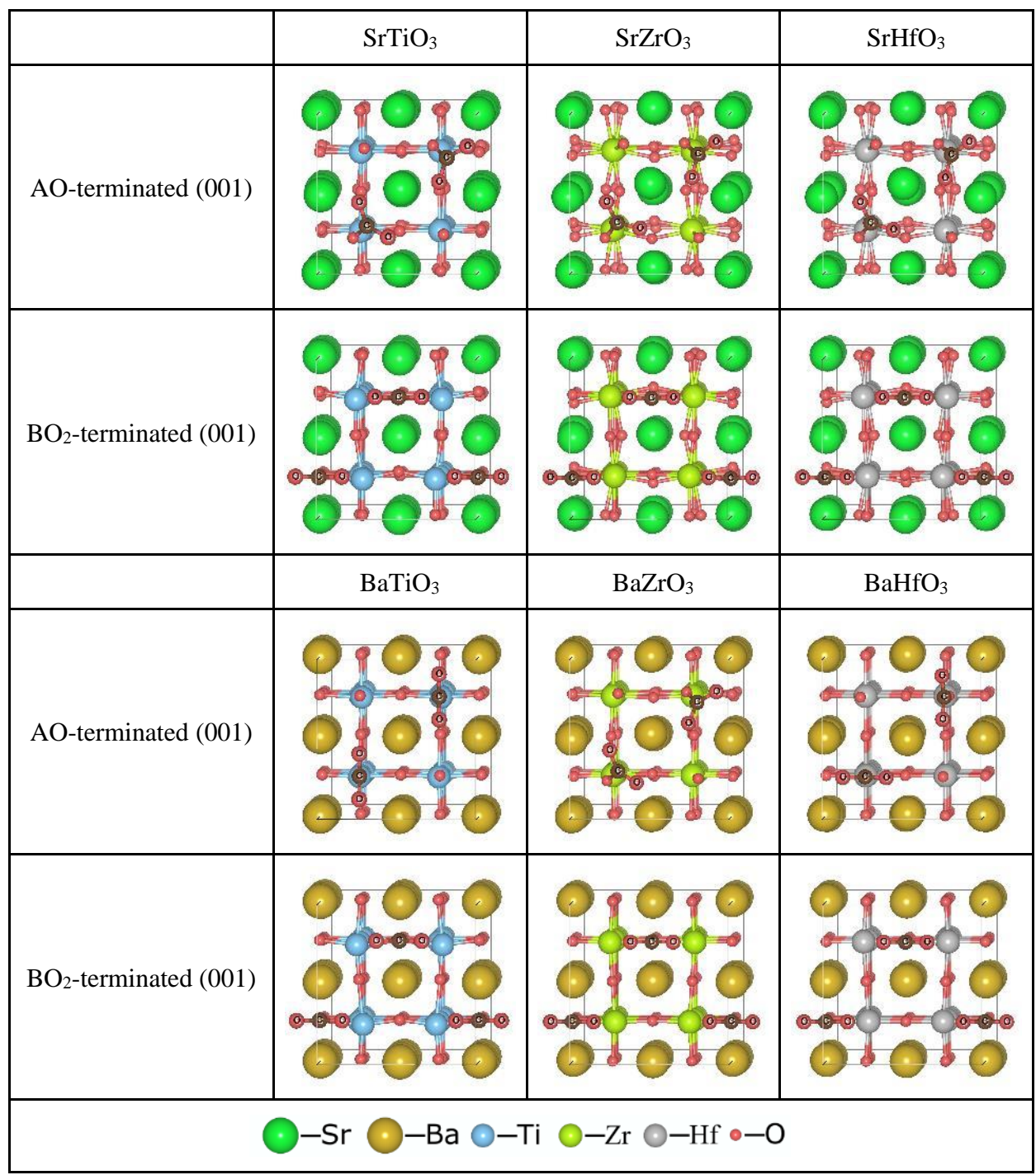

Figure S9. Surface relaxations of all considered perovskite surfaces upon $CO_2$ adsorption at $\Theta$=0.50 coverage.

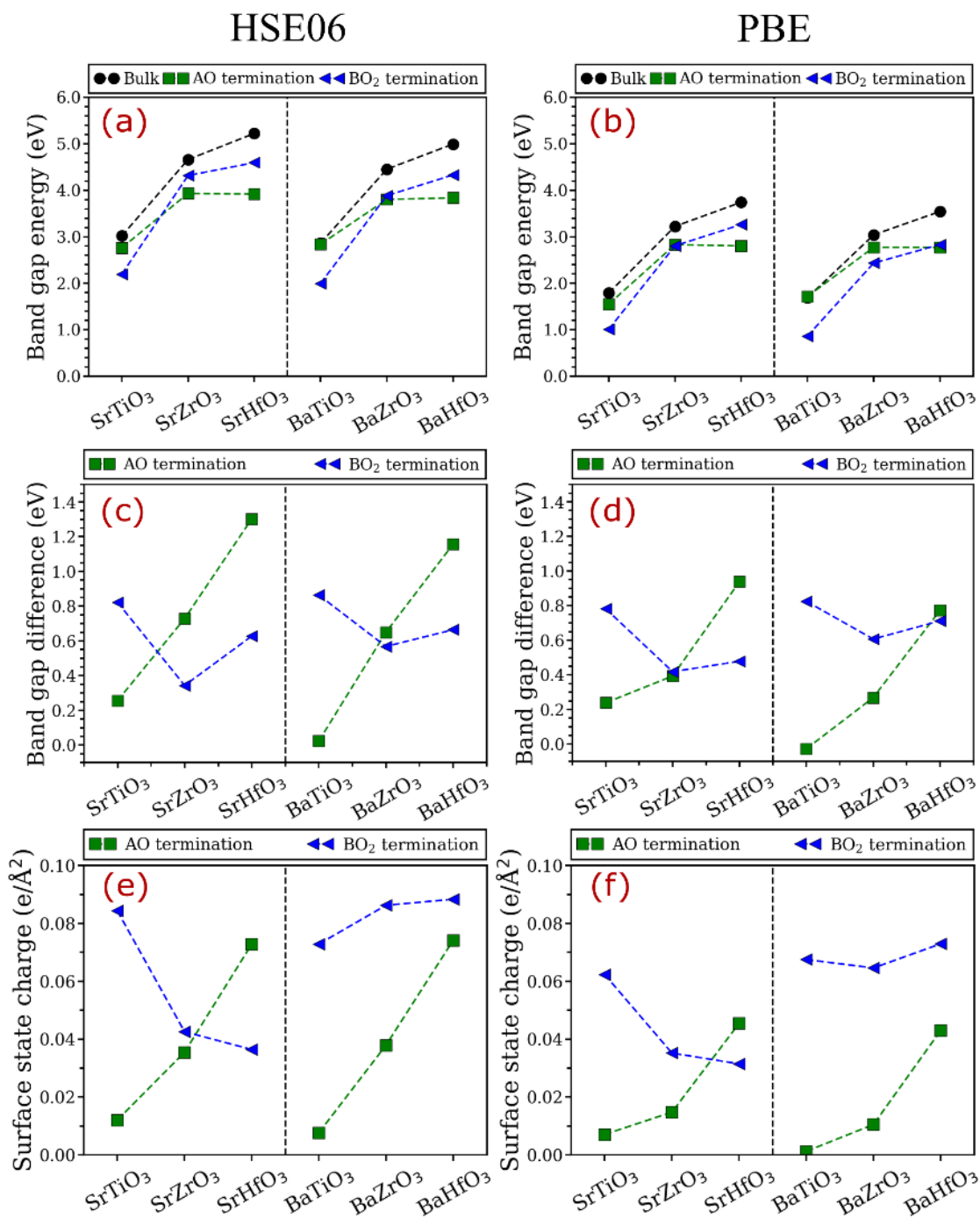

Figure S10. Comparison of (a,b) band gap energies, (c,d) band gap reductions, and (e,f) cumulative charges for the surface states at the cubic perovskites computed using (a,c,e) hybrid HSE06 and (b,d,f) PBE functionals. The HSE06 calculations were carried out using 2×2×1 Monkhorst-Pack grid on the structures optimized with PBE functional. Hafnium $5d^26s^2$, zirconium $4d^25s^2$, and titanium $3d^24s^2$ electrons treated explicitly for the HSE06 analysis. All other parameters were identical in both HSE06 and PBE calculations (see methods).

## SrO-terminated SrTiO$_3$(001)

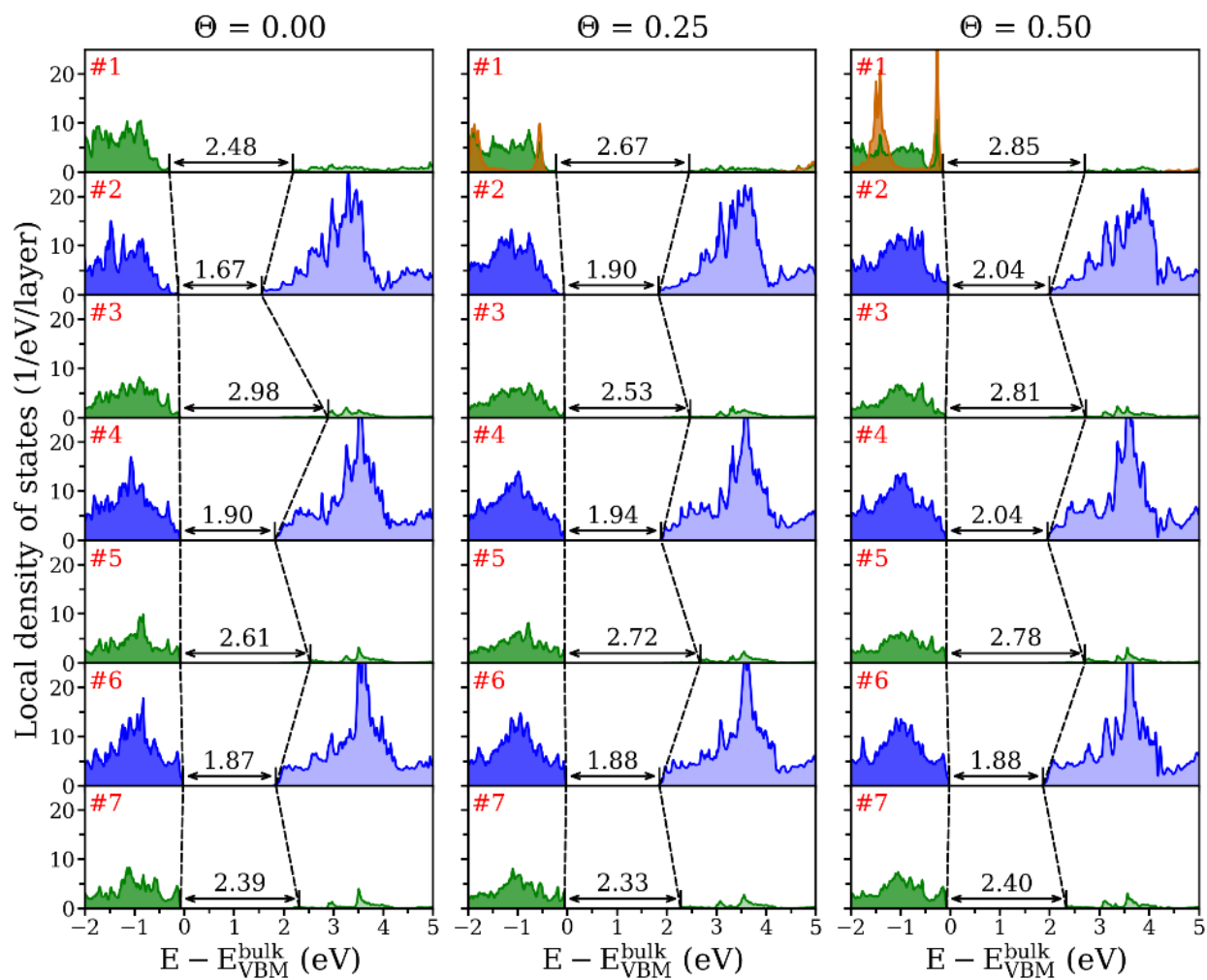

Figure S11. Layer-resolved local density of states (LDOS) for SrO-terminated SrTiO$_3$(001) at $\Theta$=0.00 (clean surface), $\Theta$=0.25, and $\Theta$=0.50 CO$_2$ coverages (only one half of the slab is presented due to the symmetry; indexing is from the surface to the middle layers; numbers represent effective band gaps for the atomic layers computed form LDOS neglecting the population densities below 0.3 1/eV/layer).

## SrO-terminated SrZrO$_3$(001)

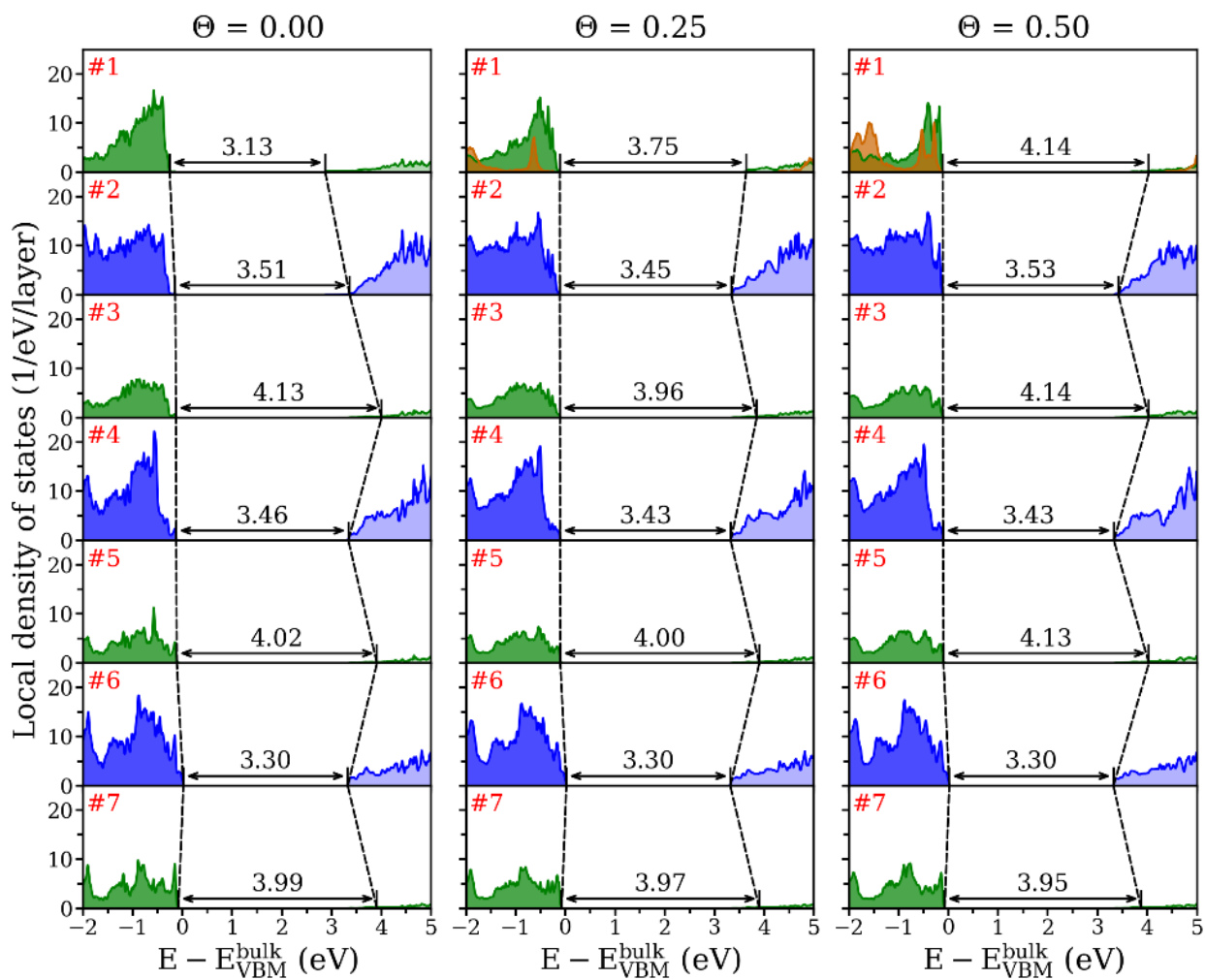

Figure S12. Layer-resolved local density of states (LDOS) for SrO-terminated SrZrO$_3$(001) at Θ=0.00 (clean surface), Θ=0.25, and Θ=0.50 CO$_2$ coverages (only one half of the slab is presented due to the symmetry; indexing is from the surface to the middle layers; numbers represent effective band gaps for the atomic layers computed form LDOS neglecting the population densities below 0.3 1/eV/layer).

## SrO-terminated SrHfO₃(001)

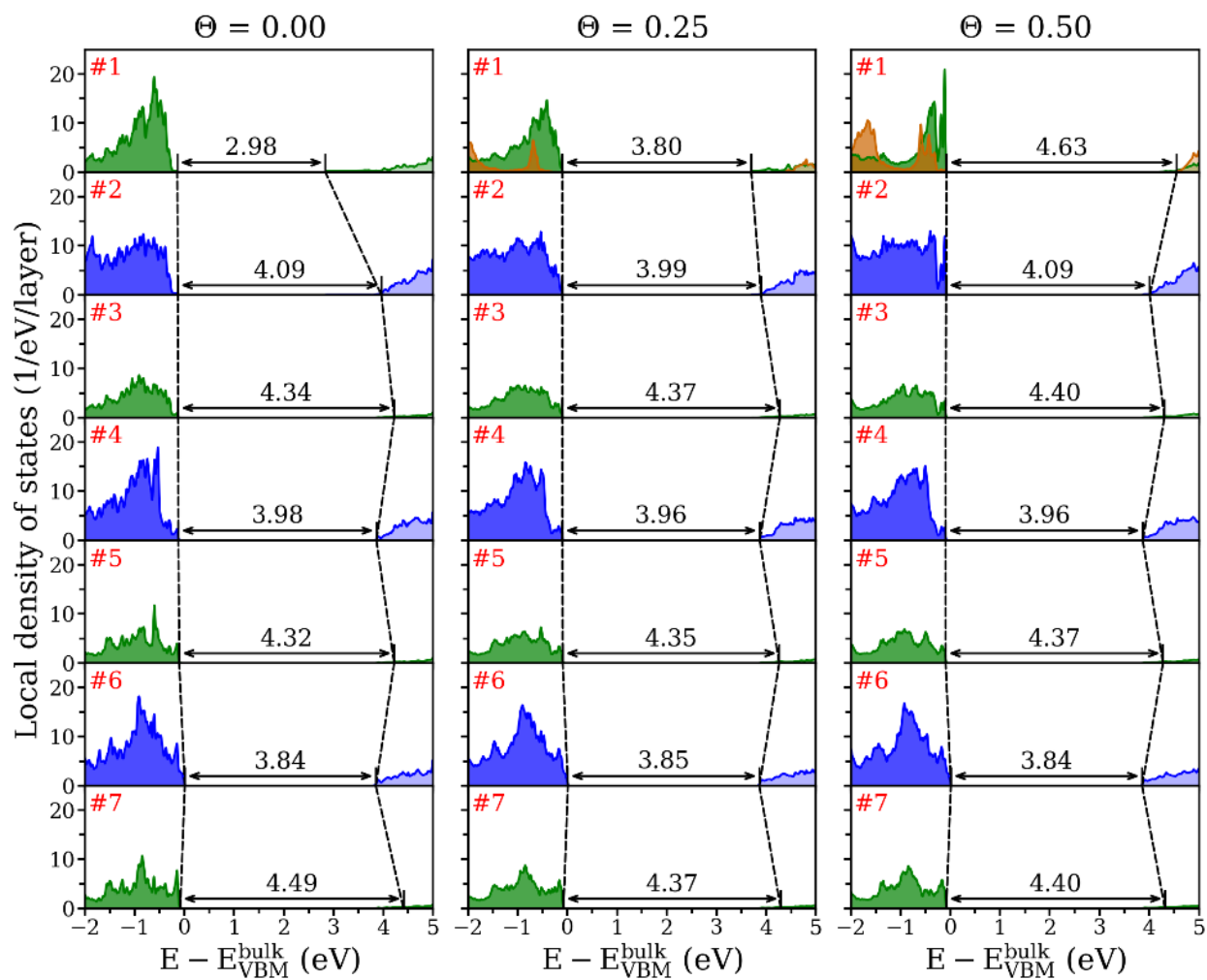

Figure S13. Layer-resolved local density of states (LDOS) for SrO-terminated SrHfO₃(001) at Θ=0.00 (clean surface), Θ=0.25, and Θ=0.50 CO₂ coverages (only one half of the slab is presented due to the symmetry; indexing is from the surface to the middle layers; numbers represent effective band gaps for the atomic layers computed form LDOS neglecting the population densities below 0.3 1/eV/layer).

# BaO-terminated BaTiO$_3$(001)

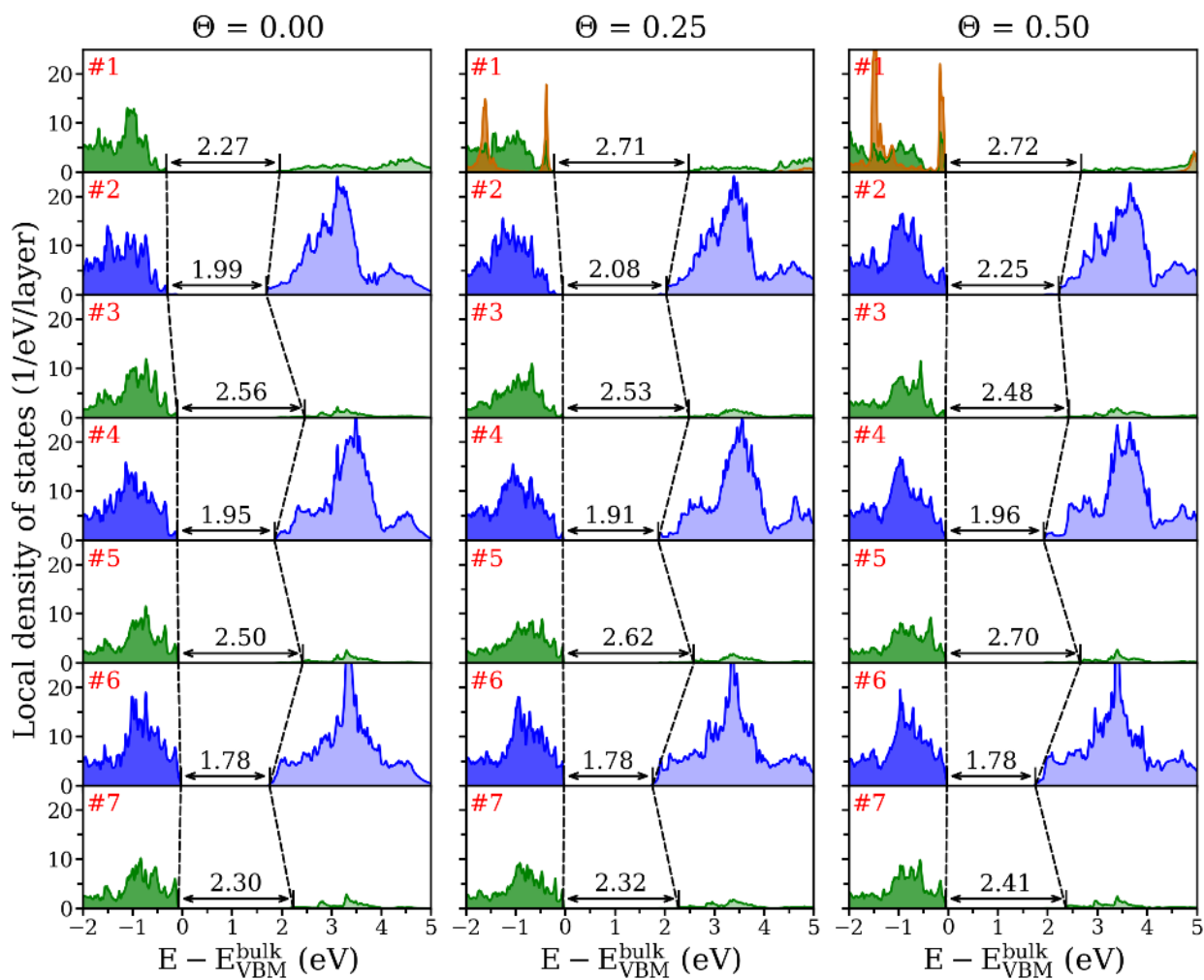

Figure S14. Layer-resolved local density of states (LDOS) for BaO-terminated BaTiO$_3$(001) at $\Theta$=0.00 (clean surface), $\Theta$=0.25, and $\Theta$=0.50 CO$_2$ coverages (only one half of the slab is presented due to the symmetry; indexing is from the surface to the middle layers; numbers represent effective band gaps for the atomic layers computed form LDOS neglecting the population densities below 0.3 1/eV/layer).

## BaO-terminated BaZrO$_3$(001)

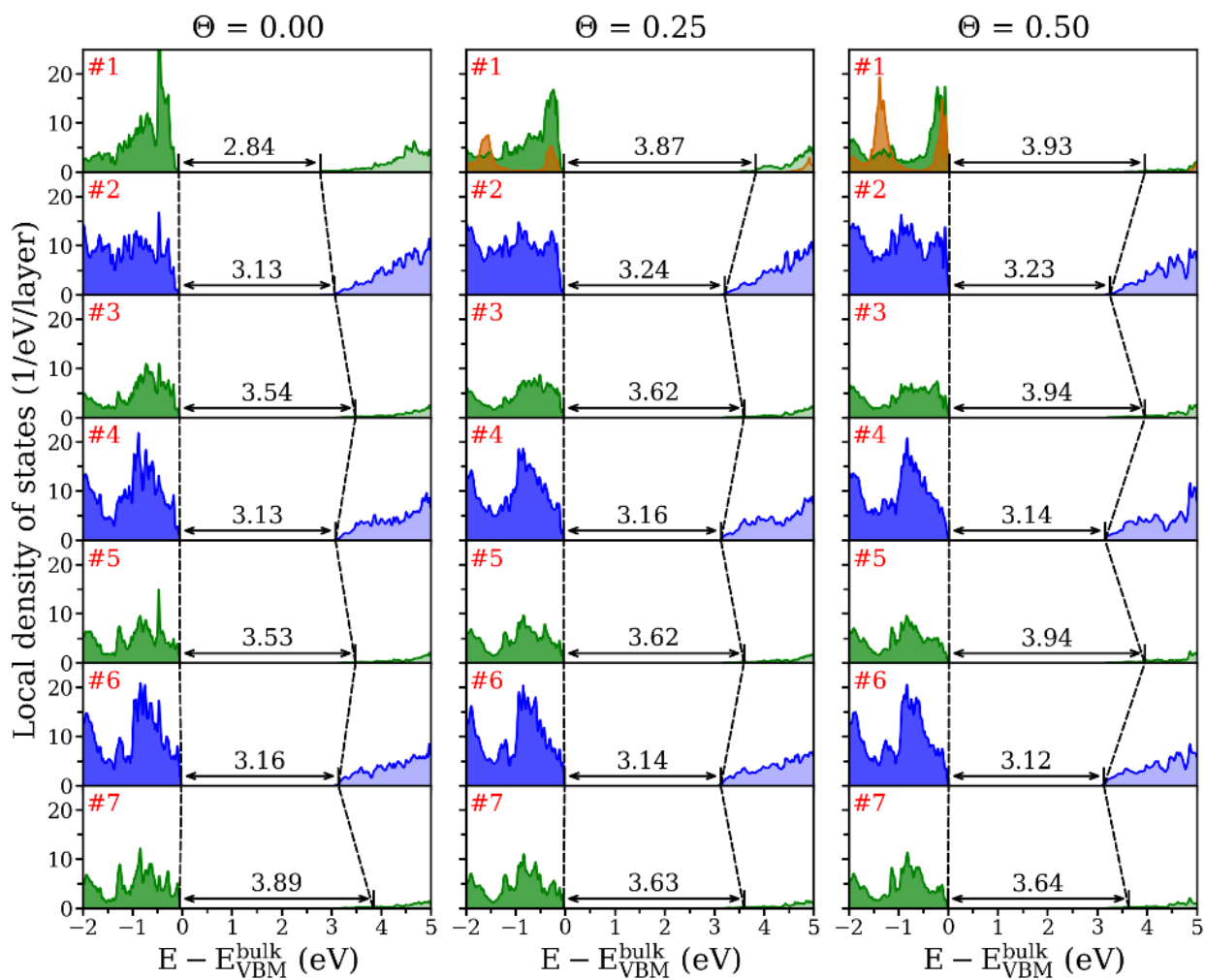

Figure S15. Layer-resolved local density of states (LDOS) for BaO-terminated BaZrO$_3$(001) at $\Theta$=0.00 (clean surface), $\Theta$=0.25, and $\Theta$=0.50 CO$_2$ coverages (only one half of the slab is presented due to the symmetry; indexing is from the surface to the middle layers; numbers represent effective band gaps for the atomic layers computed form LDOS neglecting the population densities below 0.3 1/eV/layer).

## BaO-terminated BaHfO$_3$(001)

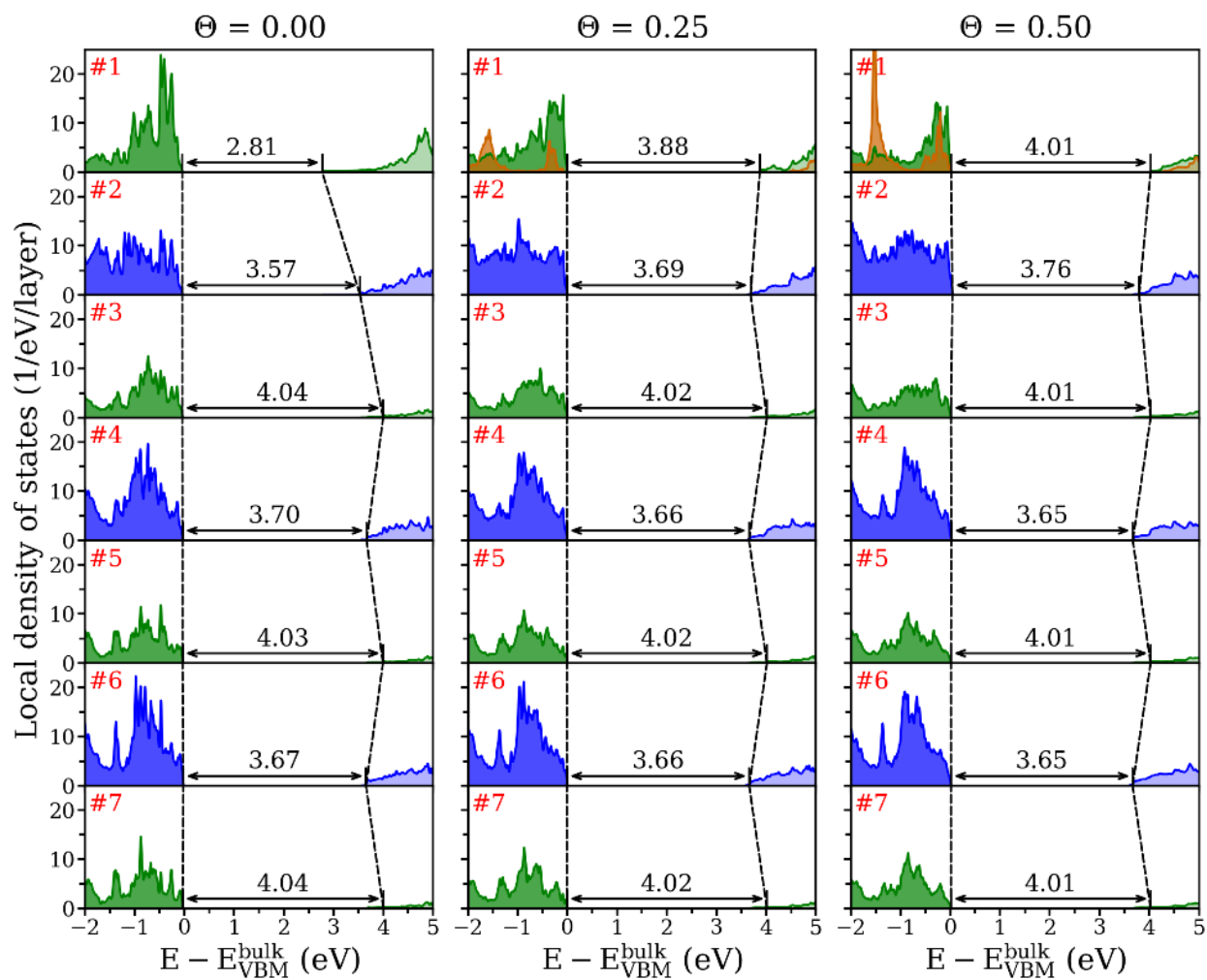

Figure S16. Layer-resolved local density of states (LDOS) for BaO-terminated BaHfO$_3$(001) at $\Theta$=0.00 (clean surface), $\Theta$=0.25, and $\Theta$=0.50 CO$_2$ coverages (only one half of the slab is presented due to the symmetry; indexing is from the surface to the middle layers; numbers represent effective band gaps for the atomic layers computed form LDOS neglecting the population densities below 0.3 1/eV/layer).

## TiO₂-terminated SrTiO₃(001)

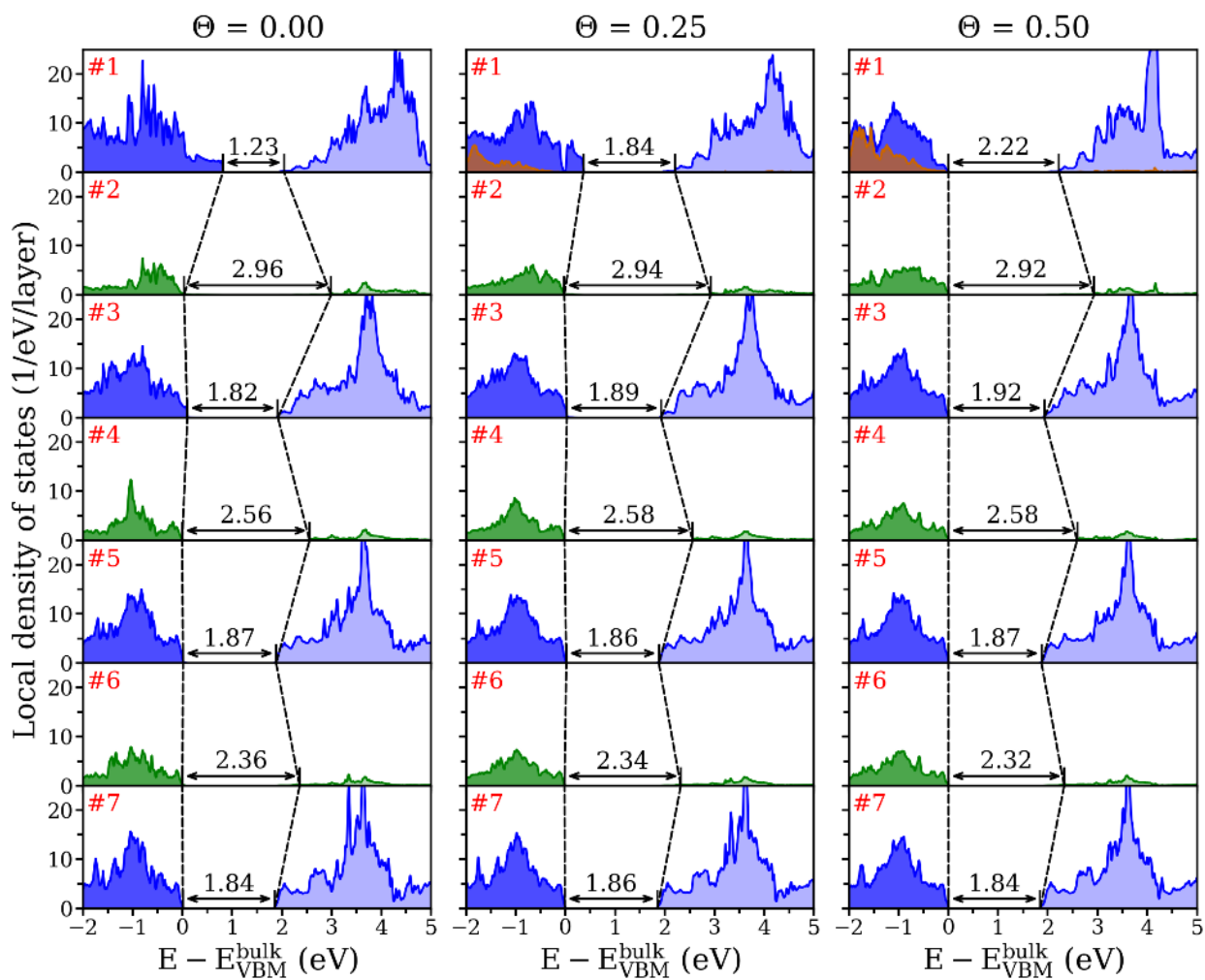

Figure S17. Layer-resolved local density of states (LDOS) for TiO$_2$-terminated SrTiO$_3$(001) at $\Theta$=0.00 (clean surface), $\Theta$=0.25, and $\Theta$=0.50 CO$_2$ coverages (only one half of the slab is presented due to the symmetry; indexing is from the surface to the middle layers; numbers represent effective band gaps for the atomic layers computed form LDOS neglecting the population densities below 0.3 1/eV/layer).

## ZrO$_2$-terminated SrZrO$_3$(001)

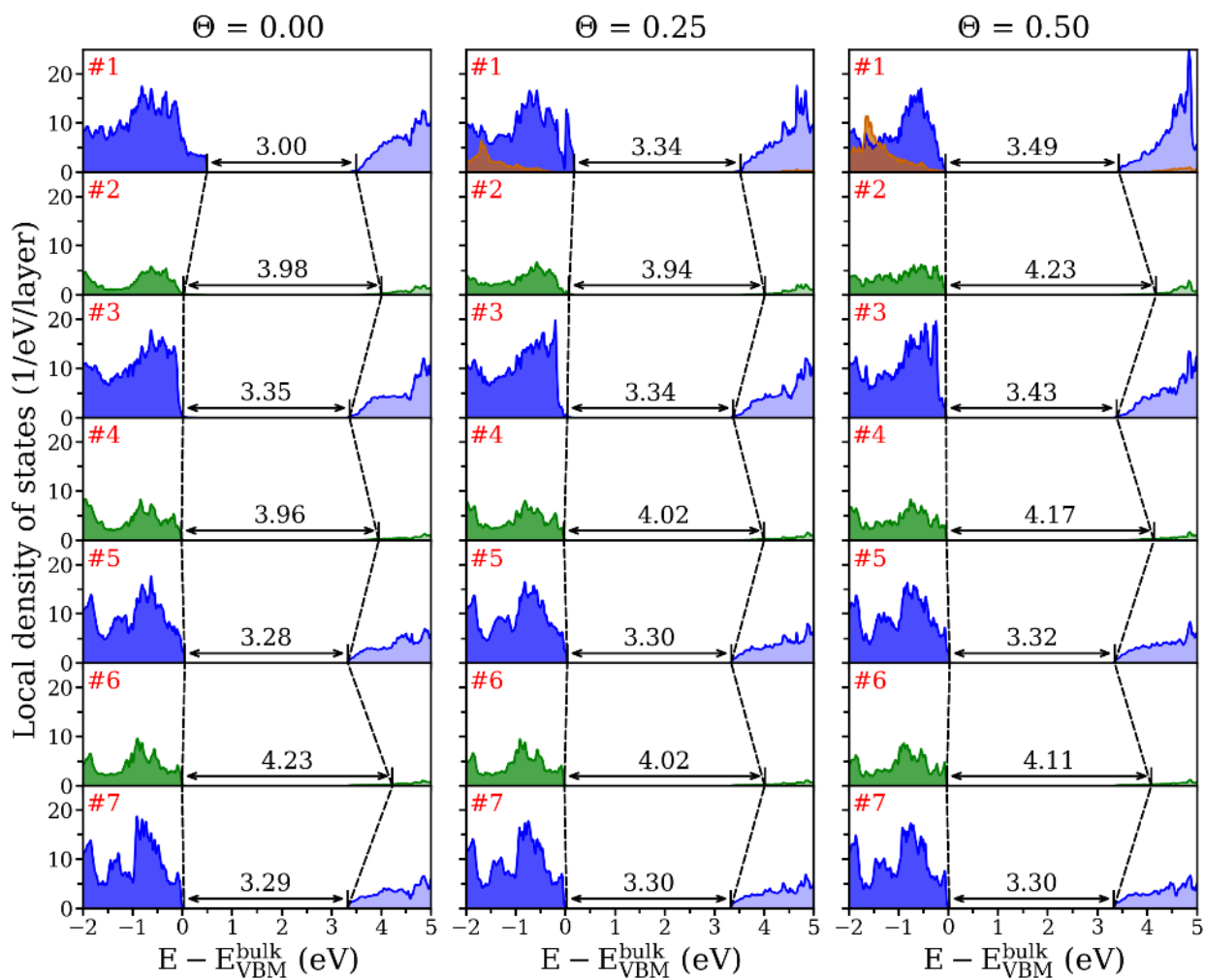

Figure S18. Layer-resolved local density of states (LDOS) for ZrO$_2$-terminated SrZrO$_3$(001) at $\Theta$=0.00 (clean surface), $\Theta$=0.25, and $\Theta$=0.50 CO$_2$ coverages (only one half of the slab is presented due to the symmetry; indexing is from the surface to the middle layers; numbers represent effective band gaps for the atomic layers computed form LDOS neglecting the population densities below 0.3 1/eV/layer).

# HfO$_2$-terminated SrHfO$_3$(001)

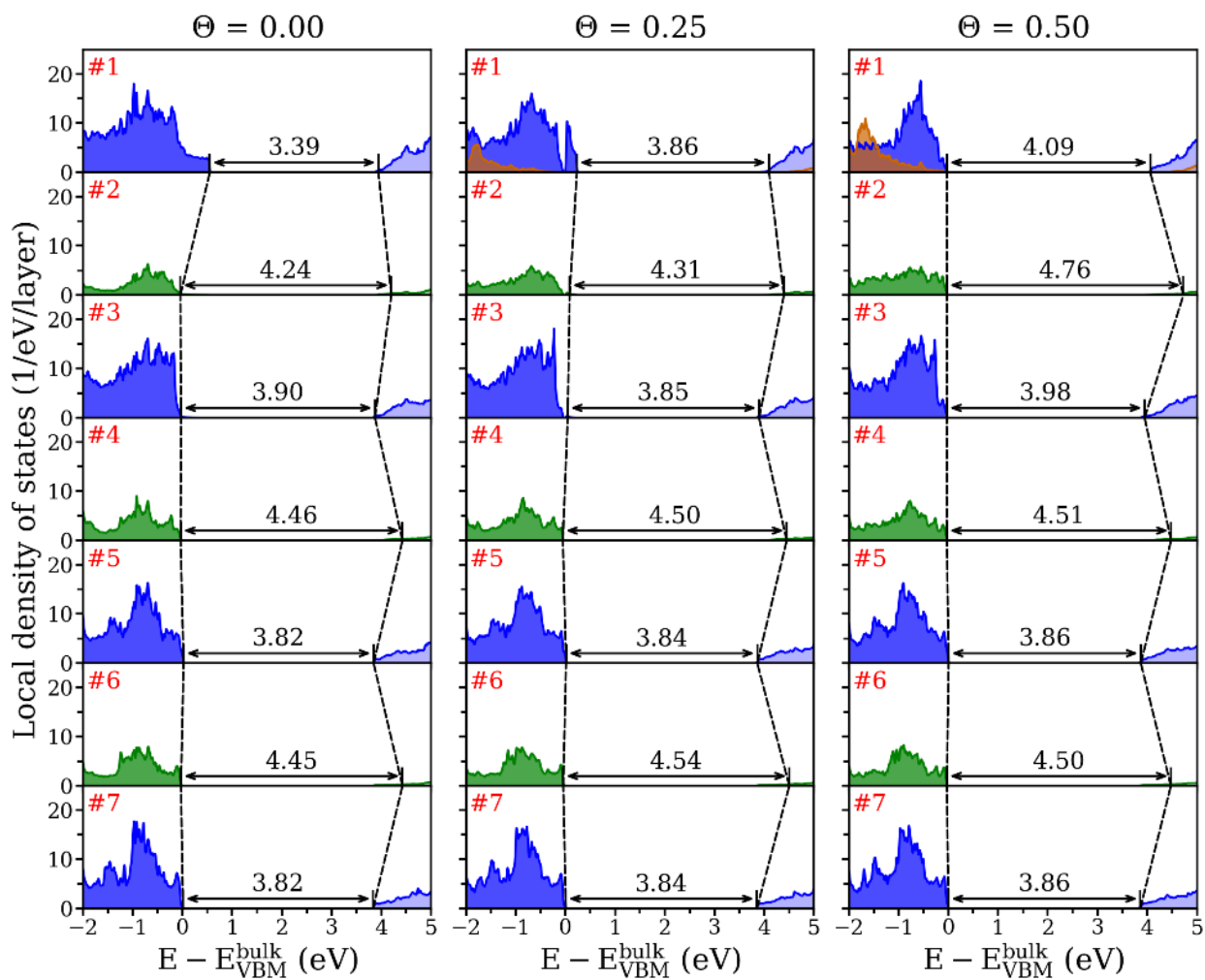

Figure S19. Layer-resolved local density of states (LDOS) for HfO$_2$-terminated SrHfO$_3$(001) at $\Theta$=0.00 (clean surface), $\Theta$=0.25, and $\Theta$=0.50 CO$_2$ coverages (only one half of the slab is presented due to the symmetry; indexing is from the surface to the middle layers; numbers represent effective band gaps for the atomic layers computed form LDOS neglecting the population densities below 0.3 1/eV/layer).

## TiO$_2$-terminated BaTiO$_3$(001)

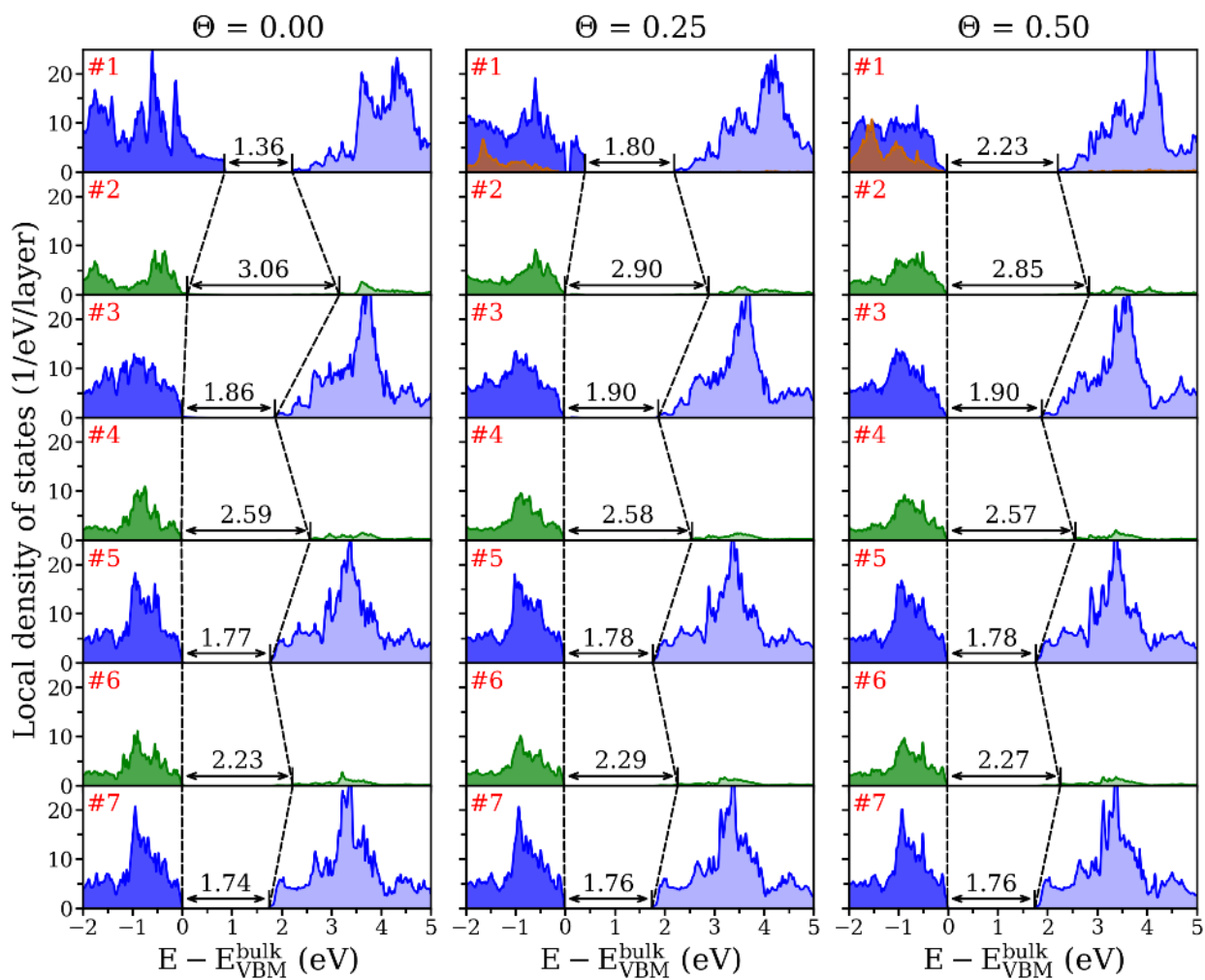

Figure S20. Layer-resolved local density of states (LDOS) for TiO$_2$-terminated BaTiO$_3$(001) at $\Theta=0.00$ (clean surface), $\Theta=0.25$, and $\Theta=0.50$ CO$_2$ coverages (only one half of the slab is presented due to the symmetry; indexing is from the surface to the middle layers; numbers represent effective band gaps for the atomic layers computed form LDOS neglecting the population densities below 0.3 1/eV/layer).

## ZrO$_2$-terminated BaZrO$_3$(001)

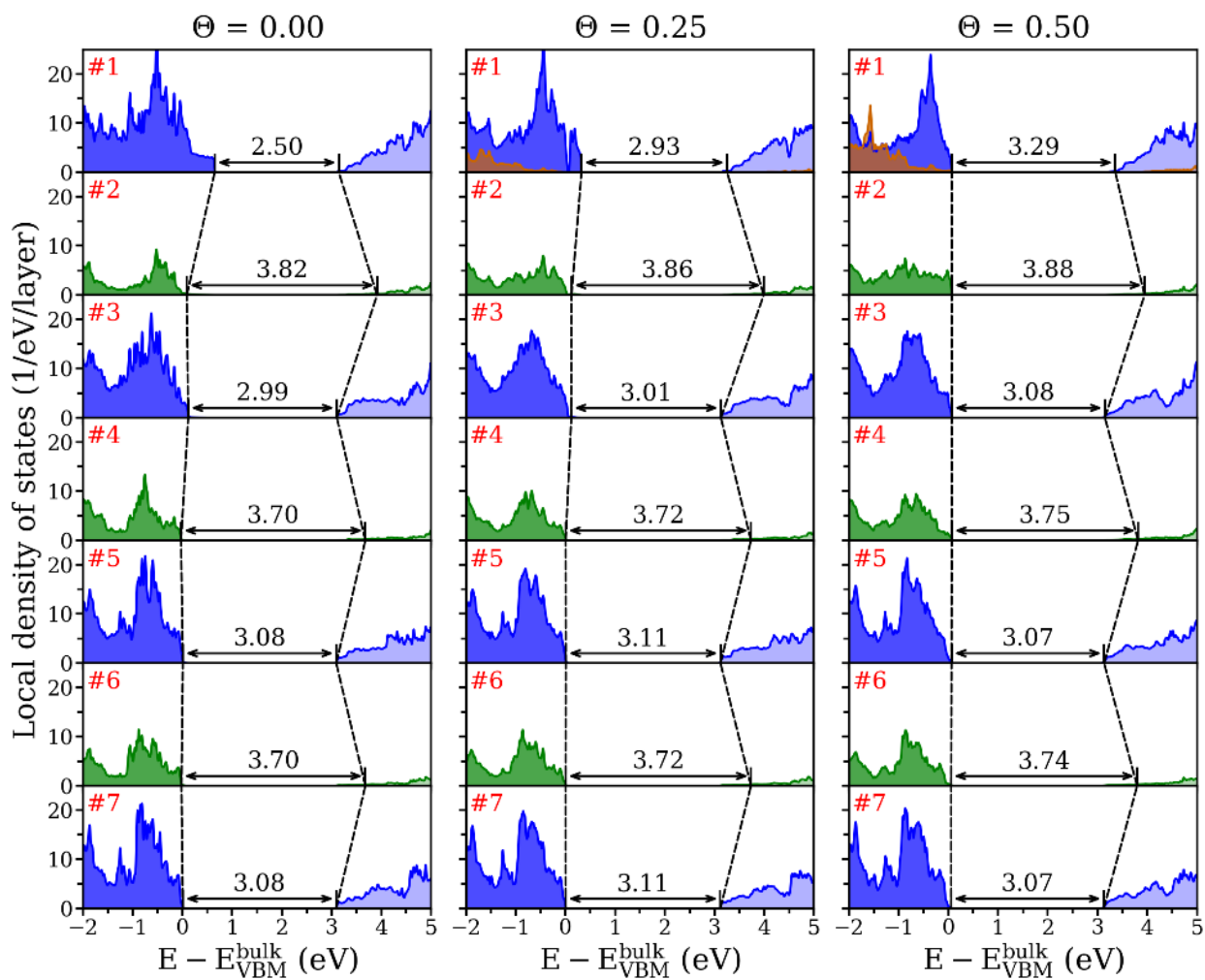

Figure S21. Layer-resolved local density of states (LDOS) for ZrO$_2$-terminated BaZrO$_3$(001) at $\Theta$=0.00 (clean surface), $\Theta$=0.25, and $\Theta$=0.50 CO$_2$ coverages (only one half of the slab is presented due to the symmetry; indexing is from the surface to the middle layers; numbers represent effective band gaps for the atomic layers computed form LDOS neglecting the population densities below 0.3 1/eV/layer).

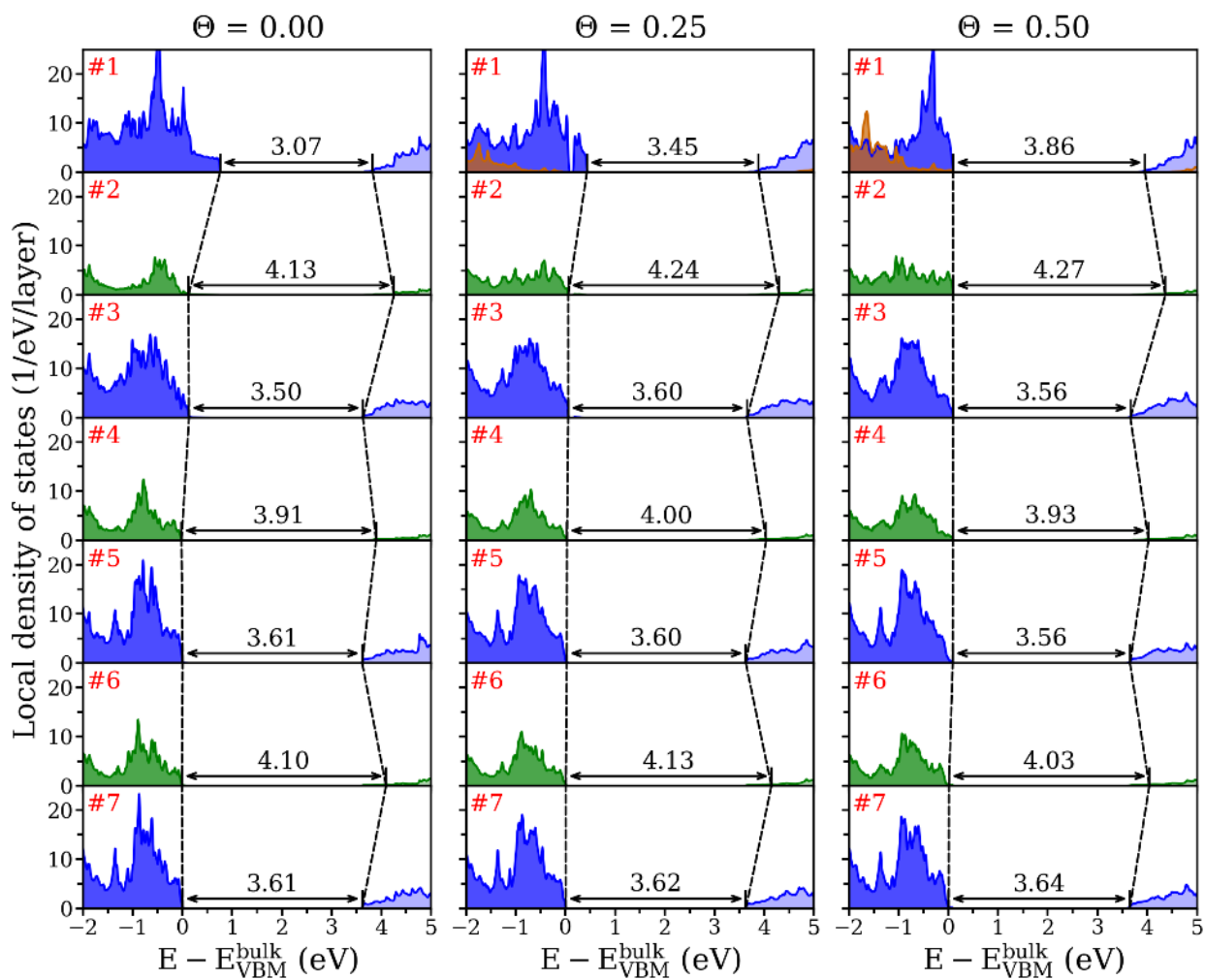

Figure S22. Layer-resolved local density of states (LDOS) for HfO$_2$-terminated BaHfO$_3$(001) at Θ=0.00 (clean surface), Θ=0.25, and Θ=0.50 CO$_2$ coverages (only one half of the slab is presented due to the symmetry; indexing is from the surface to the middle layers; numbers represent effective band gaps for the atomic layers computed form LDOS neglecting the population densities below 0.3 1/eV/layer).

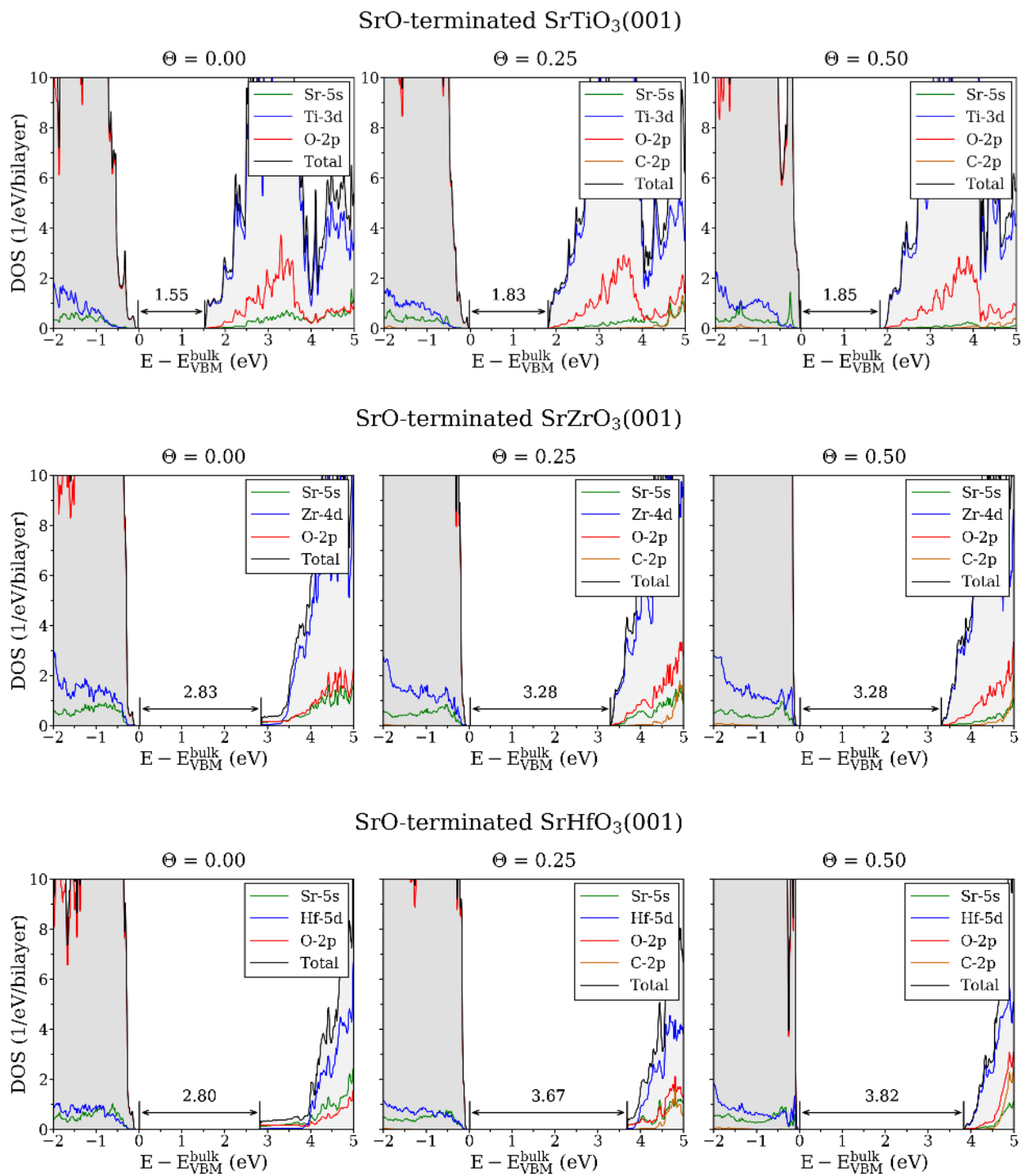

Figure S23. Projected density of states (DOS) for two surface layers of the Sr-containing AO-terminated ABO$_3$(001) at $\Theta$=0.00 (clean surface), $\Theta$=0.25, and $\Theta$=0.50 CO$_2$ coverages. The numbers represent band gap energies of the slab systems.

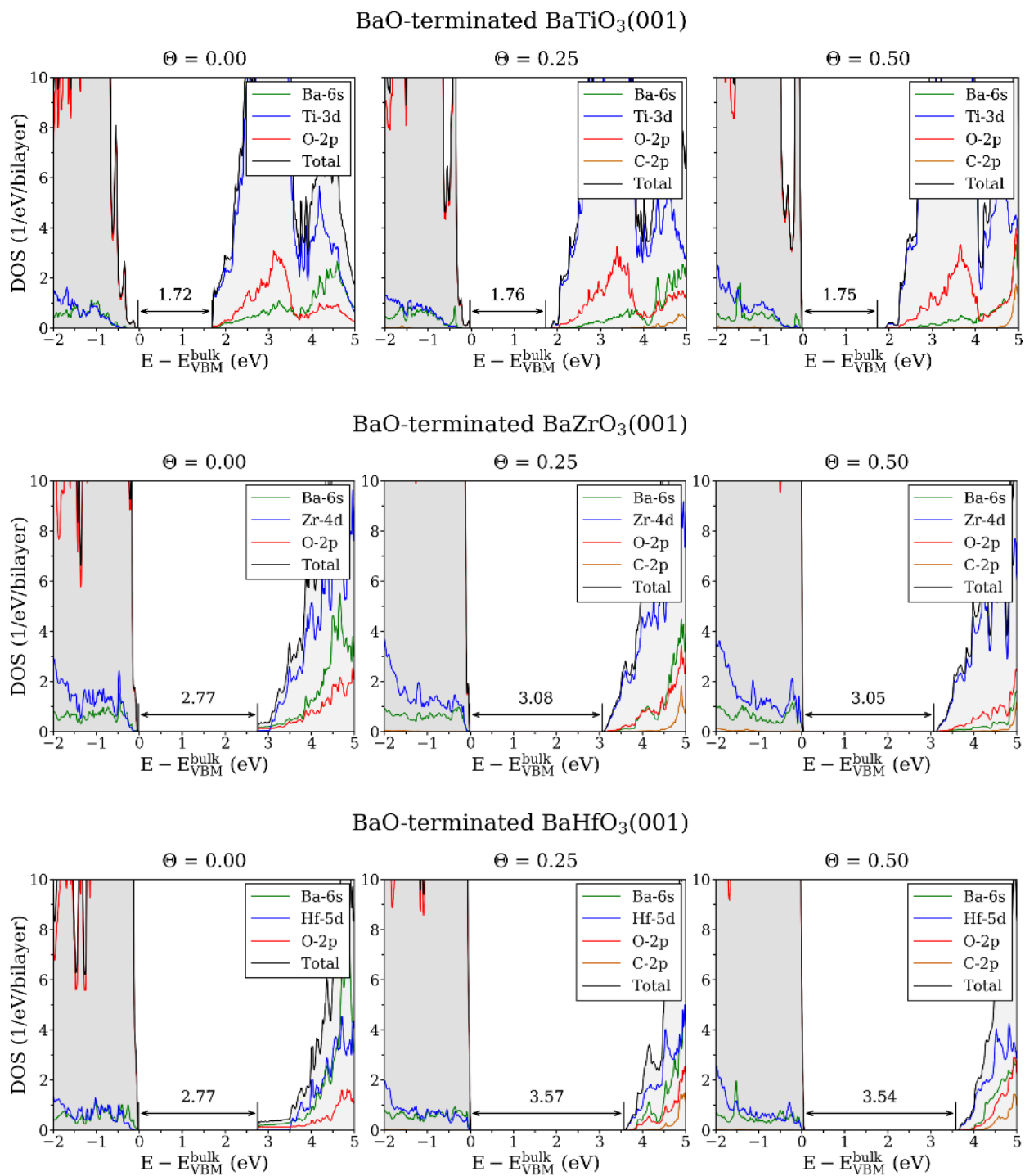

Figure S24. Projected density of states (DOS) for two surface layers of the Ba-containing AO-terminated ABO$_3$(001) at $\Theta$=0.00 (clean surface), $\Theta$=0.25, and $\Theta$=0.50 CO$_2$ coverages. The numbers represent band gap energies of the slab systems.

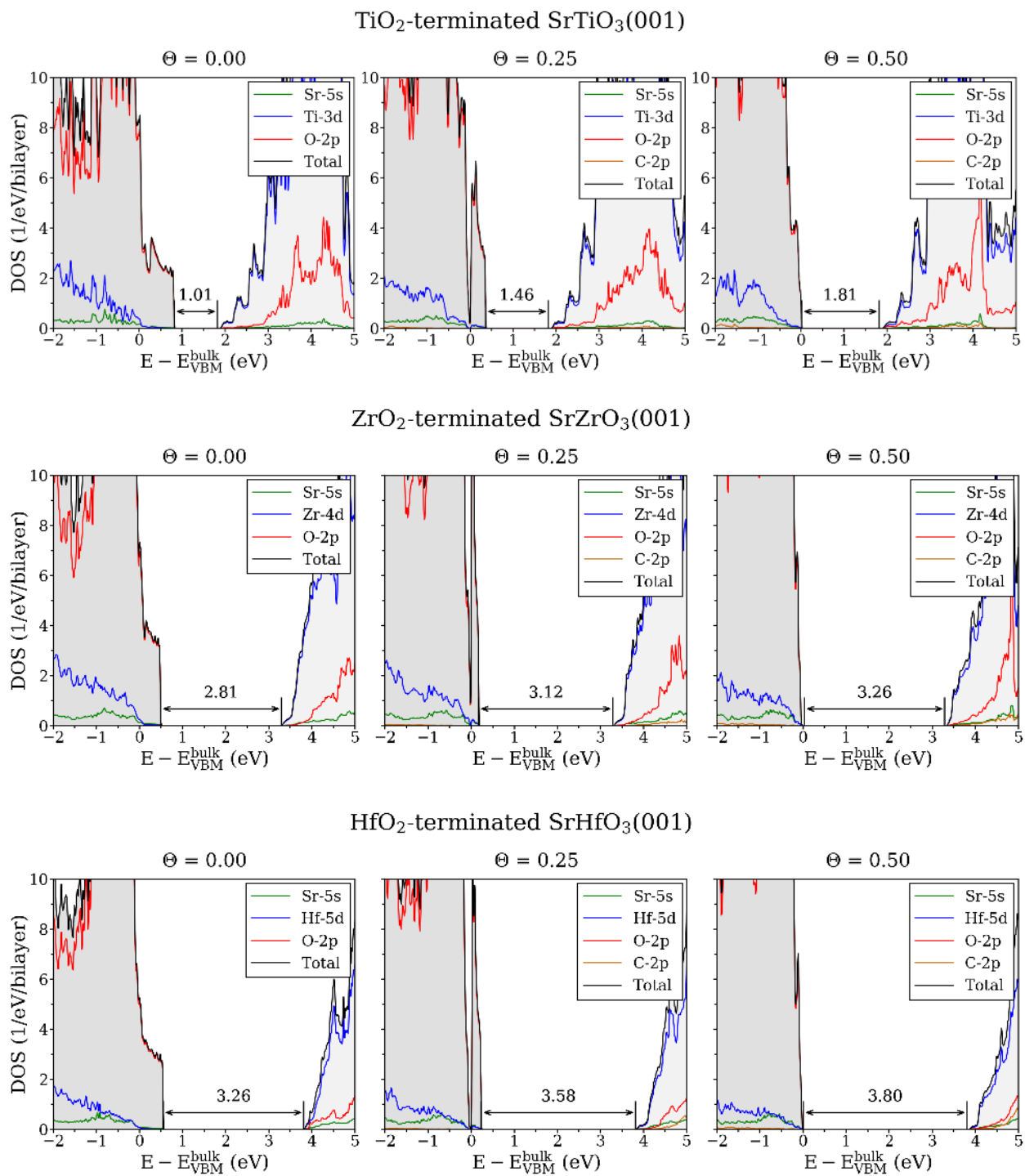

Figure S25. Projected density of states (DOS) for two surface layers of the Sr-containing $BO_2$-terminated $ABO_3$(001) at $\Theta$=0.00 (clean surface), $\Theta$=0.25, and $\Theta$=0.50 $CO_2$ coverages. The numbers represent band gap energies of the slab systems.

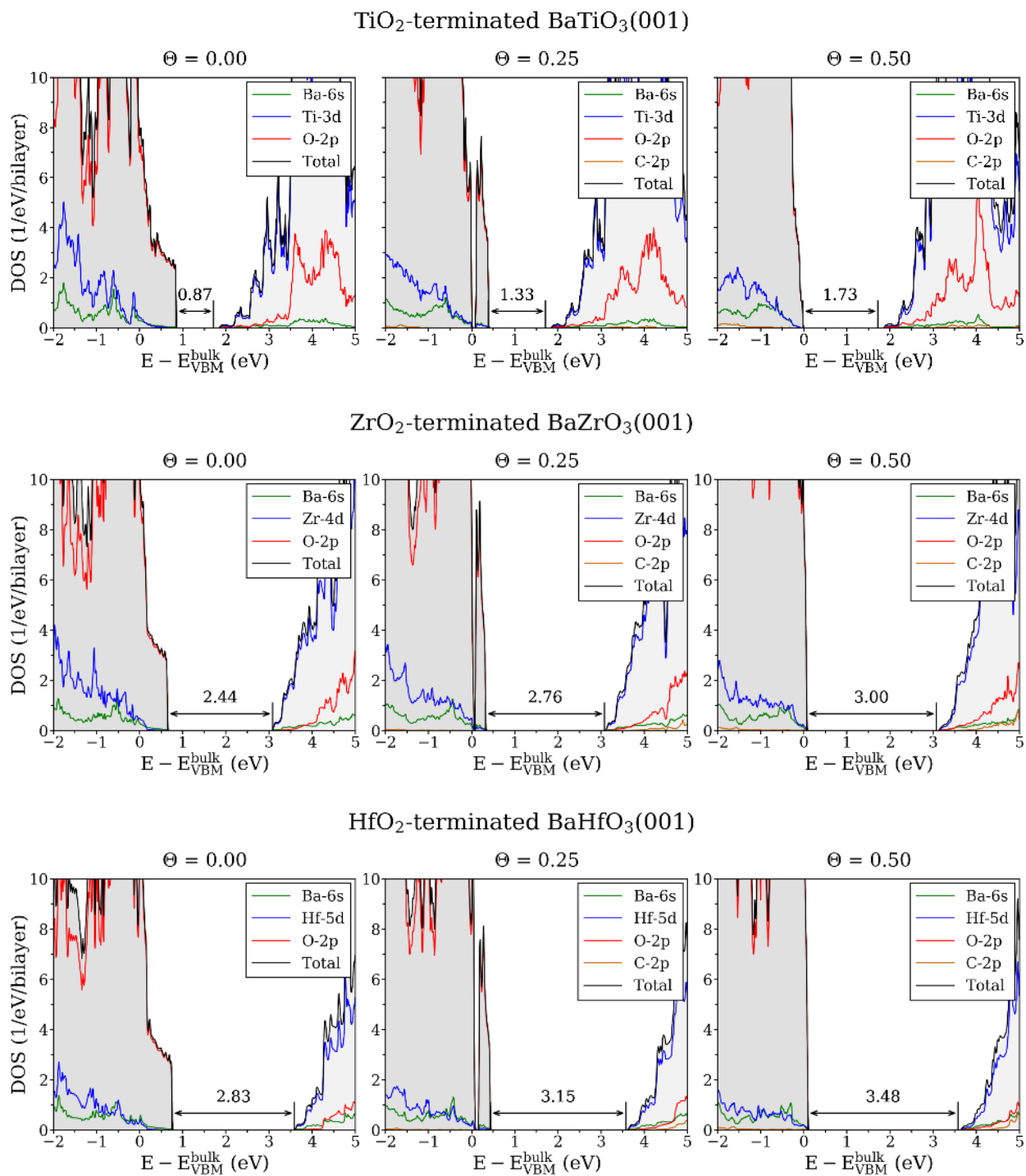

Figure S26. Projected density of states (DOS) for two surface layers of the Ba-containing $BO_2$-terminated $ABO_3$(001) at $\Theta$=0.00 (clean surface), $\Theta$=0.25, and $\Theta$=0.50 $CO_2$ coverages. The numbers represent band gap energies of the slab systems.

Structure 1. SrO-terminated SrTiO$_3$ slab at $\Theta$=0.00 CO$_2$ coverage

_cell_length_a  7.89043600
_cell_length_b  7.89043600
_cell_length_c  43.39739600
_cell_angle_alpha  90.00000000
_cell_angle_beta  90.00000000
_cell_angle_gamma  90.00000000
_symmetry_space_group_name_H-M  'P 1'
loop_
 _atom_site_type_symbol
 _atom_site_fract_x
 _atom_site_fract_y
 _atom_site_fract_z
 Sr 0.000000 0.000000 0.590881
 Sr 0.000000 0.499857 0.590881
 Sr 0.499857 0.000000 0.590881
 Sr 0.499857 0.499857 0.590881
 Sr 0.000000 0.000000 0.499998
 Sr 0.000000 0.499857 0.499998
 Sr 0.499857 0.000000 0.499998
 Sr 0.499857 0.499857 0.499998
 Sr 0.000804 0.001697 0.768100
 Sr 0.000804 0.501647 0.768107
 Sr 0.500756 0.001696 0.768107
 Sr 0.500756 0.501646 0.768113
 Sr 0.001474 0.002035 0.680663
 Sr 0.001473 0.501981 0.680665
 Sr 0.501420 0.002033 0.680665
 Sr 0.501419 0.501979 0.680666
 Sr 0.000000 0.000000 0.409119
 Sr 0.000000 0.499857 0.409119
 Sr 0.499857 0.000000 0.409119
 Sr 0.499857 0.499857 0.409119
 Sr 0.000804 0.001697 0.231900
 Sr 0.000804 0.501647 0.231893
 Sr 0.500756 0.001696 0.231893
 Sr 0.500756 0.501646 0.231887
 Sr 0.001474 0.002035 0.319337
 Sr 0.001473 0.501981 0.319335
 Sr 0.501420 0.002033 0.319335
 Sr 0.501419 0.501979 0.319334
 Ti 0.249929 0.249929 0.545440
 Ti 0.249929 0.749786 0.545440
 Ti 0.749786 0.249929 0.545440
 Ti 0.749786 0.749786 0.545440
 Ti 0.250992 0.251786 0.728941

Ti 0.250993 0.751729 0.728946
Ti 0.750932 0.251788 0.728946
Ti 0.750933 0.751731 0.728951
Ti 0.252199 0.252808 0.636662
Ti 0.252204 0.752757 0.636664
Ti 0.752146 0.252813 0.636664
Ti 0.752151 0.752762 0.636665
Ti 0.249929 0.249929 0.454560
Ti 0.249929 0.749786 0.454560
Ti 0.749786 0.249929 0.454560
Ti 0.749786 0.749786 0.454560
Ti 0.250992 0.251786 0.271059
Ti 0.250993 0.751729 0.271054
Ti 0.750932 0.251788 0.271054
Ti 0.750933 0.751731 0.271049
Ti 0.252199 0.252808 0.363338
Ti 0.252204 0.752757 0.363336
Ti 0.752146 0.252813 0.363336
Ti 0.752151 0.752762 0.363335
O 0.249929 0.000000 0.545440
O 0.249929 0.499857 0.545440
O 0.749786 0.000000 0.545440
O 0.749786 0.499857 0.545440
O 0.249929 0.249929 0.590881
O 0.249929 0.749786 0.590881
O 0.749786 0.249929 0.590881
O 0.749786 0.749786 0.590881
O 0.000000 0.249929 0.545440
O 0.000000 0.749786 0.545440
O 0.499857 0.249929 0.545440
O 0.499857 0.749786 0.545440
O 0.249929 0.249929 0.499998
O 0.249929 0.749786 0.499998
O 0.749786 0.249929 0.499998
O 0.749786 0.749786 0.499998
O 0.249555 0.249173 0.773230
O 0.249555 0.749072 0.773229
O 0.749457 0.249172 0.773229
O 0.749456 0.749072 0.773228
O 0.499703 0.249754 0.727918
O 0.499702 0.749664 0.727919
O 0.999706 0.249755 0.727919
O 0.999706 0.749665 0.727920
O 0.249294 0.249080 0.681918
O 0.249294 0.748986 0.681919
O 0.749201 0.249079 0.681919

```
O  0.749201  0.748985  0.681920
O  0.249805  0.499611  0.727918
O  0.249805  0.999614  0.727919
O  0.749717  0.499610  0.727919
O  0.749717  0.999613  0.727920
O  0.498956  0.248579  0.636482
O  0.498953  0.748483  0.636483
O  0.998968  0.248580  0.636482
O  0.998966  0.748484  0.636483
O  0.248893  0.498700  0.636484
O  0.248894  0.998711  0.636484
O  0.748799  0.498698  0.636485
O  0.748800  0.998709  0.636485
O  0.249929  0.000000  0.454560
O  0.249929  0.499857  0.454560
O  0.749786  0.000000  0.454560
O  0.749786  0.499857  0.454560
O  0.249929  0.249929  0.409119
O  0.249929  0.749786  0.409119
O  0.749786  0.249929  0.409119
O  0.749786  0.749786  0.409119
O  0.000000  0.249929  0.454560
O  0.000000  0.749786  0.454560
O  0.499857  0.249929  0.454560
O  0.499857  0.749786  0.454560
O  0.249555  0.249173  0.226770
O  0.249555  0.749072  0.226771
O  0.749457  0.249172  0.226771
O  0.749456  0.749072  0.226772
O  0.499703  0.249754  0.272082
O  0.499702  0.749664  0.272081
O  0.999706  0.249755  0.272081
O  0.999706  0.749665  0.272080
O  0.249294  0.249080  0.318082
O  0.249294  0.748986  0.318081
O  0.749201  0.249079  0.318081
O  0.749201  0.748985  0.318080
O  0.249805  0.499611  0.272082
O  0.249805  0.999614  0.272081
O  0.749717  0.499610  0.272081
O  0.749717  0.999613  0.272080
O  0.498956  0.248579  0.363518
O  0.498953  0.748483  0.363517
O  0.998968  0.248580  0.363518
O  0.998966  0.748484  0.363517
O  0.248893  0.498700  0.363516
```

O  0.248894  0.998711  0.363516
O  0.748799  0.498698  0.363515
O  0.748800  0.998709  0.363515

Structure 2. SrO-terminated SrTiO$_3$ slab at $\Theta$=0.25 CO$_2$ coverage

```
_cell_length_a  7.89043600
_cell_length_b  7.89043600
_cell_length_c  43.39739600
_cell_angle_alpha  90.00000000
_cell_angle_beta  90.00000000
_cell_angle_gamma  90.00000000
_symmetry_space_group_name_H-M  'P 1'
loop_
 _atom_site_type_symbol
 _atom_site_fract_x
 _atom_site_fract_y
 _atom_site_fract_z
 Sr  0.013161  0.489044  0.226694
 Sr  0.993750  0.000464  0.228305
 Sr  0.501160  0.506908  0.228361
 Sr  0.489349  0.011608  0.230341
 Sr  0.002269  0.008371  0.318484
 Sr  0.492689  0.007673  0.318496
 Sr  0.491811  0.498027  0.318555
 Sr  0.005809  0.494379  0.318587
 Sr  0.500000  0.000000  0.409118
 Sr  0.500000  0.499857  0.409118
 Sr  0.000143  0.000000  0.409118
 Sr  0.000143  0.499857  0.409118
 Sr  0.500000  0.000000  0.499998
 Sr  0.500000  0.499857  0.499998
 Sr  0.000143  0.000000  0.499998
 Sr  0.000143  0.499857  0.499998
 Sr  0.500000  0.000000  0.590882
 Sr  0.500000  0.499857  0.590882
 Sr  0.000143  0.000000  0.590882
 Sr  0.000143  0.499857  0.590882
 Sr  0.005809  0.494379  0.681413
 Sr  0.491811  0.498027  0.681445
 Sr  0.492689  0.007673  0.681504
 Sr  0.002269  0.008371  0.681516
 Sr  0.489349  0.011608  0.769659
 Sr  0.501160  0.506908  0.771639
 Sr  0.993750  0.000464  0.771695
 Sr  0.013161  0.489044  0.773306
 Ti  0.749972  0.254607  0.269181
 Ti  0.246156  0.750961  0.269206
 Ti  0.247512  0.253407  0.270446
 Ti  0.748356  0.752031  0.276150
 Ti  0.747169  0.255935  0.362536
```

```
Ti  0.244262  0.753075  0.362541
Ti  0.248255  0.251970  0.363544
Ti  0.747846  0.752394  0.365028
Ti  0.250071  0.249929  0.454560
Ti  0.250071  0.749786  0.454560
Ti  0.750214  0.249929  0.454560
Ti  0.750214  0.749786  0.454560
Ti  0.250071  0.749786  0.545440
Ti  0.250071  0.249929  0.545440
Ti  0.750214  0.249929  0.545440
Ti  0.750214  0.749786  0.545440
Ti  0.747846  0.752394  0.634972
Ti  0.248255  0.251970  0.636456
Ti  0.244262  0.753075  0.637459
Ti  0.747169  0.255935  0.637464
Ti  0.748356  0.752031  0.723850
Ti  0.247512  0.253407  0.729554
Ti  0.246156  0.750961  0.730794
Ti  0.749972  0.254607  0.730819
C   0.793181  0.707221  0.197485
C   0.793181  0.707221  0.802515
O   0.942299  0.760424  0.190324
O   0.741145  0.557654  0.190335
O   0.698745  0.800839  0.216359
O   0.247085  0.252757  0.226540
O   0.753884  0.220707  0.227456
O   0.278820  0.745087  0.227457
O   0.753984  0.504143  0.265726
O   0.995878  0.749596  0.265839
O   0.500659  0.252356  0.271787
O   0.251089  0.999166  0.271803
O   0.000333  0.249683  0.272925
O   0.246586  0.499480  0.272958
O   0.506299  0.745280  0.274438
O   0.750642  0.993635  0.274497
O   0.769586  0.730304  0.317388
O   0.748569  0.262440  0.318152
O   0.237722  0.751355  0.318166
O   0.253192  0.246645  0.318358
O   0.752617  0.995804  0.361848
O   0.504094  0.748936  0.361860
O   0.253041  0.498423  0.363345
O   0.001430  0.248447  0.363347
O   0.501403  0.246625  0.363871
O   0.251502  0.998497  0.363879
O   0.000266  0.747465  0.364723
```

| | | | |
|---|---|---|---|
| O | 0.750590 | 0.499653 | 0.364725 |
| O | 0.250071 | 0.249929 | 0.409118 |
| O | 0.250071 | 0.749786 | 0.409118 |
| O | 0.750214 | 0.249929 | 0.409118 |
| O | 0.750214 | 0.749786 | 0.409118 |
| O | 0.250071 | 0.000000 | 0.454560 |
| O | 0.250071 | 0.499857 | 0.454560 |
| O | 0.750214 | 0.000000 | 0.454560 |
| O | 0.750214 | 0.499857 | 0.454560 |
| O | 0.500000 | 0.749786 | 0.454560 |
| O | 0.500000 | 0.249929 | 0.454560 |
| O | 0.000143 | 0.749786 | 0.454560 |
| O | 0.000143 | 0.249929 | 0.454560 |
| O | 0.250071 | 0.249929 | 0.499998 |
| O | 0.250071 | 0.749786 | 0.499998 |
| O | 0.750214 | 0.249929 | 0.499998 |
| O | 0.750214 | 0.749786 | 0.499998 |
| O | 0.250071 | 0.000000 | 0.545440 |
| O | 0.250071 | 0.499857 | 0.545440 |
| O | 0.750214 | 0.000000 | 0.545440 |
| O | 0.750214 | 0.499857 | 0.545440 |
| O | 0.500000 | 0.249929 | 0.545440 |
| O | 0.500000 | 0.749786 | 0.545440 |
| O | 0.000143 | 0.249929 | 0.545440 |
| O | 0.000143 | 0.749786 | 0.545440 |
| O | 0.250071 | 0.249929 | 0.590882 |
| O | 0.250071 | 0.749786 | 0.590882 |
| O | 0.750214 | 0.249929 | 0.590882 |
| O | 0.750214 | 0.749786 | 0.590882 |
| O | 0.750590 | 0.499653 | 0.635275 |
| O | 0.000266 | 0.747465 | 0.635277 |
| O | 0.251502 | 0.998497 | 0.636121 |
| O | 0.501403 | 0.246625 | 0.636129 |
| O | 0.001430 | 0.248447 | 0.636653 |
| O | 0.253041 | 0.498423 | 0.636655 |
| O | 0.504094 | 0.748936 | 0.638140 |
| O | 0.752617 | 0.995804 | 0.638152 |
| O | 0.253192 | 0.246645 | 0.681642 |
| O | 0.237722 | 0.751355 | 0.681834 |
| O | 0.748569 | 0.262440 | 0.681848 |
| O | 0.769586 | 0.730304 | 0.682612 |
| O | 0.750642 | 0.993635 | 0.725503 |
| O | 0.506299 | 0.745280 | 0.725562 |
| O | 0.246586 | 0.499480 | 0.727042 |
| O | 0.000333 | 0.249683 | 0.727075 |
| O | 0.251089 | 0.999166 | 0.728197 |

O 0.500659 0.252356 0.728213
O 0.995878 0.749596 0.734161
O 0.753984 0.504143 0.734274
O 0.278820 0.745087 0.772543
O 0.753884 0.220707 0.772544
O 0.247085 0.252757 0.773460
O 0.698745 0.800839 0.783641
O 0.741145 0.557654 0.809666
O 0.942299 0.760424 0.809676

Structure 3. SrO-terminated SrTiO$_3$ slab at $\Theta$=0.50 CO$_2$ coverage

_cell_length_a  7.89043600
_cell_length_b  7.89043600
_cell_length_c  43.39739600
_cell_angle_alpha  90.00000000
_cell_angle_beta  90.00000000
_cell_angle_gamma  90.00000000
_symmetry_space_group_name_H-M  'P 1'
loop_
 _atom_site_type_symbol
 _atom_site_fract_x
 _atom_site_fract_y
 _atom_site_fract_z
 Sr 0.506888 0.517683 0.776161
 Sr 0.005550 0.478892 0.775658
 Sr 0.493023 0.007957 0.772575
 Sr 0.993277 0.989533 0.772324
 Sr 0.502627 0.498592 0.682868
 Sr 0.500478 0.996493 0.682463
 Sr 0.000915 0.494365 0.682291
 Sr 0.001787 0.995832 0.682158
 Sr 0.000000 0.000000 0.590881
 Sr 0.000000 0.499857 0.590881
 Sr 0.499857 0.000000 0.590881
 Sr 0.499857 0.499857 0.590881
 Sr 0.000000 0.000000 0.499998
 Sr 0.000000 0.499857 0.499998
 Sr 0.499857 0.499857 0.499998
 Sr 0.499857 0.000000 0.499998
 Sr 0.499857 0.499857 0.409119
 Sr 0.499857 0.000000 0.409119
 Sr 0.000000 0.499857 0.409119
 Sr 0.000000 0.000000 0.409119
 Sr 0.001787 0.995832 0.317842
 Sr 0.000915 0.494365 0.317709
 Sr 0.500478 0.996493 0.317537
 Sr 0.502627 0.498592 0.317132
 Sr 0.993277 0.989533 0.227676
 Sr 0.493023 0.007957 0.227425
 Sr 0.005550 0.478892 0.224342
 Sr 0.506888 0.517683 0.223839
 Ti 0.250132 0.744449 0.731882
 Ti 0.749600 0.250648 0.731745
 Ti 0.749344 0.749499 0.725011
 Ti 0.248482 0.246580 0.724914
 Ti 0.252121 0.742821 0.637989

| | | | |
|----|----------|----------|----------|
| Ti | 0.752350 | 0.244568 | 0.637976 |
| Ti | 0.256278 | 0.246458 | 0.635136 |
| Ti | 0.756156 | 0.744251 | 0.635101 |
| Ti | 0.249929 | 0.249929 | 0.545440 |
| Ti | 0.249929 | 0.749786 | 0.545440 |
| Ti | 0.749786 | 0.249929 | 0.545440 |
| Ti | 0.749786 | 0.749786 | 0.545440 |
| Ti | 0.749786 | 0.749786 | 0.454560 |
| Ti | 0.749786 | 0.249929 | 0.454560 |
| Ti | 0.249929 | 0.749786 | 0.454560 |
| Ti | 0.249929 | 0.249929 | 0.454560 |
| Ti | 0.756156 | 0.744251 | 0.364899 |
| Ti | 0.256278 | 0.246458 | 0.364864 |
| Ti | 0.752350 | 0.244568 | 0.362024 |
| Ti | 0.252121 | 0.742821 | 0.362011 |
| Ti | 0.248482 | 0.246580 | 0.275086 |
| Ti | 0.749344 | 0.749499 | 0.274989 |
| Ti | 0.749600 | 0.250648 | 0.268255 |
| Ti | 0.250132 | 0.744449 | 0.268118 |
| C  | 0.797421 | 0.713648 | 0.803373 |
| C  | 0.296717 | 0.286027 | 0.803274 |
| C  | 0.296717 | 0.286027 | 0.196726 |
| C  | 0.797421 | 0.713648 | 0.196627 |
| O  | 0.209282 | 0.209090 | 0.780115 |
| O  | 0.709609 | 0.788564 | 0.779966 |
| O  | 0.277534 | 0.769762 | 0.772455 |
| O  | 0.776741 | 0.229830 | 0.772406 |
| O  | 0.995806 | 0.743747 | 0.733145 |
| O  | 0.495941 | 0.247593 | 0.732659 |
| O  | 0.253484 | 0.495785 | 0.732032 |
| O  | 0.744948 | 0.503493 | 0.731804 |
| O  | 0.245468 | 0.003291 | 0.726005 |
| O  | 0.753824 | 0.996356 | 0.725937 |
| O  | 0.003577 | 0.256823 | 0.725212 |
| O  | 0.503563 | 0.751000 | 0.724944 |
| O  | 0.766419 | 0.739359 | 0.682717 |
| O  | 0.265009 | 0.261359 | 0.682668 |
| O  | 0.235104 | 0.742527 | 0.681624 |
| O  | 0.736264 | 0.259775 | 0.681532 |
| O  | 0.499304 | 0.732614 | 0.638123 |
| O  | 0.999222 | 0.232137 | 0.638076 |
| O  | 0.727387 | 0.000533 | 0.637459 |
| O  | 0.268037 | 0.004434 | 0.637434 |
| O  | 0.226226 | 0.500493 | 0.635924 |
| O  | 0.769028 | 0.503868 | 0.635844 |
| O  | 0.996519 | 0.774708 | 0.635310 |
| O | 0.496502 | 0.273381 | 0.635291 |
| O | 0.249929 | 0.249929 | 0.590881 |
| O | 0.749786 | 0.749786 | 0.590881 |
| O | 0.249929 | 0.749786 | 0.590881 |
| O | 0.749786 | 0.249929 | 0.590881 |
| O | 0.249929 | 0.000000 | 0.545440 |
| O | 0.249929 | 0.499857 | 0.545440 |
| O | 0.749786 | 0.000000 | 0.545440 |
| O | 0.749786 | 0.499857 | 0.545440 |
| O | 0.000000 | 0.749786 | 0.545440 |
| O | 0.000000 | 0.249929 | 0.545440 |
| O | 0.499857 | 0.249929 | 0.545440 |
| O | 0.499857 | 0.749786 | 0.545440 |
| O | 0.249929 | 0.249929 | 0.499998 |
| O | 0.249929 | 0.749786 | 0.499998 |
| O | 0.749786 | 0.249929 | 0.499998 |
| O | 0.749786 | 0.749786 | 0.499998 |
| O | 0.499857 | 0.249929 | 0.454560 |
| O | 0.499857 | 0.749786 | 0.454560 |
| O | 0.000000 | 0.749786 | 0.454560 |
| O | 0.000000 | 0.249929 | 0.454560 |
| O | 0.249929 | 0.000000 | 0.454560 |
| O | 0.749786 | 0.000000 | 0.454560 |
| O | 0.749786 | 0.499857 | 0.454560 |
| O | 0.249929 | 0.499857 | 0.454560 |
| O | 0.749786 | 0.749786 | 0.409119 |
| O | 0.249929 | 0.749786 | 0.409119 |
| O | 0.749786 | 0.249929 | 0.409119 |
| O | 0.249929 | 0.249929 | 0.409119 |
| O | 0.496502 | 0.273381 | 0.364709 |
| O | 0.996519 | 0.774708 | 0.364690 |
| O | 0.769028 | 0.503868 | 0.364156 |
| O | 0.226226 | 0.500493 | 0.364076 |
| O | 0.268037 | 0.004434 | 0.362566 |
| O | 0.727387 | 0.000533 | 0.362541 |
| O | 0.999222 | 0.232137 | 0.361924 |
| O | 0.499304 | 0.732614 | 0.361877 |
| O | 0.736264 | 0.259775 | 0.318468 |
| O | 0.235104 | 0.742527 | 0.318376 |
| O | 0.265009 | 0.261359 | 0.317332 |
| O | 0.766419 | 0.739359 | 0.317283 |
| O | 0.503563 | 0.751000 | 0.275056 |
| O | 0.003577 | 0.256823 | 0.274788 |
| O | 0.753824 | 0.996356 | 0.274063 |
| O | 0.245468 | 0.003291 | 0.273995 |
| O | 0.744948 | 0.503493 | 0.268196 |

O  0.253484  0.495785  0.267968
O  0.495941  0.247593  0.267341
O  0.995806  0.743747  0.266855
O  0.776741  0.229830  0.227594
O  0.277534  0.769762  0.227545
O  0.709609  0.788564  0.220034
O  0.209282  0.209090  0.219885
O  0.248498  0.435209  0.189066
O  0.751135  0.563681  0.189032
O  0.431161  0.215962  0.186932
O  0.929143  0.786809  0.186506
O  0.929143  0.786809  0.813494
O  0.431161  0.215962  0.813068
O  0.751135  0.563681  0.810968
O  0.248498  0.435209  0.810934

Structure 4. SrO-terminated SrZrO$_3$ slab at $\Theta$=0.00 CO$_2$ coverage

```
_cell_length_a   8.39465800
_cell_length_b   8.39465800
_cell_length_c   46.17061600
_cell_angle_alpha   90.00000000
_cell_angle_beta   90.00000000
_cell_angle_gamma   90.00000000
_symmetry_space_group_name_H-M   'P 1'
loop_
 _atom_site_type_symbol
 _atom_site_fract_x
 _atom_site_fract_y
 _atom_site_fract_z
  Sr 0.000000 0.000000 0.500000
  Sr 0.000000 0.500000 0.500000
  Sr 0.500000 0.000000 0.500000
  Sr 0.500000 0.500000 0.500000
  Sr 0.500000 0.500000 0.590909
  Sr 0.505364 0.495239 0.680163
  Sr 0.465410 0.489142 0.766726
  Sr 0.500000 0.000000 0.590909
  Sr 0.493708 0.003240 0.680169
  Sr 0.489626 0.964777 0.766721
  Sr 0.000000 0.500000 0.590909
  Sr 0.993590 0.503513 0.680183
  Sr 0.989357 0.465093 0.766661
  Sr 0.000000 0.000000 0.590909
  Sr 0.005599 0.995257 0.680219
  Sr 0.965637 0.989622 0.766745
  Sr 0.500000 0.500000 0.409091
  Sr 0.505364 0.495239 0.319837
  Sr 0.465410 0.489142 0.233274
  Sr 0.500000 0.000000 0.409091
  Sr 0.493708 0.003240 0.319831
  Sr 0.489626 0.964777 0.233279
  Sr 0.000000 0.500000 0.409091
  Sr 0.993590 0.503513 0.319817
  Sr 0.989357 0.465093 0.233339
  Sr 0.000000 0.000000 0.409091
  Sr 0.005599 0.995257 0.319781
  Sr 0.965637 0.989622 0.233255
  Zr 0.750000 0.750000 0.545455
  Zr 0.752429 0.749369 0.636200
  Zr 0.744661 0.744933 0.727276
  Zr 0.750000 0.250000 0.545455
  Zr 0.749345 0.252285 0.636205
```

| | | | |
|---|---|---|---|
| Zr | 0.745320 | 0.243827 | 0.727252 |
| Zr | 0.250000 | 0.750000 | 0.545455 |
| Zr | 0.249409 | 0.752306 | 0.636197 |
| Zr | 0.245499 | 0.743820 | 0.727239 |
| Zr | 0.250000 | 0.250000 | 0.545455 |
| Zr | 0.252480 | 0.249339 | 0.636209 |
| Zr | 0.244565 | 0.244874 | 0.727273 |
| Zr | 0.750000 | 0.750000 | 0.454545 |
| Zr | 0.752429 | 0.749369 | 0.363800 |
| Zr | 0.744661 | 0.744933 | 0.272724 |
| Zr | 0.750000 | 0.250000 | 0.454545 |
| Zr | 0.749345 | 0.252285 | 0.363795 |
| Zr | 0.745320 | 0.243827 | 0.272748 |
| Zr | 0.250000 | 0.750000 | 0.454545 |
| Zr | 0.249409 | 0.752306 | 0.363803 |
| Zr | 0.245499 | 0.743820 | 0.272761 |
| Zr | 0.250000 | 0.250000 | 0.454545 |
| Zr | 0.252480 | 0.249339 | 0.363791 |
| Zr | 0.244565 | 0.244874 | 0.272727 |
| O | 0.250000 | 0.250000 | 0.500000 |
| O | 0.250000 | 0.750000 | 0.500000 |
| O | 0.750000 | 0.250000 | 0.500000 |
| O | 0.750000 | 0.750000 | 0.500000 |
| O | 0.500000 | 0.750000 | 0.545455 |
| O | 0.500728 | 0.700015 | 0.637816 |
| O | 0.501773 | 0.801961 | 0.721460 |
| O | 0.500000 | 0.250000 | 0.545455 |
| O | 0.500338 | 0.307767 | 0.634070 |
| O | 0.500090 | 0.199881 | 0.731772 |
| O | 0.000000 | 0.750000 | 0.545455 |
| O | 0.000368 | 0.807686 | 0.634130 |
| O | 0.000198 | 0.699620 | 0.731544 |
| O | 0.000000 | 0.250000 | 0.545455 |
| O | 0.000715 | 0.200069 | 0.637897 |
| O | 0.001672 | 0.301793 | 0.721373 |
| O | 0.750000 | 0.750000 | 0.590909 |
| O | 0.774332 | 0.723556 | 0.681350 |
| O | 0.733638 | 0.802240 | 0.770683 |
| O | 0.750000 | 0.250000 | 0.590909 |
| O | 0.726136 | 0.277089 | 0.681377 |
| O | 0.802410 | 0.232305 | 0.770664 |
| O | 0.250000 | 0.750000 | 0.590909 |
| O | 0.227111 | 0.777511 | 0.681366 |
| O | 0.302182 | 0.732166 | 0.770663 |
| O | 0.250000 | 0.250000 | 0.590909 |
| O | 0.275022 | 0.223442 | 0.681359 |

```
O   0.232933   0.302252   0.770668
O   0.750000   0.500000   0.545455
O   0.807622   0.500332   0.633891
O   0.701082   0.499503   0.732011
O   0.750000   0.000000   0.545455
O   0.699898   0.000749   0.637977
O   0.802974   0.001018   0.721287
O   0.250000   0.500000   0.545455
O   0.199969   0.500769   0.638032
O   0.303015   0.500975   0.721270
O   0.250000   0.000000   0.545455
O   0.307587   0.000380   0.633857
O   0.200919   0.999478   0.731983
O   0.500000   0.750000   0.454545
O   0.500728   0.700015   0.362184
O   0.501773   0.801961   0.278540
O   0.500000   0.250000   0.454545
O   0.500338   0.307767   0.365930
O   0.500090   0.199881   0.268228
O   0.000000   0.750000   0.454545
O   0.000368   0.807686   0.365870
O   0.000198   0.699620   0.268456
O   0.000000   0.250000   0.454545
O   0.000715   0.200069   0.362103
O   0.001672   0.301793   0.278627
O   0.750000   0.750000   0.409091
O   0.774332   0.723556   0.318650
O   0.733638   0.802240   0.229317
O   0.750000   0.250000   0.409091
O   0.726136   0.277089   0.318623
O   0.802410   0.232305   0.229336
O   0.250000   0.750000   0.409091
O   0.227111   0.777511   0.318634
O   0.302182   0.732166   0.229337
O   0.250000   0.250000   0.409091
O   0.275022   0.223442   0.318641
O   0.232933   0.302252   0.229332
O   0.750000   0.500000   0.454545
O   0.807622   0.500332   0.366109
O   0.701082   0.499503   0.267989
O   0.750000   0.000000   0.454545
O   0.699898   0.000749   0.362023
O   0.802974   0.001018   0.278713
O   0.250000   0.500000   0.454545
O   0.199969   0.500769   0.361968
O   0.303015   0.500975   0.278730
```

O  0.250000  0.000000  0.454545
O  0.307587  0.000380  0.366143
O  0.200919  0.999478  0.268017

Structure 5. SrO-terminated SrZrO$_3$ slab at $\Theta$=0.25 CO$_2$ coverage

_cell_length_a   8.39465800
_cell_length_b   8.39465800
_cell_length_c   46.17061600
_cell_angle_alpha   90.00000000
_cell_angle_beta   90.00000000
_cell_angle_gamma   90.00000000
_symmetry_space_group_name_H-M   'P 1'
loop_
_atom_site_type_symbol
_atom_site_fract_x
_atom_site_fract_y
_atom_site_fract_z
 Sr  0.009646  0.496240  0.682449
 Sr  0.063690  0.510220  0.774130
 Sr  0.512162  0.494854  0.681737
 Sr  0.516630  0.488118  0.773239
 Sr  0.999200  0.002524  0.678629
 Sr  0.970324  0.015458  0.766356
 Sr  0.481917  0.004313  0.679541
 Sr  0.491216  0.015510  0.765120
 Sr  0.500000  0.000000  0.500000
 Sr  0.000000  0.000000  0.500000
 Sr  0.500000  0.500000  0.500000
 Sr  0.000000  0.500000  0.500000
 Sr  0.000000  0.500000  0.590909
 Sr  0.500000  0.500000  0.590909
 Sr  0.000000  0.000000  0.590909
 Sr  0.500000  0.000000  0.590909
 Sr  0.009646  0.496240  0.317551
 Sr  0.063690  0.510220  0.225870
 Sr  0.512162  0.494854  0.318263
 Sr  0.516630  0.488118  0.226761
 Sr  0.999200  0.002524  0.321371
 Sr  0.970324  0.015458  0.233644
 Sr  0.481917  0.004313  0.320459
 Sr  0.491216  0.015510  0.234880
 Sr  0.000000  0.500000  0.409091
 Sr  0.500000  0.500000  0.409091
 Sr  0.000000  0.000000  0.409091
 Sr  0.500000  0.000000  0.409091
 Zr  0.251103  0.256048  0.727479
 Zr  0.749584  0.257189  0.727529
 Zr  0.255317  0.750570  0.728193
 Zr  0.752177  0.751062  0.724260
 Zr  0.250000  0.250000  0.545455

Zr 0.247782 0.247756 0.636226
Zr 0.750000 0.250000 0.545455
Zr 0.753238 0.252106 0.636404
Zr 0.250000 0.750000 0.545455
Zr 0.248659 0.752075 0.636607
Zr 0.750000 0.750000 0.545455
Zr 0.752702 0.747353 0.635671
Zr 0.251103 0.256048 0.272521
Zr 0.749584 0.257189 0.272471
Zr 0.255317 0.750570 0.271807
Zr 0.752177 0.751062 0.275740
Zr 0.250000 0.250000 0.454545
Zr 0.247782 0.247756 0.363774
Zr 0.750000 0.250000 0.454545
Zr 0.753238 0.252106 0.363596
Zr 0.250000 0.750000 0.454545
Zr 0.248659 0.752075 0.363393
Zr 0.750000 0.750000 0.454545
Zr 0.752702 0.747353 0.364329
C 0.808719 0.719018 0.792852
C 0.808719 0.719018 0.207148
O 0.274401 0.501145 0.716502
O 0.694103 0.505054 0.729886
O 0.219101 0.001328 0.735689
O 0.800655 0.998446 0.721737
O 0.221425 0.213268 0.681202
O 0.779000 0.281346 0.681267
O 0.714687 0.202857 0.769695
O 0.226594 0.787390 0.681484
O 0.298800 0.688785 0.769714
O 0.779779 0.716081 0.681549
O 0.000314 0.294061 0.733460
O 0.499408 0.209603 0.720153
O 0.997309 0.713506 0.734450
O 0.505235 0.804482 0.721925
O 0.290582 0.293512 0.770900
O 0.710706 0.798128 0.774784
O 0.776388 0.573088 0.799038
O 0.937418 0.786937 0.801324
O 0.750000 0.750000 0.500000
O 0.250000 0.750000 0.500000
O 0.750000 0.250000 0.500000
O 0.250000 0.250000 0.500000
O 0.250000 0.500000 0.545455
O 0.200338 0.499662 0.638192
O 0.750000 0.500000 0.545455

```
O  0.802619  0.500316  0.632928
O  0.250000  0.000000  0.545455
O  0.307631  0.999805  0.633231
O  0.750000  0.000000  0.545455
O  0.702280  0.998840  0.639239
O  0.250000  0.250000  0.590909
O  0.750000  0.250000  0.590909
O  0.250000  0.750000  0.590909
O  0.750000  0.750000  0.590909
O  0.000000  0.250000  0.545455
O  0.999806  0.193997  0.633527
O  0.500000  0.250000  0.545455
O  0.500011  0.298187  0.638515
O  0.000000  0.750000  0.545455
O  0.999599  0.804935  0.633990
O  0.500000  0.750000  0.545455
O  0.501137  0.699229  0.638386
O  0.274401  0.501145  0.283498
O  0.694103  0.505054  0.270114
O  0.219101  0.001328  0.264311
O  0.800655  0.998446  0.278263
O  0.221425  0.213268  0.318798
O  0.779000  0.281346  0.318733
O  0.714687  0.202857  0.230305
O  0.226594  0.787390  0.318516
O  0.298800  0.688785  0.230286
O  0.779779  0.716081  0.318451
O  0.000314  0.294061  0.266540
O  0.499408  0.209603  0.279847
O  0.997309  0.713506  0.265550
O  0.505235  0.804482  0.278075
O  0.290582  0.293512  0.229100
O  0.710706  0.798128  0.225216
O  0.776388  0.573088  0.200962
O  0.937418  0.786937  0.198676
O  0.250000  0.500000  0.454545
O  0.200338  0.499662  0.361808
O  0.750000  0.500000  0.454545
O  0.802619  0.500316  0.367072
O  0.250000  0.000000  0.454545
O  0.307631  0.999805  0.366769
O  0.750000  0.000000  0.454545
O  0.702280  0.998840  0.360761
O  0.250000  0.250000  0.409091
O  0.750000  0.250000  0.409091
O  0.250000  0.750000  0.409091
```

```
O  0.750000  0.750000  0.409091
O  0.000000  0.250000  0.454545
O  0.999806  0.193997  0.366473
O  0.500000  0.250000  0.454545
O  0.500011  0.298187  0.361485
O  0.000000  0.750000  0.454545
O  0.999599  0.804935  0.366010
O  0.500000  0.750000  0.454545
O  0.501137  0.699229  0.361614
```

Structure 6. SrO-terminated SrZrO$_3$ slab at $\Theta$=0.50 CO$_2$ coverage

_cell_length_a  8.39465800
_cell_length_b  8.39465800
_cell_length_c  46.17061600
_cell_angle_alpha  90.00000000
_cell_angle_beta  90.00000000
_cell_angle_gamma  90.00000000
_symmetry_space_group_name_H-M  'P 1'
loop_
 _atom_site_type_symbol
 _atom_site_fract_x
 _atom_site_fract_y
 _atom_site_fract_z
 Sr 0.488259 0.552294 0.223789
 Sr 0.504799 0.026993 0.226473
 Sr 0.969224 0.464750 0.228811
 Sr 0.992058 0.994157 0.232747
 Sr 0.496803 0.505929 0.316756
 Sr 0.501055 0.014954 0.317392
 Sr 0.000573 0.985463 0.320300
 Sr 0.002099 0.493132 0.321582
 Sr 0.500000 0.000000 0.409091
 Sr 0.000000 0.500000 0.409091
 Sr 0.500000 0.500000 0.409091
 Sr 0.000000 0.000000 0.409091
 Sr 0.000000 0.000000 0.500000
 Sr 0.000000 0.500000 0.500000
 Sr 0.500000 0.000000 0.500000
 Sr 0.500000 0.500000 0.500000
 Sr 0.500000 0.000000 0.590909
 Sr 0.500000 0.500000 0.590909
 Sr 0.000000 0.000000 0.590909
 Sr 0.000000 0.500000 0.590909
 Sr 0.002099 0.493132 0.678418
 Sr 0.000573 0.985463 0.679700
 Sr 0.501055 0.014954 0.682608
 Sr 0.496803 0.505929 0.683244
 Sr 0.992058 0.994157 0.767253
 Sr 0.969224 0.464750 0.771189
 Sr 0.504799 0.026993 0.773527
 Sr 0.488259 0.552294 0.776210
 Zr 0.249419 0.752535 0.270993
 Zr 0.743973 0.250541 0.272363
 Zr 0.245534 0.253564 0.275415
 Zr 0.749118 0.748563 0.275817
 Zr 0.248547 0.748457 0.363175

```
Zr  0.747256  0.252858  0.363432
Zr  0.752660  0.747450  0.364246
Zr  0.253144  0.252053  0.364429
Zr  0.750000  0.750000  0.454545
Zr  0.750000  0.250000  0.454545
Zr  0.250000  0.750000  0.454545
Zr  0.250000  0.250000  0.454545
Zr  0.750000  0.750000  0.545455
Zr  0.750000  0.250000  0.545455
Zr  0.250000  0.750000  0.545455
Zr  0.250000  0.250000  0.545455
Zr  0.253144  0.252053  0.635571
Zr  0.752660  0.747450  0.635754
Zr  0.747256  0.252858  0.636568
Zr  0.248547  0.748457  0.636825
Zr  0.749118  0.748563  0.724183
Zr  0.245534  0.253564  0.724585
Zr  0.743973  0.250541  0.727637
Zr  0.249419  0.752535  0.729007
C   0.789373  0.735991  0.204379
C   0.281323  0.305875  0.206683
C   0.281323  0.305875  0.793317
C   0.789373  0.735991  0.795621
O   0.902687  0.814131  0.193441
O   0.751835  0.591832  0.197661
O   0.215509  0.436338  0.199088
O   0.424792  0.270518  0.199571
O   0.204210  0.208603  0.225313
O   0.702888  0.803044  0.226370
O   0.791914  0.209875  0.230664
O   0.319991  0.797525  0.230737
O   0.995202  0.720926  0.262651
O   0.707429  0.503405  0.266476
O   0.290670  0.497802  0.267250
O   0.494194  0.199306  0.270193
O   0.000037  0.298510  0.278129
O   0.201635  0.004252  0.278424
O   0.785766  0.996618  0.280362
O   0.503903  0.774991  0.283524
O   0.213202  0.726423  0.318496
O   0.279367  0.278002  0.318569
O   0.794412  0.717073  0.318664
O   0.717830  0.279669  0.318886
O   0.000972  0.202962  0.360753
O   0.701816  0.998867  0.361292
O   0.501655  0.699747  0.361550
```

| | | | |
|---|---|---|---|
| O | 0.300673 | 0.000958 | 0.361745 |
| O | 0.196107 | 0.499509 | 0.366107 |
| O | 0.806721 | 0.500101 | 0.366608 |
| O | 0.000123 | 0.808154 | 0.366727 |
| O | 0.499361 | 0.302442 | 0.367105 |
| O | 0.750000 | 0.750000 | 0.409091 |
| O | 0.750000 | 0.250000 | 0.409091 |
| O | 0.250000 | 0.250000 | 0.409091 |
| O | 0.250000 | 0.750000 | 0.409091 |
| O | 0.500000 | 0.750000 | 0.454545 |
| O | 0.500000 | 0.250000 | 0.454545 |
| O | 0.750000 | 0.500000 | 0.454545 |
| O | 0.750000 | 0.000000 | 0.454545 |
| O | 0.250000 | 0.000000 | 0.454545 |
| O | 0.250000 | 0.500000 | 0.454545 |
| O | 0.000000 | 0.250000 | 0.454545 |
| O | 0.000000 | 0.750000 | 0.454545 |
| O | 0.250000 | 0.250000 | 0.500000 |
| O | 0.250000 | 0.750000 | 0.500000 |
| O | 0.750000 | 0.250000 | 0.500000 |
| O | 0.750000 | 0.750000 | 0.500000 |
| O | 0.500000 | 0.750000 | 0.545455 |
| O | 0.500000 | 0.250000 | 0.545455 |
| O | 0.000000 | 0.750000 | 0.545455 |
| O | 0.000000 | 0.250000 | 0.545455 |
| O | 0.750000 | 0.500000 | 0.545455 |
| O | 0.750000 | 0.000000 | 0.545455 |
| O | 0.250000 | 0.500000 | 0.545455 |
| O | 0.250000 | 0.000000 | 0.545455 |
| O | 0.750000 | 0.750000 | 0.590909 |
| O | 0.750000 | 0.250000 | 0.590909 |
| O | 0.250000 | 0.750000 | 0.590909 |
| O | 0.250000 | 0.250000 | 0.590909 |
| O | 0.499361 | 0.302442 | 0.632895 |
| O | 0.000123 | 0.808154 | 0.633273 |
| O | 0.806721 | 0.500101 | 0.633392 |
| O | 0.196107 | 0.499509 | 0.633893 |
| O | 0.300673 | 0.000958 | 0.638255 |
| O | 0.501655 | 0.699747 | 0.638450 |
| O | 0.701816 | 0.998867 | 0.638708 |
| O | 0.000972 | 0.202962 | 0.639247 |
| O | 0.717830 | 0.279669 | 0.681114 |
| O | 0.794412 | 0.717073 | 0.681336 |
| O | 0.279367 | 0.278002 | 0.681431 |
| O | 0.213202 | 0.726423 | 0.681504 |
| O | 0.503903 | 0.774991 | 0.716476 |

O 0.785766 0.996618 0.719638
O 0.201635 0.004252 0.721576
O 0.000037 0.298510 0.721871
O 0.494194 0.199306 0.729807
O 0.290670 0.497802 0.732750
O 0.707429 0.503405 0.733524
O 0.995202 0.720926 0.737349
O 0.319991 0.797525 0.769263
O 0.791914 0.209875 0.769336
O 0.702888 0.803044 0.773630
O 0.204210 0.208603 0.774687
O 0.424792 0.270518 0.800429
O 0.215509 0.436338 0.800912
O 0.751835 0.591832 0.802339
O 0.902687 0.814131 0.806559

Structure 7. SrO-terminated SrHfO$_3$ slab at $\Theta$=0.00 CO$_2$ coverage

_cell_length_a  8.28657000
_cell_length_b  8.28657000
_cell_length_c  45.57613800
_cell_angle_alpha  90.00000000
_cell_angle_beta  90.00000000
_cell_angle_gamma  90.00000000
_symmetry_space_group_name_H-M  'P 1'
loop_
 _atom_site_type_symbol
 _atom_site_fract_x
 _atom_site_fract_y
 _atom_site_fract_z
 Sr  0.000000  0.000000  0.500000
 Sr  0.000000  0.500000  0.500000
 Sr  0.500000  0.000000  0.500000
 Sr  0.500000  0.500000  0.500000
 Sr  0.500000  0.500000  0.590909
 Sr  0.504750  0.497550  0.680668
 Sr  0.469417  0.491731  0.767919
 Sr  0.500000  0.000000  0.590909
 Sr  0.497121  0.003618  0.680698
 Sr  0.492496  0.968909  0.767943
 Sr  0.000000  0.500000  0.590909
 Sr  0.996949  0.503899  0.680690
 Sr  0.992365  0.469238  0.767883
 Sr  0.000000  0.000000  0.590909
 Sr  0.004990  0.997502  0.680716
 Sr  0.969626  0.992375  0.767940
 Sr  0.500000  0.500000  0.409091
 Sr  0.504750  0.497550  0.319332
 Sr  0.469417  0.491731  0.232082
 Sr  0.500000  0.000000  0.409091
 Sr  0.497121  0.003618  0.319302
 Sr  0.492496  0.968909  0.232057
 Sr  0.000000  0.500000  0.409091
 Sr  0.996949  0.503899  0.319310
 Sr  0.992365  0.469238  0.232117
 Sr  0.000000  0.000000  0.409091
 Sr  0.004990  0.997502  0.319284
 Sr  0.969626  0.992375  0.232060
 Hf  0.750000  0.750000  0.545455
 Hf  0.752155  0.749302  0.636299
 Hf  0.746753  0.746352  0.727657
 Hf  0.750000  0.250000  0.545455
 Hf  0.749390  0.252031  0.636306

```
Hf 0.746702 0.246186 0.727641
Hf 0.250000 0.750000 0.545455
Hf 0.249462 0.752035 0.636299
Hf 0.246815 0.746125 0.727647
Hf 0.250000 0.250000 0.545455
Hf 0.252206 0.249264 0.636307
Hf 0.246693 0.246229 0.727644
Hf 0.750000 0.750000 0.454545
Hf 0.752155 0.749302 0.363701
Hf 0.746753 0.746352 0.272343
Hf 0.750000 0.250000 0.454545
Hf 0.749390 0.252031 0.363694
Hf 0.746702 0.246186 0.272359
Hf 0.250000 0.750000 0.454545
Hf 0.249462 0.752035 0.363701
Hf 0.246815 0.746125 0.272353
Hf 0.250000 0.250000 0.454545
Hf 0.252206 0.249264 0.363693
Hf 0.246693 0.246229 0.272356
O 0.250000 0.250000 0.500000
O 0.250000 0.750000 0.500000
O 0.750000 0.250000 0.500000
O 0.750000 0.750000 0.500000
O 0.500000 0.750000 0.545455
O 0.500485 0.705403 0.637678
O 0.500701 0.796519 0.722634
O 0.500000 0.250000 0.545455
O 0.500237 0.299704 0.634515
O 0.499698 0.203092 0.731302
O 0.000000 0.750000 0.545455
O 0.000261 0.799612 0.634573
O 0.999758 0.702953 0.731135
O 0.000000 0.250000 0.545455
O 0.000470 0.205460 0.637746
O 0.000646 0.296370 0.722541
O 0.750000 0.750000 0.590909
O 0.768869 0.728215 0.681618
O 0.735181 0.794060 0.771591
O 0.750000 0.250000 0.590909
O 0.729343 0.270250 0.681628
O 0.794212 0.234083 0.771581
O 0.250000 0.750000 0.590909
O 0.230176 0.770347 0.681625
O 0.293974 0.733761 0.771590
O 0.250000 0.250000 0.590909
O 0.269550 0.227998 0.681622
```

```
O  0.234497  0.293810  0.771580
O  0.750000  0.500000  0.545455
O  0.799778  0.500139  0.634444
O  0.703823  0.499256  0.731381
O  0.750000  0.000000  0.545455
O  0.705533  0.000374  0.637781
O  0.797065  0.000188  0.722491
O  0.250000  0.500000  0.545455
O  0.205563  0.500359  0.637815
O  0.297083  0.500078  0.722495
O  0.250000  0.000000  0.545455
O  0.299737  0.000133  0.634421
O  0.203732  0.999159  0.731369
O  0.500000  0.750000  0.454545
O  0.500485  0.705403  0.362322
O  0.500701  0.796519  0.277366
O  0.500000  0.250000  0.454545
O  0.500237  0.299704  0.365485
O  0.499698  0.203092  0.268698
O  0.000000  0.750000  0.454545
O  0.000261  0.799612  0.365427
O  0.999758  0.702953  0.268865
O  0.000000  0.250000  0.454545
O  0.000470  0.205460  0.362254
O  0.000646  0.296370  0.277459
O  0.750000  0.750000  0.409091
O  0.768869  0.728215  0.318382
O  0.735181  0.794060  0.228409
O  0.750000  0.250000  0.409091
O  0.729343  0.270250  0.318372
O  0.794212  0.234083  0.228419
O  0.250000  0.750000  0.409091
O  0.230176  0.770347  0.318375
O  0.293974  0.733761  0.228410
O  0.250000  0.250000  0.409091
O  0.269550  0.227998  0.318378
O  0.234497  0.293810  0.228420
O  0.750000  0.500000  0.454545
O  0.799778  0.500139  0.365556
O  0.703823  0.499256  0.268619
O  0.750000  0.000000  0.454545
O  0.705533  0.000374  0.362219
O  0.797065  0.000188  0.277509
O  0.250000  0.500000  0.454545
O  0.205563  0.500359  0.362185
O  0.297083  0.500078  0.277505
```

O  0.250000  0.000000  0.454545
O  0.299737  0.000133  0.365579
O  0.203732  0.999159  0.268631

Structure 8. SrO-terminated SrHfO$_3$ slab at $\Theta$=0.25 CO$_2$ coverage

```
_cell_length_a   8.28657000
_cell_length_b   8.28657000
_cell_length_c   45.57613800
_cell_angle_alpha   90.00000000
_cell_angle_beta   90.00000000
_cell_angle_gamma   90.00000000
_symmetry_space_group_name_H-M   'P 1'
loop_
_atom_site_type_symbol
_atom_site_fract_x
_atom_site_fract_y
_atom_site_fract_z
 Sr  0.060407  0.508983  0.774546
 Sr  0.511355  0.487681  0.773020
 Sr  0.972755  0.017601  0.767632
 Sr  0.487616  0.015430  0.766816
 Sr  0.500000  0.000000  0.500000
 Sr  0.000000  0.000000  0.500000
 Sr  0.500000  0.500000  0.500000
 Sr  0.000000  0.500000  0.500000
 Sr  0.000000  0.500000  0.590909
 Sr  0.007771  0.497052  0.682679
 Sr  0.500000  0.500000  0.590909
 Sr  0.508485  0.495875  0.681855
 Sr  0.000000  0.000000  0.590909
 Sr  0.000767  0.002255  0.679517
 Sr  0.500000  0.000000  0.590909
 Sr  0.487364  0.004655  0.679986
 Sr  0.060407  0.508983  0.225454
 Sr  0.511355  0.487681  0.226980
 Sr  0.972755  0.017601  0.232368
 Sr  0.487616  0.015430  0.233184
 Sr  0.000000  0.500000  0.409091
 Sr  0.007771  0.497052  0.317321
 Sr  0.500000  0.500000  0.409091
 Sr  0.508485  0.495875  0.318145
 Sr  0.000000  0.000000  0.409091
 Sr  0.000767  0.002255  0.320483
 Sr  0.500000  0.000000  0.409091
 Sr  0.487364  0.004655  0.320014
 Hf  0.249505  0.254644  0.727560
 Hf  0.748519  0.255270  0.727792
 Hf  0.254314  0.750685  0.728391
 Hf  0.750831  0.751531  0.724828
 Hf  0.250000  0.250000  0.545455
```

```
Hf  0.248002  0.248332  0.636221
Hf  0.750000  0.250000  0.545455
Hf  0.752636  0.252107  0.636426
Hf  0.250000  0.750000  0.545455
Hf  0.248674  0.751810  0.636621
Hf  0.750000  0.750000  0.545455
Hf  0.752526  0.747792  0.635841
Hf  0.249505  0.254644  0.272440
Hf  0.748519  0.255270  0.272208
Hf  0.254314  0.750685  0.271609
Hf  0.750831  0.751531  0.275172
Hf  0.250000  0.250000  0.454545
Hf  0.248002  0.248332  0.363779
Hf  0.750000  0.250000  0.454545
Hf  0.752636  0.252107  0.363574
Hf  0.250000  0.750000  0.454545
Hf  0.248674  0.751810  0.363379
Hf  0.750000  0.750000  0.454545
Hf  0.752526  0.747792  0.364159
C   0.805828  0.717744  0.794062
C   0.805828  0.717744  0.205938
O   0.275030  0.501232  0.718943
O   0.700697  0.505010  0.730023
O   0.217409  0.001665  0.733873
O   0.795473  0.999214  0.722330
O   0.717612  0.204832  0.770430
O   0.294724  0.698957  0.770728
O   0.999060  0.291565  0.732293
O   0.498136  0.213854  0.721563
O   0.996913  0.716064  0.733954
O   0.503866  0.799494  0.722570
O   0.281754  0.285020  0.771747
O   0.710892  0.796425  0.774670
O   0.770104  0.571493  0.800677
O   0.934889  0.786911  0.803057
O   0.750000  0.750000  0.500000
O   0.250000  0.750000  0.500000
O   0.750000  0.250000  0.500000
O   0.250000  0.250000  0.500000
O   0.250000  0.500000  0.545455
O   0.205956  0.499865  0.638081
O   0.750000  0.500000  0.545455
O   0.796714  0.500346  0.633674
O   0.250000  0.000000  0.545455
O   0.299826  0.999990  0.633954
O   0.750000  0.000000  0.545455
```

O  0.706900  0.999223  0.638644
O  0.250000  0.250000  0.590909
O  0.225975  0.221563  0.681410
O  0.750000  0.250000  0.590909
O  0.770921  0.275328  0.681473
O  0.250000  0.750000  0.590909
O  0.228743  0.778718  0.681727
O  0.750000  0.750000  0.590909
O  0.774676  0.723245  0.681670
O  0.000000  0.250000  0.545455
O  0.999753  0.201412  0.634101
O  0.500000  0.250000  0.545455
O  0.499701  0.293333  0.638095
O  0.000000  0.750000  0.545455
O  0.999751  0.798493  0.634340
O  0.500000  0.750000  0.545455
O  0.500849  0.705263  0.638210
O  0.275030  0.501232  0.281057
O  0.700697  0.505010  0.269977
O  0.217409  0.001665  0.266127
O  0.795473  0.999214  0.277670
O  0.717612  0.204832  0.229570
O  0.294724  0.698957  0.229272
O  0.999060  0.291565  0.267707
O  0.498136  0.213854  0.278437
O  0.996913  0.716064  0.266046
O  0.503866  0.799494  0.277430
O  0.281754  0.285020  0.228253
O  0.710892  0.796425  0.225330
O  0.770104  0.571493  0.199323
O  0.934889  0.786911  0.196943
O  0.250000  0.500000  0.454545
O  0.205956  0.499865  0.361919
O  0.750000  0.500000  0.454545
O  0.796714  0.500346  0.366326
O  0.250000  0.000000  0.454545
O  0.299826  0.999990  0.366046
O  0.750000  0.000000  0.454545
O  0.706900  0.999223  0.361356
O  0.250000  0.250000  0.409091
O  0.225975  0.221563  0.318590
O  0.750000  0.250000  0.409091
O  0.770921  0.275328  0.318527
O  0.250000  0.750000  0.409091
O  0.228743  0.778718  0.318273
O  0.750000  0.750000  0.409091

O   0.774676   0.723245   0.318330
O   0.000000   0.250000   0.454545
O   0.999753   0.201412   0.365899
O   0.500000   0.250000   0.454545
O   0.499701   0.293333   0.361905
O   0.000000   0.750000   0.454545
O   0.999751   0.798493   0.365660
O   0.500000   0.750000   0.454545
O   0.500849   0.705263   0.361790

Structure 9. SrO-terminated SrHfO$_3$ slab at $\Theta$=0.50 CO$_2$ coverage

_cell_length_a  8.28657000
_cell_length_b  8.28657000
_cell_length_c  45.57613800
_cell_angle_alpha  90.00000000
_cell_angle_beta  90.00000000
_cell_angle_gamma  90.00000000
_symmetry_space_group_name_H-M  'P 1'
loop_
 _atom_site_type_symbol
 _atom_site_fract_x
 _atom_site_fract_y
 _atom_site_fract_z
  Sr  0.487469  0.548858  0.223367
  Sr  0.503946  0.022376  0.226542
  Sr  0.970930  0.462635  0.227991
  Sr  0.993480  0.992132  0.231467
  Sr  0.497591  0.504559  0.316649
  Sr  0.500267  0.010658  0.317448
  Sr  0.000134  0.989868  0.319877
  Sr  0.001300  0.494139  0.320701
  Sr  0.500000  0.500000  0.409091
  Sr  0.500000  0.000000  0.409091
  Sr  0.000000  0.000000  0.409091
  Sr  0.000000  0.500000  0.409091
  Sr  0.000000  0.000000  0.500000
  Sr  0.000000  0.500000  0.500000
  Sr  0.500000  0.000000  0.500000
  Sr  0.500000  0.500000  0.500000
  Sr  0.500000  0.500000  0.590909
  Sr  0.500000  0.000000  0.590909
  Sr  0.000000  0.500000  0.590909
  Sr  0.000000  0.000000  0.590909
  Sr  0.001300  0.494139  0.679299
  Sr  0.000134  0.989868  0.680123
  Sr  0.500267  0.010658  0.682552
  Sr  0.497591  0.504559  0.683351
  Sr  0.993480  0.992132  0.768533
  Sr  0.970930  0.462635  0.772008
  Sr  0.503946  0.022376  0.773458
  Sr  0.487469  0.548858  0.776633
  Hf  0.249707  0.751817  0.270847
  Hf  0.745476  0.248590  0.272150
  Hf  0.246012  0.251971  0.275020
  Hf  0.749342  0.747605  0.275194
  Hf  0.248296  0.748448  0.363154

```
Hf  0.747355  0.252412  0.363448
Hf  0.752343  0.747559  0.364108
Hf  0.252545  0.251879  0.364282
Hf  0.750000  0.750000  0.454545
Hf  0.750000  0.250000  0.454545
Hf  0.250000  0.750000  0.454545
Hf  0.250000  0.250000  0.454545
Hf  0.750000  0.750000  0.545455
Hf  0.750000  0.250000  0.545455
Hf  0.250000  0.750000  0.545455
Hf  0.250000  0.250000  0.545455
Hf  0.252545  0.251879  0.635718
Hf  0.752343  0.747559  0.635892
Hf  0.747355  0.252412  0.636552
Hf  0.248296  0.748448  0.636846
Hf  0.749342  0.747605  0.724806
Hf  0.246012  0.251971  0.724980
Hf  0.745476  0.248590  0.727850
Hf  0.249707  0.751817  0.729153
C   0.789515  0.732186  0.202807
C   0.282585  0.302947  0.205103
C   0.282585  0.302947  0.794897
C   0.789515  0.732186  0.797193
O   0.901429  0.813283  0.191345
O   0.749993  0.587204  0.195820
O   0.216426  0.434382  0.196990
O   0.425230  0.262230  0.197403
O   0.205871  0.210584  0.225379
O   0.706337  0.797521  0.226088
O   0.311846  0.793390  0.229732
O   0.789995  0.211001  0.230044
O   0.995200  0.720653  0.263594
O   0.711158  0.501812  0.267018
O   0.285569  0.497100  0.267534
O   0.494251  0.205317  0.270146
O   0.206371  0.003073  0.277615
O   0.999685  0.291794  0.277651
O   0.781863  0.995718  0.279267
O   0.502905  0.773664  0.281520
O   0.218633  0.729444  0.318189
O   0.786604  0.720884  0.318461
O   0.273593  0.272968  0.318487
O   0.724383  0.274165  0.318655
O   0.000660  0.207658  0.361347
O   0.501105  0.705861  0.361483
O   0.707027  0.998849  0.361541
```

| | | | |
|---|---|---|---|
| O | 0.294453 | 0.000627 | 0.361904 |
| O | 0.202439 | 0.499583 | 0.365688 |
| O | 0.799120 | 0.499936 | 0.366179 |
| O | 0.999968 | 0.800343 | 0.366211 |
| O | 0.499401 | 0.296180 | 0.366390 |
| O | 0.750000 | 0.750000 | 0.409091 |
| O | 0.750000 | 0.250000 | 0.409091 |
| O | 0.250000 | 0.250000 | 0.409091 |
| O | 0.250000 | 0.750000 | 0.409091 |
| O | 0.000000 | 0.750000 | 0.454545 |
| O | 0.000000 | 0.250000 | 0.454545 |
| O | 0.500000 | 0.250000 | 0.454545 |
| O | 0.500000 | 0.750000 | 0.454545 |
| O | 0.750000 | 0.500000 | 0.454545 |
| O | 0.750000 | 0.000000 | 0.454545 |
| O | 0.250000 | 0.000000 | 0.454545 |
| O | 0.250000 | 0.500000 | 0.454545 |
| O | 0.250000 | 0.250000 | 0.500000 |
| O | 0.250000 | 0.750000 | 0.500000 |
| O | 0.750000 | 0.250000 | 0.500000 |
| O | 0.750000 | 0.750000 | 0.500000 |
| O | 0.500000 | 0.750000 | 0.545455 |
| O | 0.500000 | 0.250000 | 0.545455 |
| O | 0.000000 | 0.750000 | 0.545455 |
| O | 0.000000 | 0.250000 | 0.545455 |
| O | 0.750000 | 0.500000 | 0.545455 |
| O | 0.750000 | 0.000000 | 0.545455 |
| O | 0.250000 | 0.500000 | 0.545455 |
| O | 0.250000 | 0.000000 | 0.545455 |
| O | 0.750000 | 0.750000 | 0.590909 |
| O | 0.750000 | 0.250000 | 0.590909 |
| O | 0.250000 | 0.750000 | 0.590909 |
| O | 0.250000 | 0.250000 | 0.590909 |
| O | 0.499401 | 0.296180 | 0.633610 |
| O | 0.999968 | 0.800343 | 0.633789 |
| O | 0.799120 | 0.499936 | 0.633821 |
| O | 0.202439 | 0.499583 | 0.634312 |
| O | 0.294453 | 0.000627 | 0.638096 |
| O | 0.707027 | 0.998849 | 0.638459 |
| O | 0.501105 | 0.705861 | 0.638517 |
| O | 0.000660 | 0.207658 | 0.638653 |
| O | 0.724383 | 0.274165 | 0.681345 |
| O | 0.273593 | 0.272968 | 0.681513 |
| O | 0.786604 | 0.720884 | 0.681539 |
| O | 0.218633 | 0.729444 | 0.681811 |
| O | 0.502905 | 0.773664 | 0.718480 |

O 0.781863 0.995718 0.720733
O 0.999685 0.291794 0.722349
O 0.206371 0.003073 0.722385
O 0.494251 0.205317 0.729854
O 0.285569 0.497100 0.732466
O 0.711158 0.501812 0.732982
O 0.995200 0.720653 0.736406
O 0.789995 0.211001 0.769956
O 0.311846 0.793390 0.770268
O 0.706337 0.797521 0.773912
O 0.205871 0.210584 0.774621
O 0.425230 0.262230 0.802597
O 0.216426 0.434382 0.803010
O 0.749993 0.587204 0.804180
O 0.901429 0.813283 0.808655

Structure 10. BaO-terminated BaTiO$_3$ slab at $\Theta$=0.00 CO$_2$ coverage

_cell_length_a  8.07335600
_cell_length_b  8.07335600
_cell_length_c  44.40346100
_cell_angle_alpha  90.00000000
_cell_angle_beta  90.00000000
_cell_angle_gamma  90.00000000
_symmetry_space_group_name_H-M  'P 1'

loop_
_atom_site_type_symbol
_atom_site_fract_x
_atom_site_fract_y
_atom_site_fract_z
 Ba 0.000000 0.000000 0.500000
 Ba 0.000000 0.000000 0.590909
 Ba 0.001708 0.996694 0.681867
 Ba 0.000233 0.996620 0.771461
 Ba 0.000000 0.500000 0.500000
 Ba 0.000000 0.500000 0.590909
 Ba 0.001250 0.497026 0.681845
 Ba 0.998711 0.496738 0.771474
 Ba 0.500000 0.000000 0.500000
 Ba 0.500000 0.000000 0.590909
 Ba 0.501717 0.996723 0.681849
 Ba 0.500129 0.996622 0.771462
 Ba 0.500000 0.500000 0.500000
 Ba 0.500000 0.500000 0.590909
 Ba 0.501331 0.497004 0.681825
 Ba 0.498617 0.496728 0.771475
 Ba 0.498617 0.496728 0.228525
 Ba 0.501331 0.497004 0.318175
 Ba 0.500000 0.500000 0.409091
 Ba 0.500129 0.996622 0.228538
 Ba 0.501717 0.996723 0.318151
 Ba 0.500000 0.000000 0.409091
 Ba 0.998711 0.496738 0.228526
 Ba 0.001250 0.497026 0.318155
 Ba 0.000000 0.500000 0.409091
 Ba 0.000233 0.996620 0.228539
 Ba 0.001708 0.996694 0.318133
 Ba 0.000000 0.000000 0.409091
 Ti 0.250000 0.250000 0.545455
 Ti 0.259551 0.239921 0.636913
 Ti 0.255841 0.241878 0.729152
 Ti 0.250000 0.750000 0.545455

```
Ti  0.259839  0.739943  0.636901
Ti  0.245406  0.741847  0.729200
Ti  0.750000  0.250000  0.545455
Ti  0.759585  0.239925  0.636921
Ti  0.755816  0.241870  0.729175
Ti  0.750000  0.750000  0.545455
Ti  0.759852  0.739958  0.636909
Ti  0.745385  0.741845  0.729217
Ti  0.745385  0.741845  0.270783
Ti  0.759852  0.739958  0.363091
Ti  0.750000  0.750000  0.454545
Ti  0.755816  0.241870  0.270825
Ti  0.759585  0.239925  0.363079
Ti  0.750000  0.250000  0.454545
Ti  0.245406  0.741847  0.270800
Ti  0.259839  0.739943  0.363099
Ti  0.250000  0.750000  0.454545
Ti  0.255841  0.241878  0.270848
Ti  0.259551  0.239921  0.363087
Ti  0.250000  0.250000  0.454545
O   0.250000  0.000000  0.545455
O   0.246202  0.007700  0.636812
O   0.248710  0.005889  0.728778
O   0.250000  0.500000  0.545455
O   0.246229  0.507767  0.636813
O   0.248628  0.506010  0.728766
O   0.750000  0.000000  0.545455
O   0.746191  0.007693  0.636807
O   0.748738  0.005877  0.728785
O   0.750000  0.500000  0.545455
O   0.746233  0.507745  0.636808
O   0.748610  0.506010  0.728774
O   0.250000  0.250000  0.500000
O   0.250000  0.250000  0.590909
O   0.246778  0.253028  0.682254
O   0.246421  0.253785  0.772492
O   0.250000  0.750000  0.500000
O   0.250000  0.750000  0.590909
O   0.248538  0.753037  0.682245
O   0.250221  0.753624  0.772521
O   0.750000  0.250000  0.500000
O   0.750000  0.250000  0.590909
O   0.746755  0.253034  0.682236
O   0.746451  0.253776  0.772498
O   0.750000  0.750000  0.500000
O   0.750000  0.750000  0.590909
```

```
O  0.748557  0.753020  0.682225
O  0.750189  0.753624  0.772522
O  0.000000  0.250000  0.545455
O  0.992086  0.253620  0.636805
O  0.994827  0.253000  0.728777
O  0.000000  0.750000  0.545455
O  0.992554  0.753601  0.636783
O  0.001109  0.752883  0.728818
O  0.500000  0.250000  0.545455
O  0.492023  0.253621  0.636803
O  0.494866  0.252976  0.728781
O  0.500000  0.750000  0.545455
O  0.492522  0.753594  0.636784
O  0.501126  0.752910  0.728814
O  0.501126  0.752910  0.271186
O  0.492522  0.753594  0.363216
O  0.500000  0.750000  0.454545
O  0.494866  0.252976  0.271219
O  0.492023  0.253621  0.363197
O  0.500000  0.250000  0.454545
O  0.001109  0.752883  0.271182
O  0.992554  0.753601  0.363217
O  0.000000  0.750000  0.454545
O  0.994827  0.253000  0.271223
O  0.992086  0.253620  0.363195
O  0.000000  0.250000  0.454545
O  0.750189  0.753624  0.227478
O  0.748557  0.753020  0.317775
O  0.750000  0.750000  0.409091
O  0.746451  0.253776  0.227502
O  0.746755  0.253034  0.317764
O  0.750000  0.250000  0.409091
O  0.250221  0.753624  0.227479
O  0.248538  0.753037  0.317755
O  0.250000  0.750000  0.409091
O  0.246421  0.253785  0.227508
O  0.246778  0.253028  0.317746
O  0.250000  0.250000  0.409091
O  0.748610  0.506010  0.271226
O  0.746233  0.507745  0.363192
O  0.750000  0.500000  0.454545
O  0.748738  0.005877  0.271215
O  0.746191  0.007693  0.363193
O  0.750000  0.000000  0.454545
O  0.248628  0.506010  0.271234
O  0.246229  0.507767  0.363187
```

O  0.250000  0.500000  0.454545
O  0.248710  0.005889  0.271222
O  0.246202  0.007700  0.363188
O  0.250000  0.000000  0.454545

Structure 11. BaO-terminated BaTiO$_3$ slab at $\Theta$=0.25 CO$_2$ coverage

_cell_length_a  8.07335600
_cell_length_b  8.07335600
_cell_length_c  44.40346100
_cell_angle_alpha  90.00000000
_cell_angle_beta  90.00000000
_cell_angle_gamma  90.00000000
_symmetry_space_group_name_H-M  'P 1'
loop_
 _atom_site_type_symbol
 _atom_site_fract_x
 _atom_site_fract_y
 _atom_site_fract_z
 Ba 0.000000 0.500000 0.500000
 Ba 0.000000 0.500000 0.590909
 Ba 0.006546 0.500191 0.682518
 Ba 0.013002 0.493897 0.775977
 Ba 0.000000 0.000000 0.500000
 Ba 0.000000 0.000000 0.590909
 Ba 0.006029 0.006035 0.682421
 Ba 0.998907 0.996853 0.774923
 Ba 0.500000 0.500000 0.500000
 Ba 0.500000 0.500000 0.590909
 Ba 0.499357 0.499371 0.682573
 Ba 0.506804 0.503125 0.776005
 Ba 0.500000 0.000000 0.500000
 Ba 0.500000 0.000000 0.590909
 Ba 0.499654 0.006977 0.682398
 Ba 0.494286 0.011085 0.773640
 Ba 0.494286 0.011085 0.226360
 Ba 0.499654 0.006977 0.317602
 Ba 0.500000 0.000000 0.409091
 Ba 0.506804 0.503125 0.223995
 Ba 0.499357 0.499371 0.317427
 Ba 0.500000 0.500000 0.409091
 Ba 0.998907 0.996853 0.225077
 Ba 0.006029 0.006035 0.317579
 Ba 0.000000 0.000000 0.409091
 Ba 0.013002 0.493897 0.224022
 Ba 0.006546 0.500191 0.317482
 Ba 0.000000 0.500000 0.409091
 Ti 0.250000 0.250000 0.545455
 Ti 0.259705 0.259747 0.636998
 Ti 0.257313 0.260069 0.730189
 Ti 0.250000 0.750000 0.545455
 Ti 0.261156 0.759576 0.637789

| | | | |
|---|---|---|---|
| Ti | 0.259772 | 0.755772 | 0.731357 |
| Ti | 0.750000 | 0.250000 | 0.545455 |
| Ti | 0.759328 | 0.261683 | 0.637783 |
| Ti | 0.758803 | 0.258994 | 0.731267 |
| Ti | 0.750000 | 0.750000 | 0.545455 |
| Ti | 0.759453 | 0.759080 | 0.634421 |
| Ti | 0.753992 | 0.755717 | 0.724352 |
| Ti | 0.753992 | 0.755717 | 0.275648 |
| Ti | 0.759453 | 0.759080 | 0.365579 |
| Ti | 0.750000 | 0.750000 | 0.454545 |
| Ti | 0.758803 | 0.258994 | 0.268733 |
| Ti | 0.759328 | 0.261683 | 0.362217 |
| Ti | 0.750000 | 0.250000 | 0.454545 |
| Ti | 0.259772 | 0.755772 | 0.268643 |
| Ti | 0.261156 | 0.759576 | 0.362211 |
| Ti | 0.250000 | 0.750000 | 0.454545 |
| Ti | 0.257313 | 0.260069 | 0.269811 |
| Ti | 0.259705 | 0.259747 | 0.363002 |
| Ti | 0.250000 | 0.250000 | 0.454545 |
| C | 0.795818 | 0.705538 | 0.810114 |
| C | 0.795818 | 0.705538 | 0.189886 |
| O | 0.711820 | 0.786076 | 0.788471 |
| O | 0.741402 | 0.563113 | 0.818534 |
| O | 0.935983 | 0.764726 | 0.818498 |
| O | 0.711820 | 0.786076 | 0.211529 |
| O | 0.741402 | 0.563113 | 0.181466 |
| O | 0.935983 | 0.764726 | 0.181502 |
| O | 0.250000 | 0.500000 | 0.545455 |
| O | 0.246627 | 0.492728 | 0.636742 |
| O | 0.245983 | 0.494818 | 0.728415 |
| O | 0.250000 | 0.000000 | 0.545455 |
| O | 0.246656 | 0.992533 | 0.636684 |
| O | 0.247170 | 0.991748 | 0.729126 |
| O | 0.750000 | 0.500000 | 0.545455 |
| O | 0.746031 | 0.495575 | 0.636940 |
| O | 0.745538 | 0.500849 | 0.732015 |
| O | 0.750000 | 0.000000 | 0.545455 |
| O | 0.746065 | 0.989593 | 0.637042 |
| O | 0.745053 | 0.988770 | 0.729924 |
| O | 0.250000 | 0.250000 | 0.500000 |
| O | 0.250000 | 0.250000 | 0.590909 |
| O | 0.247482 | 0.246893 | 0.681742 |
| O | 0.244971 | 0.249750 | 0.772761 |
| O | 0.250000 | 0.750000 | 0.500000 |
| O | 0.250000 | 0.750000 | 0.590909 |
| O | 0.246185 | 0.746933 | 0.681514 |

O 0.255498 0.743393 0.772462
O 0.750000 0.250000 0.500000
O 0.750000 0.250000 0.590909
O 0.746042 0.249271 0.681597
O 0.747285 0.239575 0.772279
O 0.750000 0.750000 0.500000
O 0.750000 0.750000 0.590909
O 0.750934 0.746168 0.683761
O 0.000000 0.250000 0.545455
O 0.992245 0.246819 0.636815
O 0.992879 0.248859 0.728510
O 0.000000 0.750000 0.545455
O 0.989846 0.746132 0.636764
O 0.986532 0.744615 0.732238
O 0.500000 0.250000 0.545455
O 0.492807 0.246708 0.636648
O 0.493111 0.247874 0.728901
O 0.500000 0.750000 0.545455
O 0.495382 0.746202 0.637190
O 0.500900 0.745586 0.730098
O 0.500900 0.745586 0.269902
O 0.495382 0.746202 0.362810
O 0.500000 0.750000 0.454545
O 0.493111 0.247874 0.271099
O 0.492807 0.246708 0.363352
O 0.500000 0.250000 0.454545
O 0.986532 0.744615 0.267762
O 0.989846 0.746132 0.363236
O 0.000000 0.750000 0.454545
O 0.992879 0.248859 0.271490
O 0.992245 0.246819 0.363185
O 0.000000 0.250000 0.454545
O 0.750934 0.746168 0.316239
O 0.750000 0.750000 0.409091
O 0.747285 0.239575 0.227721
O 0.746042 0.249271 0.318403
O 0.750000 0.250000 0.409091
O 0.255498 0.743393 0.227538
O 0.246185 0.746933 0.318486
O 0.250000 0.750000 0.409091
O 0.244971 0.249750 0.227239
O 0.247482 0.246893 0.318258
O 0.250000 0.250000 0.409091
O 0.745053 0.988770 0.270076
O 0.746065 0.989593 0.362958
O 0.750000 0.000000 0.454545

```
O  0.745538  0.500849  0.267985
O  0.746031  0.495575  0.363060
O  0.750000  0.500000  0.454545
O  0.247170  0.991748  0.270874
O  0.246656  0.992533  0.363316
O  0.250000  0.000000  0.454545
O  0.245983  0.494818  0.271585
O  0.246627  0.492728  0.363258
O  0.250000  0.500000  0.454545
```

Structure 12. BaO-terminated BaTiO$_3$ slab at $\Theta$=0.50 CO$_2$ coverage

```
_cell_length_a  8.07335600
_cell_length_b  8.07335600
_cell_length_c  44.40346100
_cell_angle_alpha  90.00000000
_cell_angle_beta  90.00000000
_cell_angle_gamma  90.00000000
_symmetry_space_group_name_H-M  'P 1'
loop_
_atom_site_type_symbol
_atom_site_fract_x
_atom_site_fract_y
_atom_site_fract_z
 Ba  0.502979  0.499342  0.777915
 Ba  0.499218  0.998860  0.777914
 Ba  0.000799  0.500082  0.777426
 Ba  0.000764  0.000385  0.777340
 Ba  0.498627  0.996760  0.683367
 Ba  0.499012  0.498078  0.683285
 Ba  0.999333  0.496718  0.683198
 Ba  0.998499  0.997610  0.683123
 Ba  0.000000  0.000000  0.590909
 Ba  0.000000  0.500000  0.590909
 Ba  0.500000  0.000000  0.590909
 Ba  0.500000  0.500000  0.590909
 Ba  0.000000  0.000000  0.500000
 Ba  0.000000  0.500000  0.500000
 Ba  0.500000  0.500000  0.500000
 Ba  0.500000  0.000000  0.500000
 Ba  0.500000  0.500000  0.409091
 Ba  0.500000  0.000000  0.409091
 Ba  0.000000  0.500000  0.409091
 Ba  0.000000  0.000000  0.409091
 Ba  0.998499  0.997610  0.316877
 Ba  0.999333  0.496718  0.316802
 Ba  0.499012  0.498078  0.316715
 Ba  0.498627  0.996760  0.316633
 Ba  0.000764  0.000385  0.222660
 Ba  0.000799  0.500082  0.222574
 Ba  0.499218  0.998860  0.222086
 Ba  0.502979  0.499342  0.222085
 Ti  0.257586  0.741982  0.732559
 Ti  0.742141  0.242074  0.732468
 Ti  0.244257  0.241983  0.726113
 Ti  0.756053  0.742087  0.726099
 Ti  0.739550  0.239457  0.638301
```

```
Ti  0.239669  0.739437  0.638281
Ti  0.241697  0.241213  0.634986
Ti  0.741591  0.741199  0.634961
Ti  0.750000  0.750000  0.545455
Ti  0.750000  0.250000  0.545455
Ti  0.250000  0.750000  0.545455
Ti  0.250000  0.250000  0.545455
Ti  0.250000  0.250000  0.454545
Ti  0.250000  0.750000  0.454545
Ti  0.750000  0.250000  0.454545
Ti  0.750000  0.750000  0.454545
Ti  0.741591  0.741199  0.365039
Ti  0.241697  0.241213  0.365014
Ti  0.239669  0.739437  0.361719
Ti  0.739550  0.239457  0.361699
Ti  0.756053  0.742087  0.273901
Ti  0.244257  0.241983  0.273887
Ti  0.742141  0.242074  0.267532
Ti  0.257586  0.741982  0.267441
C   0.751240  0.752535  0.807153
C   0.251193  0.252450  0.807060
C   0.251193  0.252450  0.192940
C   0.751240  0.752535  0.192847
O   0.749903  0.753876  0.776036
O   0.254136  0.254158  0.775950
O   0.249841  0.754620  0.773247
O   0.753334  0.254880  0.773178
O   0.501390  0.755708  0.730572
O   0.502323  0.255889  0.730505
O   0.008738  0.255796  0.730381
O   0.995005  0.755859  0.730373
O   0.251935  0.507941  0.729648
O   0.750754  0.007809  0.729641
O   0.251845  0.009303  0.729484
O   0.751015  0.509404  0.729440
O   0.751512  0.752575  0.684330
O   0.252214  0.252565  0.684320
O   0.753398  0.254122  0.680931
O   0.252315  0.754155  0.680895
O   0.754030  0.005288  0.637021
O   0.254090  0.505218  0.637013
O   0.505158  0.254148  0.637010
O   0.004869  0.754108  0.636966
O   0.510528  0.754225  0.636906
O   0.010729  0.254174  0.636895
O   0.754300  0.510723  0.636876
```

```
O  0.254331  0.010709  0.636859
O  0.750000  0.750000  0.590909
O  0.250000  0.750000  0.590909
O  0.750000  0.250000  0.590909
O  0.250000  0.250000  0.590909
O  0.750000  0.000000  0.545455
O  0.750000  0.500000  0.545455
O  0.250000  0.000000  0.545455
O  0.250000  0.500000  0.545455
O  0.000000  0.250000  0.545455
O  0.000000  0.750000  0.545455
O  0.500000  0.250000  0.545455
O  0.500000  0.750000  0.545455
O  0.750000  0.750000  0.500000
O  0.750000  0.250000  0.500000
O  0.250000  0.750000  0.500000
O  0.250000  0.250000  0.500000
O  0.500000  0.250000  0.454545
O  0.500000  0.750000  0.454545
O  0.000000  0.750000  0.454545
O  0.000000  0.250000  0.454545
O  0.250000  0.000000  0.454545
O  0.750000  0.500000  0.454545
O  0.750000  0.000000  0.454545
O  0.250000  0.500000  0.454545
O  0.250000  0.250000  0.409091
O  0.750000  0.250000  0.409091
O  0.250000  0.750000  0.409091
O  0.750000  0.750000  0.409091
O  0.254331  0.010709  0.363141
O  0.754300  0.510723  0.363124
O  0.010729  0.254174  0.363105
O  0.510528  0.754225  0.363094
O  0.004869  0.754108  0.363034
O  0.505158  0.254148  0.362990
O  0.254090  0.505218  0.362987
O  0.754030  0.005288  0.362979
O  0.252315  0.754155  0.319105
O  0.753398  0.254122  0.319069
O  0.252214  0.252565  0.315680
O  0.751512  0.752575  0.315670
O  0.751015  0.509404  0.270560
O  0.251845  0.009303  0.270516
O  0.750754  0.007809  0.270359
O  0.251935  0.507941  0.270352
O  0.995005  0.755859  0.269627
```

```
O  0.008738  0.255796  0.269619
O  0.502323  0.255889  0.269495
O  0.501390  0.755708  0.269428
O  0.753334  0.254880  0.226822
O  0.249841  0.754620  0.226753
O  0.254136  0.254158  0.224050
O  0.749903  0.753876  0.223964
O  0.249731  0.109366  0.180925
O  0.752086  0.609566  0.180784
O  0.250705  0.394305  0.180490
O  0.751033  0.894543  0.180454
O  0.751033  0.894543  0.819546
O  0.250705  0.394305  0.819510
O  0.752086  0.609566  0.819216
O  0.249731  0.109366  0.819075
```

Structure 13. BaO-terminated BaZrO$_3$ slab at $\Theta$=0.00 CO$_2$ coverage

_cell_length_a  8.51149400
_cell_length_b  8.51149400
_cell_length_c  46.81321300
_cell_angle_alpha  90.00000000
_cell_angle_beta  90.00000000
_cell_angle_gamma  90.00000000
_symmetry_space_group_name_H-M  'P 1'
loop_
 _atom_site_type_symbol
 _atom_site_fract_x
 _atom_site_fract_y
 _atom_site_fract_z
 Ba  0.500000  0.000000  0.500000
 Ba  0.500000  0.000000  0.590909
 Ba  0.499630  0.000030  0.681044
 Ba  0.487829  0.000572  0.769982
 Ba  0.500000  0.500000  0.500000
 Ba  0.500000  0.500000  0.590909
 Ba  0.499884  0.500277  0.681039
 Ba  0.499101  0.512348  0.769988
 Ba  0.000000  0.000000  0.500000
 Ba  0.000000  0.000000  0.590909
 Ba  0.999886  0.000275  0.681038
 Ba  0.999101  0.012349  0.769988
 Ba  0.000000  0.500000  0.500000
 Ba  0.000000  0.500000  0.590909
 Ba  0.999631  0.500031  0.681044
 Ba  0.987830  0.500572  0.769982
 Ba  0.500000  0.000000  0.409091
 Ba  0.499630  0.000030  0.318956
 Ba  0.487829  0.000572  0.230018
 Ba  0.500000  0.500000  0.409091
 Ba  0.499884  0.500277  0.318961
 Ba  0.499101  0.512348  0.230012
 Ba  0.000000  0.000000  0.409091
 Ba  0.999886  0.000275  0.318962
 Ba  0.999101  0.012349  0.230012
 Ba  0.000000  0.500000  0.409091
 Ba  0.999631  0.500031  0.318956
 Ba  0.987830  0.500572  0.230018
 Zr  0.750000  0.250000  0.545455
 Zr  0.749500  0.249489  0.636561
 Zr  0.747803  0.251627  0.728504
 Zr  0.750000  0.750000  0.545455
 Zr  0.750541  0.750481  0.636562

| | | | |
|----|----------|----------|----------|
| Zr | 0.748262 | 0.752058 | 0.728508 |
| Zr | 0.250000 | 0.250000 | 0.545455 |
| Zr | 0.250541 | 0.250480 | 0.636562 |
| Zr | 0.248261 | 0.252058 | 0.728508 |
| Zr | 0.250000 | 0.750000 | 0.545455 |
| Zr | 0.249502 | 0.749488 | 0.636561 |
| Zr | 0.247803 | 0.751627 | 0.728504 |
| Zr | 0.750000 | 0.250000 | 0.454545 |
| Zr | 0.749500 | 0.249489 | 0.363439 |
| Zr | 0.747803 | 0.251627 | 0.271496 |
| Zr | 0.750000 | 0.750000 | 0.454545 |
| Zr | 0.750541 | 0.750481 | 0.363438 |
| Zr | 0.748262 | 0.752058 | 0.271492 |
| Zr | 0.250000 | 0.250000 | 0.454545 |
| Zr | 0.250541 | 0.250480 | 0.363438 |
| Zr | 0.248261 | 0.252058 | 0.271492 |
| Zr | 0.250000 | 0.750000 | 0.454545 |
| Zr | 0.249502 | 0.749488 | 0.363439 |
| Zr | 0.247803 | 0.751627 | 0.271496 |
| O | 0.750000 | 0.000000 | 0.545455 |
| O | 0.761754 | 0.999886 | 0.635943 |
| O | 0.721537 | 0.000429 | 0.729452 |
| O | 0.750000 | 0.500000 | 0.545455 |
| O | 0.739082 | 0.499875 | 0.636924 |
| O | 0.777708 | 0.500174 | 0.726446 |
| O | 0.250000 | 0.000000 | 0.545455 |
| O | 0.239081 | 0.999874 | 0.636924 |
| O | 0.277708 | 0.000174 | 0.726446 |
| O | 0.250000 | 0.500000 | 0.545455 |
| O | 0.261754 | 0.499885 | 0.635943 |
| O | 0.221537 | 0.500429 | 0.729452 |
| O | 0.750000 | 0.250000 | 0.500000 |
| O | 0.750000 | 0.250000 | 0.590909 |
| O | 0.744239 | 0.244626 | 0.681948 |
| O | 0.767860 | 0.259071 | 0.772523 |
| O | 0.750000 | 0.750000 | 0.500000 |
| O | 0.750000 | 0.750000 | 0.590909 |
| O | 0.755528 | 0.755425 | 0.681948 |
| O | 0.740209 | 0.732517 | 0.772528 |
| O | 0.250000 | 0.250000 | 0.500000 |
| O | 0.250000 | 0.250000 | 0.590909 |
| O | 0.255528 | 0.255427 | 0.681948 |
| O | 0.240210 | 0.232516 | 0.772528 |
| O | 0.250000 | 0.750000 | 0.500000 |
| O | 0.250000 | 0.750000 | 0.590909 |
| O | 0.244237 | 0.744626 | 0.681948 |

```
O   0.267860   0.759072   0.772523
O   0.500000   0.250000   0.545455
O   0.500066   0.238122   0.635926
O   0.499436   0.278285   0.729511
O   0.500000   0.750000   0.545455
O   0.500073   0.760789   0.636948
O   0.499689   0.722161   0.726377
O   0.000000   0.250000   0.545455
O   0.000073   0.260789   0.636948
O   0.999689   0.222161   0.726377
O   0.000000   0.750000   0.545455
O   0.000067   0.738122   0.635926
O   0.999436   0.778285   0.729511
O   0.750000   0.000000   0.454545
O   0.761754   0.999886   0.364057
O   0.721537   0.000429   0.270548
O   0.750000   0.500000   0.454545
O   0.739082   0.499875   0.363076
O   0.777708   0.500174   0.273554
O   0.250000   0.000000   0.454545
O   0.239081   0.999874   0.363076
O   0.277708   0.000174   0.273554
O   0.250000   0.500000   0.454545
O   0.261754   0.499885   0.364057
O   0.221537   0.500429   0.270548
O   0.750000   0.250000   0.409091
O   0.744239   0.244626   0.318052
O   0.767860   0.259071   0.227477
O   0.750000   0.750000   0.409091
O   0.755528   0.755425   0.318052
O   0.740209   0.732517   0.227472
O   0.250000   0.250000   0.409091
O   0.255528   0.255427   0.318052
O   0.240210   0.232516   0.227472
O   0.250000   0.750000   0.409091
O   0.244237   0.744626   0.318052
O   0.267860   0.759072   0.227477
O   0.500000   0.250000   0.454545
O   0.500066   0.238122   0.364074
O   0.499436   0.278285   0.270489
O   0.500000   0.750000   0.454545
O   0.500073   0.760789   0.363052
O   0.499689   0.722161   0.273623
O   0.000000   0.250000   0.454545
O   0.000073   0.260789   0.363052
O   0.999689   0.222161   0.273623
```

O  0.000000  0.750000  0.454545
O  0.000067  0.738122  0.364074
O  0.999436  0.778285  0.270489

Structure 14. BaO-terminated BaZrO$_3$ slab at $\Theta$=0.25 CO$_2$ coverage

```
_cell_length_a  8.51149400
_cell_length_b  8.51149400
_cell_length_c  46.81321300
_cell_angle_alpha  90.00000000
_cell_angle_beta  90.00000000
_cell_angle_gamma  90.00000000
_symmetry_space_group_name_H-M  'P 1'
loop_
_atom_site_type_symbol
_atom_site_fract_x
_atom_site_fract_y
_atom_site_fract_z
Ba  0.024347  0.496789  0.226304
Ba  0.002389  0.011395  0.227529
Ba  0.498800  0.504631  0.228366
Ba  0.497355  0.016792  0.229395
Ba  0.003275  0.002844  0.318301
Ba  0.497418  0.003868  0.318374
Ba  0.004184  0.496092  0.318547
Ba  0.496478  0.497134  0.318662
Ba  0.500000  0.000000  0.409091
Ba  0.500000  0.500000  0.409091
Ba  0.000000  0.500000  0.409091
Ba  0.000000  0.000000  0.409091
Ba  0.500000  0.000000  0.500000
Ba  0.500000  0.500000  0.500000
Ba  0.000000  0.000000  0.500000
Ba  0.000000  0.500000  0.500000
Ba  0.500000  0.000000  0.590909
Ba  0.500000  0.500000  0.590909
Ba  0.000000  0.000000  0.590909
Ba  0.000000  0.500000  0.590909
Ba  0.496478  0.497134  0.681338
Ba  0.004184  0.496092  0.681453
Ba  0.497418  0.003868  0.681626
Ba  0.003275  0.002844  0.681699
Ba  0.497355  0.016792  0.770605
Ba  0.498800  0.504631  0.771634
Ba  0.002389  0.011395  0.772471
Ba  0.024347  0.496789  0.773696
Zr  0.753209  0.252464  0.270649
Zr  0.251671  0.750630  0.270799
Zr  0.250492  0.252478  0.271762
Zr  0.751062  0.751393  0.274398
Zr  0.749732  0.250661  0.363107
```

| | | | |
|---|---|---|---|
| Zr | 0.249426 | 0.750511 | 0.363153 |
| Zr | 0.250642 | 0.249028 | 0.363691 |
| Zr | 0.751180 | 0.748696 | 0.364101 |
| Zr | 0.250000 | 0.250000 | 0.454545 |
| Zr | 0.250000 | 0.750000 | 0.454545 |
| Zr | 0.750000 | 0.250000 | 0.454545 |
| Zr | 0.750000 | 0.750000 | 0.454545 |
| Zr | 0.250000 | 0.250000 | 0.545455 |
| Zr | 0.250000 | 0.750000 | 0.545455 |
| Zr | 0.750000 | 0.250000 | 0.545455 |
| Zr | 0.750000 | 0.750000 | 0.545455 |
| Zr | 0.751180 | 0.748696 | 0.635899 |
| Zr | 0.250642 | 0.249028 | 0.636309 |
| Zr | 0.249426 | 0.750511 | 0.636847 |
| Zr | 0.749732 | 0.250661 | 0.636893 |
| Zr | 0.751062 | 0.751393 | 0.725602 |
| Zr | 0.250492 | 0.252478 | 0.728238 |
| Zr | 0.251671 | 0.750630 | 0.729201 |
| Zr | 0.753209 | 0.252464 | 0.729351 |
| C | 0.790187 | 0.711833 | 0.201878 |
| C | 0.790187 | 0.711833 | 0.798122 |
| O | 0.736689 | 0.579748 | 0.193065 |
| O | 0.920661 | 0.769852 | 0.193149 |
| O | 0.713606 | 0.785659 | 0.223307 |
| O | 0.244491 | 0.256649 | 0.227516 |
| O | 0.273220 | 0.730841 | 0.228087 |
| O | 0.754964 | 0.213291 | 0.228166 |
| O | 0.772093 | 0.503844 | 0.266610 |
| O | 0.997359 | 0.771311 | 0.268034 |
| O | 0.272134 | 0.000083 | 0.271358 |
| O | 0.500979 | 0.269931 | 0.271555 |
| O | 0.000999 | 0.230113 | 0.273961 |
| O | 0.504184 | 0.726799 | 0.274127 |
| O | 0.227463 | 0.499475 | 0.274286 |
| O | 0.727674 | 0.996083 | 0.274821 |
| O | 0.763212 | 0.735584 | 0.317547 |
| O | 0.746359 | 0.260985 | 0.318084 |
| O | 0.241204 | 0.754977 | 0.318184 |
| O | 0.255666 | 0.243887 | 0.318417 |
| O | 0.761770 | 0.998633 | 0.362364 |
| O | 0.501232 | 0.761164 | 0.362544 |
| O | 0.261685 | 0.499526 | 0.363126 |
| O | 0.000250 | 0.260858 | 0.363206 |
| O | 0.500221 | 0.237137 | 0.364124 |
| O | 0.237898 | 0.999678 | 0.364244 |
| O | 0.999520 | 0.737324 | 0.364415 |

```
O  0.737372  0.500309  0.364507
O  0.250000  0.250000  0.409091
O  0.250000  0.750000  0.409091
O  0.750000  0.750000  0.409091
O  0.750000  0.250000  0.409091
O  0.250000  0.000000  0.454545
O  0.250000  0.500000  0.454545
O  0.750000  0.000000  0.454545
O  0.750000  0.500000  0.454545
O  0.000000  0.250000  0.454545
O  0.000000  0.750000  0.454545
O  0.500000  0.750000  0.454545
O  0.500000  0.250000  0.454545
O  0.250000  0.250000  0.500000
O  0.250000  0.750000  0.500000
O  0.750000  0.250000  0.500000
O  0.750000  0.750000  0.500000
O  0.250000  0.000000  0.545455
O  0.250000  0.500000  0.545455
O  0.750000  0.000000  0.545455
O  0.750000  0.500000  0.545455
O  0.500000  0.250000  0.545455
O  0.500000  0.750000  0.545455
O  0.000000  0.250000  0.545455
O  0.000000  0.750000  0.545455
O  0.250000  0.250000  0.590909
O  0.250000  0.750000  0.590909
O  0.750000  0.250000  0.590909
O  0.750000  0.750000  0.590909
O  0.737372  0.500309  0.635493
O  0.999520  0.737324  0.635585
O  0.237898  0.999678  0.635756
O  0.500221  0.237137  0.635876
O  0.000250  0.260858  0.636794
O  0.261685  0.499526  0.636874
O  0.501232  0.761164  0.637456
O  0.761770  0.998633  0.637636
O  0.255666  0.243887  0.681583
O  0.241204  0.754977  0.681816
O  0.746359  0.260985  0.681916
O  0.763212  0.735584  0.682453
O  0.727674  0.996083  0.725179
O  0.227463  0.499475  0.725714
O  0.504184  0.726799  0.725873
O  0.000999  0.230113  0.726039
O  0.500979  0.269931  0.728445
```

O 0.272134 0.000083 0.728642
O 0.997359 0.771311 0.731966
O 0.772093 0.503844 0.733390
O 0.754964 0.213291 0.771834
O 0.273220 0.730841 0.771913
O 0.244491 0.256649 0.772484
O 0.713606 0.785659 0.776693
O 0.920661 0.769852 0.806851
O 0.736689 0.579748 0.806935

Structure 15. BaO-terminated BaZrO$_3$ slab at $\Theta$=0.50 CO$_2$ coverage

_cell_length_a  8.51149400
_cell_length_b  8.51149400
_cell_length_c  46.81321300
_cell_angle_alpha  90.00000000
_cell_angle_beta  90.00000000
_cell_angle_gamma  90.00000000
_symmetry_space_group_name_H-M  'P 1'
loop_
_atom_site_type_symbol
_atom_site_fract_x
_atom_site_fract_y
_atom_site_fract_z
Ba  0.993741  0.483549  0.224502
Ba  0.492218  0.519362  0.225127
Ba  0.993454  0.997285  0.227568
Ba  0.492797  0.004722  0.228162
Ba  0.999079  0.498820  0.317825
Ba  0.998971  0.000543  0.318043
Ba  0.499853  0.501933  0.318292
Ba  0.498910  0.999886  0.318293
Ba  0.000000  0.000000  0.409091
Ba  0.500000  0.000000  0.409091
Ba  0.500000  0.500000  0.409091
Ba  0.000000  0.500000  0.409091
Ba  0.500000  0.000000  0.500000
Ba  0.000000  0.000000  0.500000
Ba  0.000000  0.500000  0.500000
Ba  0.500000  0.500000  0.500000
Ba  0.500000  0.000000  0.590909
Ba  0.000000  0.000000  0.590909
Ba  0.000000  0.500000  0.590909
Ba  0.500000  0.500000  0.590909
Ba  0.498910  0.999886  0.681707
Ba  0.499853  0.501933  0.681708
Ba  0.998971  0.000543  0.681957
Ba  0.999079  0.498820  0.682175
Ba  0.492797  0.004722  0.771838
Ba  0.993454  0.997285  0.772432
Ba  0.492218  0.519362  0.774873
Ba  0.993741  0.483549  0.775498
Zr  0.748362  0.252006  0.270378
Zr  0.247792  0.749924  0.270535
Zr  0.247947  0.251153  0.274563
Zr  0.747459  0.750972  0.274567
Zr  0.748660  0.250432  0.362970

```
Zr  0.248498  0.749592  0.363004
Zr  0.251243  0.250766  0.364219
Zr  0.751262  0.749766  0.364249
Zr  0.750000  0.750000  0.454545
Zr  0.250000  0.750000  0.454545
Zr  0.250000  0.250000  0.454545
Zr  0.750000  0.250000  0.454545
Zr  0.750000  0.250000  0.545455
Zr  0.250000  0.250000  0.545455
Zr  0.250000  0.750000  0.545455
Zr  0.750000  0.750000  0.545455
Zr  0.751262  0.749766  0.635751
Zr  0.251243  0.250766  0.635782
Zr  0.248498  0.749592  0.636996
Zr  0.748660  0.250432  0.637030
Zr  0.747459  0.750972  0.725433
Zr  0.247947  0.251153  0.725437
Zr  0.247792  0.749924  0.729465
Zr  0.748362  0.252006  0.729622
C   0.286217  0.283111  0.200301
C   0.784565  0.716639  0.200337
C   0.784565  0.716639  0.799663
C   0.286217  0.283111  0.799699
O   0.404550  0.212471  0.190040
O   0.905793  0.783989  0.190340
O   0.730979  0.585539  0.191326
O   0.235783  0.416122  0.191521
O   0.215996  0.221989  0.224195
O   0.715732  0.779079  0.224215
O   0.787993  0.235350  0.228735
O   0.284632  0.768167  0.228813
O   0.495472  0.261095  0.267493
O   0.995422  0.760758  0.268434
O   0.237145  0.497515  0.269750
O   0.757639  0.504526  0.270193
O   0.737862  0.998546  0.273901
O   0.258267  0.003644  0.274399
O   0.502111  0.738609  0.275989
O   0.002564  0.243553  0.276486
O   0.266428  0.259273  0.317997
O   0.764579  0.744895  0.318019
O   0.737164  0.255893  0.318292
O   0.238039  0.743588  0.318341
O   0.001018  0.260167  0.362294
O   0.500947  0.760771  0.362385
O   0.239314  0.001018  0.362937
```

| | | | |
|---|---|---|---|
| O | 0.760946 | 0.999286 | 0.363172 |
| O | 0.738947 | 0.500844 | 0.363970 |
| O | 0.261326 | 0.499443 | 0.364226 |
| O | 0.999392 | 0.739271 | 0.364899 |
| O | 0.499473 | 0.237819 | 0.364920 |
| O | 0.750000 | 0.750000 | 0.409091 |
| O | 0.250000 | 0.750000 | 0.409091 |
| O | 0.750000 | 0.250000 | 0.409091 |
| O | 0.250000 | 0.250000 | 0.409091 |
| O | 0.000000 | 0.250000 | 0.454545 |
| O | 0.500000 | 0.250000 | 0.454545 |
| O | 0.500000 | 0.750000 | 0.454545 |
| O | 0.750000 | 0.000000 | 0.454545 |
| O | 0.000000 | 0.750000 | 0.454545 |
| O | 0.250000 | 0.000000 | 0.454545 |
| O | 0.250000 | 0.500000 | 0.454545 |
| O | 0.750000 | 0.500000 | 0.454545 |
| O | 0.750000 | 0.750000 | 0.500000 |
| O | 0.250000 | 0.750000 | 0.500000 |
| O | 0.750000 | 0.250000 | 0.500000 |
| O | 0.250000 | 0.250000 | 0.500000 |
| O | 0.000000 | 0.750000 | 0.545455 |
| O | 0.500000 | 0.250000 | 0.545455 |
| O | 0.750000 | 0.000000 | 0.545455 |
| O | 0.000000 | 0.250000 | 0.545455 |
| O | 0.500000 | 0.750000 | 0.545455 |
| O | 0.250000 | 0.000000 | 0.545455 |
| O | 0.750000 | 0.500000 | 0.545455 |
| O | 0.250000 | 0.500000 | 0.545455 |
| O | 0.250000 | 0.750000 | 0.590909 |
| O | 0.750000 | 0.750000 | 0.590909 |
| O | 0.750000 | 0.250000 | 0.590909 |
| O | 0.250000 | 0.250000 | 0.590909 |
| O | 0.499473 | 0.237819 | 0.635080 |
| O | 0.999392 | 0.739271 | 0.635101 |
| O | 0.261326 | 0.499443 | 0.635774 |
| O | 0.738947 | 0.500844 | 0.636030 |
| O | 0.760946 | 0.999286 | 0.636828 |
| O | 0.239314 | 0.001018 | 0.637063 |
| O | 0.500947 | 0.760771 | 0.637615 |
| O | 0.001018 | 0.260167 | 0.637706 |
| O | 0.238039 | 0.743588 | 0.681659 |
| O | 0.737164 | 0.255893 | 0.681708 |
| O | 0.764579 | 0.744895 | 0.681981 |
| O | 0.266428 | 0.259273 | 0.682003 |
| O | 0.002564 | 0.243553 | 0.723514 |

```
O  0.502111  0.738609  0.724011
O  0.258267  0.003644  0.725601
O  0.737862  0.998546  0.726099
O  0.757639  0.504526  0.729807
O  0.237145  0.497515  0.730250
O  0.995422  0.760758  0.731566
O  0.495472  0.261095  0.732507
O  0.284632  0.768167  0.771187
O  0.787993  0.235350  0.771265
O  0.715732  0.779079  0.775784
O  0.215996  0.221989  0.775805
O  0.235783  0.416122  0.808479
O  0.730979  0.585539  0.808674
O  0.905793  0.783989  0.809660
O  0.404550  0.212471  0.809960
```

Structure 16. BaO-terminated BaHfO$_3$ slab at $\Theta$=0.00 CO$_2$ coverage

```
_cell_length_a  8.41088800
_cell_length_b  8.41088800
_cell_length_c  46.25988800
_cell_angle_alpha  90.00000000
_cell_angle_beta  90.00000000
_cell_angle_gamma  90.00000000
_symmetry_space_group_name_H-M  'P 1'
loop_
_atom_site_type_symbol
_atom_site_fract_x
_atom_site_fract_y
_atom_site_fract_z
 Ba  0.500000  0.000000  0.500000
 Ba  0.500000  0.000000  0.590909
 Ba  0.500783  0.000588  0.681397
 Ba  0.503298  0.003420  0.770275
 Ba  0.500000  0.500000  0.500000
 Ba  0.500000  0.500000  0.590909
 Ba  0.500329  0.500003  0.681383
 Ba  0.498273  0.497684  0.770276
 Ba  0.000000  0.000000  0.500000
 Ba  0.000000  0.000000  0.590909
 Ba  0.000422  0.000124  0.681381
 Ba  0.998590  0.997690  0.770293
 Ba  0.000000  0.500000  0.500000
 Ba  0.000000  0.500000  0.590909
 Ba  0.000732  0.500449  0.681391
 Ba  0.003001  0.503533  0.770322
 Ba  0.500000  0.000000  0.409091
 Ba  0.500783  0.000588  0.318603
 Ba  0.503298  0.003420  0.229725
 Ba  0.500000  0.500000  0.409091
 Ba  0.500329  0.500003  0.318617
 Ba  0.498273  0.497684  0.229724
 Ba  0.000000  0.000000  0.409091
 Ba  0.000422  0.000124  0.318619
 Ba  0.998590  0.997690  0.229707
 Ba  0.000000  0.500000  0.409091
 Ba  0.000732  0.500449  0.318609
 Ba  0.003001  0.503533  0.229678
 Hf  0.750000  0.750000  0.545455
 Hf  0.750098  0.749963  0.636592
 Hf  0.750698  0.750368  0.728531
 Hf  0.750000  0.250000  0.545455
 Hf  0.750556  0.250355  0.636587
```

```
Hf  0.750896  0.250572  0.728543
Hf  0.250000  0.750000  0.545455
Hf  0.250604  0.750346  0.636588
Hf  0.250852  0.750521  0.728527
Hf  0.250000  0.250000  0.545455
Hf  0.250136  0.249943  0.636582
Hf  0.250591  0.250345  0.728537
Hf  0.750000  0.750000  0.454545
Hf  0.750098  0.749963  0.363408
Hf  0.750698  0.750368  0.271469
Hf  0.750000  0.250000  0.454545
Hf  0.750556  0.250355  0.363413
Hf  0.750896  0.250572  0.271457
Hf  0.250000  0.750000  0.454545
Hf  0.250604  0.750346  0.363412
Hf  0.250852  0.750521  0.271473
Hf  0.250000  0.250000  0.454545
Hf  0.250136  0.249943  0.363418
Hf  0.250591  0.250345  0.271463
O   0.750000  0.000000  0.545455
O   0.751624  0.000156  0.636670
O   0.750685  0.000463  0.727445
O   0.750000  0.500000  0.545455
O   0.749311  0.500150  0.636349
O   0.750770  0.500459  0.728552
O   0.250000  0.000000  0.545455
O   0.249240  0.000134  0.636337
O   0.250846  0.000407  0.728538
O   0.250000  0.500000  0.545455
O   0.251571  0.500130  0.636670
O   0.250851  0.500404  0.727434
O   0.750000  0.750000  0.500000
O   0.750000  0.750000  0.590909
O   0.748259  0.748573  0.681946
O   0.757663  0.755764  0.772732
O   0.750000  0.250000  0.500000
O   0.750000  0.250000  0.590909
O   0.752286  0.252086  0.681944
O   0.744723  0.245256  0.772741
O   0.250000  0.750000  0.500000
O   0.250000  0.750000  0.590909
O   0.252889  0.752025  0.681943
O   0.244123  0.745075  0.772728
O   0.250000  0.250000  0.500000
O   0.250000  0.250000  0.590909
O   0.248738  0.248493  0.681939
```

O 0.256755 0.255654 0.772733
O 0.500000 0.750000 0.545455
O 0.500374 0.751280 0.636292
O 0.500788 0.750428 0.728721
O 0.500000 0.250000 0.545455
O 0.500364 0.248968 0.636660
O 0.500738 0.250424 0.727396
O 0.000000 0.750000 0.545455
O 0.000377 0.748979 0.636722
O 0.000787 0.750434 0.727266
O 0.000000 0.250000 0.545455
O 0.000361 0.251274 0.636338
O 0.000736 0.250376 0.728571
O 0.750000 0.000000 0.454545
O 0.751624 0.000156 0.363330
O 0.750685 0.000463 0.272555
O 0.750000 0.500000 0.454545
O 0.749311 0.500150 0.363651
O 0.750770 0.500459 0.271448
O 0.250000 0.000000 0.454545
O 0.249240 0.000134 0.363663
O 0.250846 0.000407 0.271462
O 0.250000 0.500000 0.454545
O 0.251571 0.500130 0.363330
O 0.250851 0.500404 0.272566
O 0.750000 0.750000 0.409091
O 0.748259 0.748573 0.318054
O 0.757663 0.755764 0.227268
O 0.750000 0.250000 0.409091
O 0.752286 0.252086 0.318056
O 0.744723 0.245256 0.227259
O 0.250000 0.750000 0.409091
O 0.252889 0.752025 0.318057
O 0.244123 0.745075 0.227272
O 0.250000 0.250000 0.409091
O 0.248738 0.248493 0.318061
O 0.256755 0.255654 0.227267
O 0.500000 0.750000 0.454545
O 0.500374 0.751280 0.363708
O 0.500788 0.750428 0.271279
O 0.500000 0.250000 0.454545
O 0.500364 0.248968 0.363340
O 0.500738 0.250424 0.272604
O 0.000000 0.750000 0.454545
O 0.000377 0.748979 0.363278
O 0.000787 0.750434 0.272734

```
O  0.000000  0.250000  0.454545
O  0.000361  0.251274  0.363662
O  0.000736  0.250376  0.271429
```

Structure 17. BaO-terminated BaHfO$_3$ slab at $\Theta$=0.25 CO$_2$ coverage

```
_cell_length_a  8.41088800
_cell_length_b  8.41088800
_cell_length_c  46.25988800
_cell_angle_alpha  90.00000000
_cell_angle_beta  90.00000000
_cell_angle_gamma  90.00000000
_symmetry_space_group_name_H-M   'P 1'
loop_
 _atom_site_type_symbol
 _atom_site_fract_x
 _atom_site_fract_y
 _atom_site_fract_z
 Ba  0.009914  0.481570  0.226426
 Ba  0.001604  0.004273  0.227512
 Ba  0.492414  0.498775  0.227556
 Ba  0.489124  0.007747  0.228894
 Ba  0.497870  0.001531  0.318264
 Ba  0.003404  0.496109  0.318291
 Ba  0.497382  0.497298  0.318296
 Ba  0.002543  0.002366  0.318339
 Ba  0.500000  0.000000  0.409091
 Ba  0.500000  0.500000  0.409091
 Ba  0.000000  0.500000  0.409091
 Ba  0.000000  0.000000  0.409091
 Ba  0.500000  0.000000  0.500000
 Ba  0.500000  0.500000  0.500000
 Ba  0.000000  0.000000  0.500000
 Ba  0.000000  0.500000  0.500000
 Ba  0.500000  0.000000  0.590909
 Ba  0.500000  0.500000  0.590909
 Ba  0.000000  0.000000  0.590909
 Ba  0.000000  0.500000  0.590909
 Ba  0.002543  0.002366  0.681661
 Ba  0.497382  0.497298  0.681704
 Ba  0.003404  0.496109  0.681709
 Ba  0.497870  0.001531  0.681736
 Ba  0.489124  0.007747  0.771106
 Ba  0.492414  0.498775  0.772444
 Ba  0.001604  0.004273  0.772488
 Ba  0.009914  0.481570  0.773574
 Hf  0.250033  0.748297  0.270896
 Hf  0.751022  0.248708  0.270910
 Hf  0.249317  0.249839  0.272046
 Hf  0.749890  0.749399  0.274214
 Hf  0.249508  0.749931  0.363216
```

Hf 0.749743 0.250226 0.363217
Hf 0.250752 0.249373 0.363709
Hf 0.751119 0.748904 0.364004
Hf 0.250000 0.250000 0.454545
Hf 0.250000 0.750000 0.454545
Hf 0.750000 0.250000 0.454545
Hf 0.750000 0.750000 0.454545
Hf 0.250000 0.250000 0.545455
Hf 0.250000 0.750000 0.545455
Hf 0.750000 0.250000 0.545455
Hf 0.750000 0.750000 0.545455
Hf 0.751119 0.748904 0.635996
Hf 0.250752 0.249373 0.636291
Hf 0.749743 0.250226 0.636783
Hf 0.249508 0.749931 0.636784
Hf 0.749890 0.749399 0.725786
Hf 0.249317 0.249839 0.727954
Hf 0.751022 0.248708 0.729090
Hf 0.250033 0.748297 0.729104
C 0.787702 0.709772 0.200547
C 0.787702 0.709772 0.799453
O 0.726171 0.581624 0.190786
O 0.917737 0.769028 0.191141
O 0.718778 0.778824 0.224012
O 0.244460 0.252008 0.227614
O 0.762581 0.228069 0.227962
O 0.277268 0.746391 0.228023
O 0.997178 0.748724 0.268258
O 0.750664 0.502350 0.268565
O 0.500254 0.248861 0.271661
O 0.250731 0.999523 0.272298
O 0.248501 0.498961 0.273221
O 0.750007 0.996129 0.273663
O 0.000858 0.251337 0.273932
O 0.503514 0.748996 0.274161
O 0.760385 0.740976 0.317664
O 0.242751 0.751179 0.318180
O 0.747026 0.255896 0.318180
O 0.254404 0.247749 0.318444
O 0.501115 0.749806 0.362743
O 0.750390 0.998794 0.362857
O 0.000367 0.249557 0.363311
O 0.250583 0.499592 0.363468
O 0.250240 0.999712 0.363847
O 0.500244 0.249346 0.364034
O 0.749924 0.500201 0.364051

```
O  0.999718  0.749771  0.364196
O  0.250000  0.250000  0.409091
O  0.250000  0.750000  0.409091
O  0.750000  0.750000  0.409091
O  0.750000  0.250000  0.409091
O  0.250000  0.000000  0.454545
O  0.250000  0.500000  0.454545
O  0.750000  0.000000  0.454545
O  0.750000  0.500000  0.454545
O  0.000000  0.250000  0.454545
O  0.000000  0.750000  0.454545
O  0.500000  0.750000  0.454545
O  0.500000  0.250000  0.454545
O  0.250000  0.250000  0.500000
O  0.250000  0.750000  0.500000
O  0.750000  0.250000  0.500000
O  0.750000  0.750000  0.500000
O  0.250000  0.000000  0.545455
O  0.250000  0.500000  0.545455
O  0.750000  0.000000  0.545455
O  0.750000  0.500000  0.545455
O  0.500000  0.250000  0.545455
O  0.500000  0.750000  0.545455
O  0.000000  0.250000  0.545455
O  0.000000  0.750000  0.545455
O  0.250000  0.250000  0.590909
O  0.250000  0.750000  0.590909
O  0.750000  0.250000  0.590909
O  0.750000  0.750000  0.590909
O  0.999718  0.749771  0.635804
O  0.749924  0.500201  0.635949
O  0.500244  0.249346  0.635966
O  0.250240  0.999712  0.636153
O  0.250583  0.499592  0.636532
O  0.000367  0.249557  0.636689
O  0.750390  0.998794  0.637143
O  0.501115  0.749806  0.637257
O  0.254404  0.247749  0.681556
O  0.242751  0.751179  0.681820
O  0.747026  0.255896  0.681820
O  0.760385  0.740976  0.682336
O  0.503514  0.748996  0.725839
O  0.000858  0.251337  0.726068
O  0.750007  0.996129  0.726337
O  0.248501  0.498961  0.726779
O  0.250731  0.999523  0.727702
```

O 0.500254 0.248861 0.728339
O 0.750664 0.502350 0.731435
O 0.997178 0.748724 0.731742
O 0.277268 0.746391 0.771977
O 0.762581 0.228069 0.772038
O 0.244460 0.252008 0.772386
O 0.718778 0.778824 0.775988
O 0.917737 0.769028 0.808859
O 0.726171 0.581624 0.809214

Structure 18. BaO-terminated BaHfO$_3$ slab at $\Theta$=0.50 CO$_2$ coverage

_cell_length_a  8.41088800
_cell_length_b  8.41088800
_cell_length_c  46.25988800
_cell_angle_alpha  90.00000000
_cell_angle_beta  90.00000000
_cell_angle_gamma  90.00000000
_symmetry_space_group_name_H-M  'P 1'
loop_
_atom_site_type_symbol
_atom_site_fract_x
_atom_site_fract_y
_atom_site_fract_z
Ba  0.517434  0.015836  0.224657
Ba  0.994848  0.491801  0.224695
Ba  0.993913  0.008541  0.224813
Ba  0.516338  0.483340  0.224922
Ba  0.498214  0.998365  0.317897
Ba  0.497896  0.501414  0.317913
Ba  0.002708  0.502635  0.317933
Ba  0.002509  0.997130  0.317934
Ba  0.000000  0.000000  0.409091
Ba  0.000000  0.500000  0.409091
Ba  0.500000  0.500000  0.409091
Ba  0.500000  0.000000  0.409091
Ba  0.000000  0.000000  0.500000
Ba  0.000000  0.500000  0.500000
Ba  0.500000  0.000000  0.500000
Ba  0.500000  0.500000  0.500000
Ba  0.000000  0.000000  0.590909
Ba  0.000000  0.500000  0.590909
Ba  0.500000  0.000000  0.590909
Ba  0.500000  0.500000  0.590909
Ba  0.002509  0.997130  0.682066
Ba  0.002708  0.502635  0.682067
Ba  0.497896  0.501414  0.682087
Ba  0.498214  0.998365  0.682103
Ba  0.516338  0.483340  0.775078
Ba  0.993913  0.008541  0.775187
Ba  0.994848  0.491801  0.775305
Ba  0.517434  0.015836  0.775343
Hf  0.249067  0.749617  0.268975
Hf  0.751680  0.249693  0.272184
Hf  0.750762  0.749793  0.274463
Hf  0.250876  0.250001  0.274492
Hf  0.250920  0.750094  0.362932

| | | | |
|---|---|---|---|
| Hf | 0.750621 | 0.250025 | 0.363311 |
| Hf | 0.749167 | 0.749800 | 0.364146 |
| Hf | 0.249618 | 0.249773 | 0.364173 |
| Hf | 0.250000 | 0.250000 | 0.454545 |
| Hf | 0.250000 | 0.750000 | 0.454545 |
| Hf | 0.750000 | 0.250000 | 0.454545 |
| Hf | 0.750000 | 0.750000 | 0.454545 |
| Hf | 0.250000 | 0.250000 | 0.545455 |
| Hf | 0.250000 | 0.750000 | 0.545455 |
| Hf | 0.750000 | 0.250000 | 0.545455 |
| Hf | 0.750000 | 0.750000 | 0.545455 |
| Hf | 0.249618 | 0.249773 | 0.635827 |
| Hf | 0.749167 | 0.749800 | 0.635854 |
| Hf | 0.750621 | 0.250025 | 0.636689 |
| Hf | 0.250920 | 0.750094 | 0.637068 |
| Hf | 0.250876 | 0.250001 | 0.725508 |
| Hf | 0.750762 | 0.749793 | 0.725537 |
| Hf | 0.751680 | 0.249693 | 0.727816 |
| Hf | 0.249067 | 0.749617 | 0.731025 |
| C | 0.252639 | 0.248644 | 0.197728 |
| C | 0.749664 | 0.748889 | 0.198216 |
| C | 0.749664 | 0.748889 | 0.801784 |
| C | 0.252639 | 0.248644 | 0.802272 |
| O | 0.386629 | 0.244820 | 0.185128 |
| O | 0.737047 | 0.883719 | 0.186057 |
| O | 0.113396 | 0.246371 | 0.186940 |
| O | 0.741180 | 0.610662 | 0.187111 |
| O | 0.220468 | 0.745740 | 0.226622 |
| O | 0.259067 | 0.256281 | 0.227427 |
| O | 0.776265 | 0.752704 | 0.227517 |
| O | 0.740359 | 0.247163 | 0.228830 |
| O | 0.502807 | 0.749673 | 0.268554 |
| O | 0.249996 | 0.002526 | 0.271122 |
| O | 0.002147 | 0.249675 | 0.271208 |
| O | 0.749742 | 0.501349 | 0.271974 |
| O | 0.250253 | 0.497516 | 0.272286 |
| O | 0.749786 | 0.998134 | 0.272624 |
| O | 0.498828 | 0.249946 | 0.273435 |
| O | 0.997315 | 0.749954 | 0.274863 |
| O | 0.257551 | 0.751342 | 0.318028 |
| O | 0.739806 | 0.748761 | 0.318169 |
| O | 0.245986 | 0.248215 | 0.318188 |
| O | 0.753944 | 0.250688 | 0.318472 |
| O | 0.999319 | 0.750506 | 0.362840 |
| O | 0.499438 | 0.250536 | 0.363266 |
| O | 0.249399 | 0.499361 | 0.363376 |

O 0.749487 0.999261 0.363553
O 0.250640 0.000461 0.363611
O 0.750773 0.500526 0.363717
O 0.000692 0.249284 0.364074
O 0.500402 0.749263 0.364259
O 0.250000 0.250000 0.409091
O 0.250000 0.750000 0.409091
O 0.750000 0.750000 0.409091
O 0.750000 0.250000 0.409091
O 0.750000 0.000000 0.454545
O 0.750000 0.500000 0.454545
O 0.250000 0.500000 0.454545
O 0.250000 0.000000 0.454545
O 0.000000 0.250000 0.454545
O 0.000000 0.750000 0.454545
O 0.500000 0.750000 0.454545
O 0.500000 0.250000 0.454545
O 0.250000 0.250000 0.500000
O 0.250000 0.750000 0.500000
O 0.750000 0.250000 0.500000
O 0.750000 0.750000 0.500000
O 0.250000 0.000000 0.545455
O 0.250000 0.500000 0.545455
O 0.750000 0.000000 0.545455
O 0.750000 0.500000 0.545455
O 0.000000 0.250000 0.545455
O 0.000000 0.750000 0.545455
O 0.500000 0.250000 0.545455
O 0.500000 0.750000 0.545455
O 0.250000 0.250000 0.590909
O 0.250000 0.750000 0.590909
O 0.750000 0.250000 0.590909
O 0.750000 0.750000 0.590909
O 0.500402 0.749263 0.635741
O 0.000692 0.249284 0.635926
O 0.750773 0.500526 0.636283
O 0.250640 0.000461 0.636389
O 0.749487 0.999261 0.636447
O 0.249399 0.499361 0.636624
O 0.499438 0.250536 0.636734
O 0.999319 0.750506 0.637160
O 0.753944 0.250688 0.681528
O 0.245986 0.248215 0.681812
O 0.739806 0.748761 0.681831
O 0.257551 0.751342 0.681972
O 0.997315 0.749954 0.725137

O 0.498828 0.249946 0.726565
O 0.749786 0.998134 0.727376
O 0.250253 0.497516 0.727714
O 0.749742 0.501349 0.728026
O 0.002147 0.249675 0.728792
O 0.249996 0.002526 0.728877
O 0.502807 0.749673 0.731446
O 0.740359 0.247163 0.771170
O 0.776265 0.752704 0.772483
O 0.259067 0.256281 0.772573
O 0.220468 0.745740 0.773378
O 0.741180 0.610662 0.812889
O 0.113396 0.246371 0.813060
O 0.737047 0.883719 0.813943
O 0.386629 0.244820 0.814872

Structure 19. TiO$_2$-terminated SrTiO$_3$ slab at $\Theta$=0.00 CO$_2$ coverage

_cell_length_a   7.89043600
_cell_length_b   7.89043600
_cell_length_c   43.39739600
_cell_angle_alpha   90.00000000
_cell_angle_beta   90.00000000
_cell_angle_gamma   90.00000000
_symmetry_space_group_name_H-M   'P 1'
loop_
 _atom_site_type_symbol
 _atom_site_fract_x
 _atom_site_fract_y
 _atom_site_fract_z
 Sr  0.000000  0.000000  0.545440
 Sr  0.000000  0.499857  0.545440
 Sr  0.499857  0.000000  0.545440
 Sr  0.499857  0.499857  0.545440
 Sr  0.496676  0.496653  0.636815
 Sr  0.496674  0.996620  0.636814
 Sr  0.996643  0.496651  0.636814
 Sr  0.996641  0.996618  0.636812
 Sr  0.494971  0.494821  0.730588
 Sr  0.494969  0.994774  0.730591
 Sr  0.994923  0.494819  0.730591
 Sr  0.994921  0.994771  0.730594
 Sr  0.000000  0.000000  0.454560
 Sr  0.000000  0.499857  0.454560
 Sr  0.499857  0.000000  0.454560
 Sr  0.499857  0.499857  0.454560
 Sr  0.496676  0.496653  0.363185
 Sr  0.496674  0.996620  0.363186
 Sr  0.996643  0.496651  0.363186
 Sr  0.996641  0.996618  0.363188
 Sr  0.494971  0.494821  0.269412
 Sr  0.494969  0.994774  0.269409
 Sr  0.994923  0.494819  0.269409
 Sr  0.994921  0.994771  0.269406
 Ti  0.249929  0.249929  0.499998
 Ti  0.249929  0.749786  0.499998
 Ti  0.749786  0.249929  0.499998
 Ti  0.749786  0.749786  0.499998
 Ti  0.249929  0.249929  0.590881
 Ti  0.249929  0.749786  0.590881
 Ti  0.749786  0.249929  0.590881
 Ti  0.749786  0.749786  0.590881
 Ti  0.242119  0.242051  0.681348

```
Ti  0.242130  0.742022  0.681354
Ti  0.742089  0.242062  0.681354
Ti  0.742100  0.742033  0.681360
Ti  0.241344  0.241067  0.770353
Ti  0.241352  0.741034  0.770356
Ti  0.741310  0.241075  0.770356
Ti  0.741318  0.741042  0.770358
Ti  0.249929  0.249929  0.409119
Ti  0.249929  0.749786  0.409119
Ti  0.749786  0.249929  0.409119
Ti  0.749786  0.749786  0.409119
Ti  0.242119  0.242051  0.318652
Ti  0.242130  0.742022  0.318646
Ti  0.742089  0.242062  0.318646
Ti  0.742100  0.742033  0.318640
Ti  0.241344  0.241067  0.229647
Ti  0.241352  0.741034  0.229644
Ti  0.741310  0.241075  0.229644
Ti  0.741318  0.741042  0.229642
O  0.249929  0.000000  0.499998
O  0.249929  0.499857  0.499998
O  0.749786  0.000000  0.499998
O  0.749786  0.499857  0.499998
O  0.000000  0.249929  0.499998
O  0.000000  0.749786  0.499998
O  0.499857  0.249929  0.499998
O  0.499857  0.749786  0.499998
O  0.249929  0.000000  0.590881
O  0.249929  0.499857  0.590881
O  0.749786  0.000000  0.590881
O  0.749786  0.499857  0.590881
O  0.249929  0.249929  0.545440
O  0.249929  0.749786  0.545440
O  0.749786  0.249929  0.545440
O  0.749786  0.749786  0.545440
O  0.000000  0.249929  0.590881
O  0.000000  0.749786  0.590881
O  0.499857  0.249929  0.590881
O  0.499857  0.749786  0.590881
O  0.251603  0.251473  0.636420
O  0.251602  0.751362  0.636413
O  0.751491  0.251472  0.636413
O  0.751490  0.751360  0.636405
O  0.003539  0.253629  0.681753
O  0.003533  0.753521  0.681753
O  0.503526  0.253632  0.681754
```

```
O  0.503520  0.753524  0.681753
O  0.253066  0.252709  0.727543
O  0.253064  0.752596  0.727547
O  0.752952  0.252707  0.727547
O  0.752949  0.752594  0.727550
O  0.253899  0.003303  0.681754
O  0.253902  0.503290  0.681754
O  0.753789  0.003297  0.681753
O  0.753792  0.503284  0.681754
O  0.006863  0.255164  0.772575
O  0.006861  0.755056  0.772575
O  0.506843  0.255169  0.772581
O  0.506841  0.755061  0.772581
O  0.255568  0.006493  0.772595
O  0.255573  0.506472  0.772601
O  0.755461  0.006491  0.772595
O  0.755466  0.506470  0.772602
O  0.249929  0.000000  0.409119
O  0.249929  0.499857  0.409119
O  0.749786  0.000000  0.409119
O  0.749786  0.499857  0.409119
O  0.249929  0.249929  0.454560
O  0.249929  0.749786  0.454560
O  0.749786  0.249929  0.454560
O  0.749786  0.749786  0.454560
O  0.000000  0.249929  0.409119
O  0.000000  0.749786  0.409119
O  0.499857  0.249929  0.409119
O  0.499857  0.749786  0.409119
O  0.251603  0.251473  0.363580
O  0.251602  0.751362  0.363587
O  0.751491  0.251472  0.363587
O  0.751490  0.751360  0.363595
O  0.003539  0.253629  0.318247
O  0.003533  0.753521  0.318247
O  0.503526  0.253632  0.318246
O  0.503520  0.753524  0.318247
O  0.253066  0.252709  0.272457
O  0.253064  0.752596  0.272453
O  0.752952  0.252707  0.272453
O  0.752949  0.752594  0.272450
O  0.253899  0.003303  0.318246
O  0.253902  0.503290  0.318246
O  0.753789  0.003297  0.318247
O  0.753792  0.503284  0.318246
O  0.006863  0.255164  0.227425
```

O 0.006861 0.755056 0.227425
O 0.506843 0.255169 0.227419
O 0.506841 0.755061 0.227419
O 0.255568 0.006493 0.227405
O 0.255573 0.506472 0.227399
O 0.755461 0.006491 0.227405
O 0.755466 0.506470 0.227398

Structure 20. $TiO_2$-terminated $SrTiO_3$ slab at $\Theta=0.25$ $CO_2$ coverage

```
_cell_length_a   7.89043600
_cell_length_b   7.89043600
_cell_length_c   43.39739600
_cell_angle_alpha   90.00000000
_cell_angle_beta   90.00000000
_cell_angle_gamma   90.00000000
_symmetry_space_group_name_H-M   'P 1'
loop_
_atom_site_type_symbol
_atom_site_fract_x
_atom_site_fract_y
_atom_site_fract_z
 Sr  0.496769  0.494562  0.270389
 Sr  0.496185  0.993802  0.270704
 Sr  0.995993  0.995825  0.271026
 Sr  0.997511  0.487600  0.271976
 Sr  0.497317  0.499474  0.363342
 Sr  0.997106  0.994193  0.363384
 Sr  0.997260  0.499103  0.363413
 Sr  0.497384  0.993837  0.363420
 Sr  0.000000  0.000000  0.454560
 Sr  0.000000  0.499857  0.454560
 Sr  0.499857  0.000000  0.454560
 Sr  0.499857  0.499857  0.454560
 Sr  0.000000  0.000000  0.545440
 Sr  0.000000  0.499857  0.545440
 Sr  0.499857  0.000000  0.545440
 Sr  0.499857  0.499857  0.545440
 Sr  0.497384  0.993837  0.636580
 Sr  0.997260  0.499103  0.636587
 Sr  0.997106  0.994193  0.636616
 Sr  0.497317  0.499474  0.636658
 Sr  0.997511  0.487600  0.728024
 Sr  0.995993  0.995825  0.728974
 Sr  0.496185  0.993802  0.729296
 Sr  0.496769  0.494562  0.729611
 Ti  0.770631  0.734773  0.226846
 Ti  0.232012  0.735810  0.226971
 Ti  0.740760  0.237827  0.230364
 Ti  0.241353  0.237416  0.230408
 Ti  0.743860  0.742153  0.317458
 Ti  0.243816  0.742195  0.317536
 Ti  0.742476  0.242261  0.319103
 Ti  0.244146  0.241982  0.319117
 Ti  0.249929  0.249929  0.409118
```

```
Ti  0.249929  0.749786  0.409118
Ti  0.749786  0.749786  0.409118
Ti  0.749786  0.249929  0.409118
Ti  0.249929  0.749786  0.499998
Ti  0.749786  0.249929  0.499998
Ti  0.249929  0.249929  0.499998
Ti  0.749786  0.749786  0.499998
Ti  0.249929  0.249929  0.590882
Ti  0.749786  0.249929  0.590882
Ti  0.249929  0.749786  0.590882
Ti  0.749786  0.749786  0.590882
Ti  0.244146  0.241982  0.680883
Ti  0.742476  0.242261  0.680897
Ti  0.243816  0.742195  0.682464
Ti  0.743860  0.742153  0.682542
Ti  0.241353  0.237416  0.769592
Ti  0.740760  0.237827  0.769636
Ti  0.232012  0.735810  0.773030
Ti  0.770631  0.734773  0.773154
C   0.501524  0.759491  0.196400
C   0.501524  0.759491  0.803600
O   0.647693  0.759734  0.184192
O   0.355308  0.759845  0.184260
O   0.001533  0.763311  0.225125
O   0.505027  0.260165  0.227540
O   0.501966  0.754518  0.227850
O   0.760503  0.506802  0.228097
O   0.245944  0.507577  0.228451
O   0.005115  0.252659  0.228870
O   0.254742  0.007403  0.229763
O   0.750223  0.007729  0.229981
O   0.753664  0.749734  0.271731
O   0.250238  0.751591  0.271766
O   0.748514  0.257802  0.273160
O   0.255756  0.256256  0.273160
O   0.502919  0.754233  0.317405
O   0.753775  0.003295  0.317860
O   0.003026  0.254072  0.317893
O   0.002897  0.752762  0.317937
O   0.252500  0.003221  0.318102
O   0.253823  0.503290  0.318376
O   0.752526  0.503144  0.318626
O   0.502637  0.252640  0.318880
O   0.750331  0.752258  0.363452
O   0.251997  0.751670  0.363452
O   0.250185  0.251244  0.363681
```

O 0.752572 0.250555 0.363689
O 0.249929 0.000000 0.409118
O 0.249929 0.499857 0.409118
O 0.749786 0.499857 0.409118
O 0.749786 0.000000 0.409118
O 0.000000 0.249929 0.409118
O 0.000000 0.749786 0.409118
O 0.499857 0.249929 0.409118
O 0.499857 0.749786 0.409118
O 0.249929 0.249929 0.454560
O 0.249929 0.749786 0.454560
O 0.749786 0.249929 0.454560
O 0.749786 0.749786 0.454560
O 0.249929 0.000000 0.499998
O 0.249929 0.499857 0.499998
O 0.749786 0.000000 0.499998
O 0.749786 0.499857 0.499998
O 0.000000 0.249929 0.499998
O 0.499857 0.249929 0.499998
O 0.000000 0.749786 0.499998
O 0.499857 0.749786 0.499998
O 0.249929 0.249929 0.545440
O 0.249929 0.749786 0.545440
O 0.749786 0.249929 0.545440
O 0.749786 0.749786 0.545440
O 0.249929 0.000000 0.590882
O 0.749786 0.000000 0.590882
O 0.249929 0.499857 0.590882
O 0.749786 0.499857 0.590882
O 0.000000 0.249929 0.590882
O 0.000000 0.749786 0.590882
O 0.499857 0.749786 0.590882
O 0.499857 0.249929 0.590882
O 0.752572 0.250555 0.636311
O 0.250185 0.251244 0.636319
O 0.251997 0.751670 0.636548
O 0.750331 0.752258 0.636548
O 0.502637 0.252640 0.681120
O 0.752526 0.503144 0.681374
O 0.253823 0.503290 0.681624
O 0.252500 0.003221 0.681898
O 0.002897 0.752762 0.682063
O 0.003026 0.254072 0.682107
O 0.753775 0.003295 0.682140
O 0.502919 0.754233 0.682595
O 0.255756 0.256256 0.726840

O  0.748514  0.257802  0.726840
O  0.250238  0.751591  0.728234
O  0.753664  0.749734  0.728269
O  0.750223  0.007729  0.770019
O  0.254742  0.007403  0.770237
O  0.005115  0.252659  0.771130
O  0.245944  0.507577  0.771549
O  0.760503  0.506802  0.771903
O  0.501966  0.754518  0.772150
O  0.505027  0.260165  0.772460
O  0.001533  0.763311  0.774875
O  0.355308  0.759845  0.815740
O  0.647693  0.759734  0.815808

Structure 21. TiO$_2$-terminated SrTiO$_3$ slab at $\Theta$=0.50 CO$_2$ coverage

```
_cell_length_a  7.89043600
_cell_length_b  7.89043600
_cell_length_c  43.39739600
_cell_angle_alpha  90.00000000
_cell_angle_beta  90.00000000
_cell_angle_gamma  90.00000000
_symmetry_space_group_name_H-M  'P 1'
loop_
_atom_site_type_symbol
_atom_site_fract_x
_atom_site_fract_y
_atom_site_fract_z
 Sr 0.499472 0.496750 0.272174
 Sr 0.999482 0.996759 0.272175
 Sr 0.999539 0.486868 0.273260
 Sr 0.499526 0.986870 0.273260
 Sr 0.499193 0.497177 0.363568
 Sr 0.999206 0.997188 0.363603
 Sr 0.999193 0.496546 0.363698
 Sr 0.499177 0.996568 0.363698
 Sr 0.000000 0.499857 0.454560
 Sr 0.000000 0.000000 0.454560
 Sr 0.499857 0.000000 0.454560
 Sr 0.499857 0.499857 0.454560
 Sr 0.000000 0.000000 0.545440
 Sr 0.000000 0.499857 0.545440
 Sr 0.499857 0.000000 0.545440
 Sr 0.499857 0.499857 0.545440
 Sr 0.499177 0.996568 0.636302
 Sr 0.999193 0.496546 0.636302
 Sr 0.999206 0.997188 0.636397
 Sr 0.499193 0.497177 0.636432
 Sr 0.499526 0.986870 0.726740
 Sr 0.999539 0.486868 0.726740
 Sr 0.999482 0.996759 0.727825
 Sr 0.499472 0.496750 0.727826
 Ti 0.771271 0.733977 0.227639
 Ti 0.271307 0.233972 0.227645
 Ti 0.728519 0.233989 0.227649
 Ti 0.228509 0.733981 0.227652
 Ti 0.749494 0.742435 0.318076
 Ti 0.249477 0.242426 0.318083
 Ti 0.747527 0.242405 0.318087
 Ti 0.247523 0.742418 0.318087
 Ti 0.249929 0.749786 0.409119
```

```
Ti  0.249929  0.249929  0.409119
Ti  0.749786  0.249929  0.409119
Ti  0.749786  0.749786  0.409119
Ti  0.249929  0.249929  0.499998
Ti  0.249929  0.749786  0.499998
Ti  0.749786  0.749786  0.499998
Ti  0.749786  0.249929  0.499998
Ti  0.249929  0.249929  0.590881
Ti  0.249929  0.749786  0.590881
Ti  0.749786  0.249929  0.590881
Ti  0.749786  0.749786  0.590881
Ti  0.247523  0.742418  0.681913
Ti  0.747527  0.242405  0.681913
Ti  0.249477  0.242426  0.681917
Ti  0.749494  0.742435  0.681924
Ti  0.228509  0.733981  0.772348
Ti  0.728519  0.233989  0.772351
Ti  0.271307  0.233972  0.772355
Ti  0.771271  0.733977  0.772361
C   0.499907  0.749667  0.197732
C   0.999909  0.249710  0.197738
C   0.999909  0.249710  0.802262
C   0.499907  0.749667  0.802268
O   0.645980  0.752595  0.185582
O   0.145990  0.252651  0.185585
O   0.353822  0.752553  0.185586
O   0.853816  0.252606  0.185591
O   0.499908  0.281952  0.225314
O   0.999906  0.781930  0.225322
O   0.499918  0.740873  0.229225
O   0.999920  0.240885  0.229231
O   0.270712  0.009066  0.230356
O   0.770705  0.509068  0.230358
O   0.229077  0.509067  0.230378
O   0.729073  0.009053  0.230380
O   0.245246  0.755678  0.272509
O   0.745229  0.255682  0.272510
O   0.254995  0.255617  0.272512
O   0.754963  0.755607  0.272517
O   0.000409  0.256472  0.317453
O   0.500421  0.756470  0.317458
O   0.254022  0.503416  0.318274
O   0.754024  0.003403  0.318281
O   0.246631  0.003365  0.318293
O   0.746633  0.503381  0.318296
O   0.500345  0.249707  0.318625
```

| | | | |
|---|---|---|---|
| O | 0.000328 | 0.749709 | 0.318631 |
| O | 0.248844 | 0.251727 | 0.363567 |
| O | 0.251550 | 0.751644 | 0.363579 |
| O | 0.751509 | 0.251688 | 0.363579 |
| O | 0.748801 | 0.751678 | 0.363586 |
| O | 0.249929 | 0.000000 | 0.409119 |
| O | 0.749786 | 0.000000 | 0.409119 |
| O | 0.249929 | 0.499857 | 0.409119 |
| O | 0.749786 | 0.499857 | 0.409119 |
| O | 0.000000 | 0.249929 | 0.409119 |
| O | 0.000000 | 0.749786 | 0.409119 |
| O | 0.499857 | 0.249929 | 0.409119 |
| O | 0.499857 | 0.749786 | 0.409119 |
| O | 0.749786 | 0.749786 | 0.454560 |
| O | 0.249929 | 0.249929 | 0.454560 |
| O | 0.749786 | 0.249929 | 0.454560 |
| O | 0.249929 | 0.749786 | 0.454560 |
| O | 0.249929 | 0.000000 | 0.499998 |
| O | 0.249929 | 0.499857 | 0.499998 |
| O | 0.749786 | 0.000000 | 0.499998 |
| O | 0.749786 | 0.499857 | 0.499998 |
| O | 0.000000 | 0.749786 | 0.499998 |
| O | 0.000000 | 0.249929 | 0.499998 |
| O | 0.499857 | 0.249929 | 0.499998 |
| O | 0.499857 | 0.749786 | 0.499998 |
| O | 0.249929 | 0.249929 | 0.545440 |
| O | 0.249929 | 0.749786 | 0.545440 |
| O | 0.749786 | 0.249929 | 0.545440 |
| O | 0.749786 | 0.749786 | 0.545440 |
| O | 0.249929 | 0.000000 | 0.590881 |
| O | 0.249929 | 0.499857 | 0.590881 |
| O | 0.749786 | 0.000000 | 0.590881 |
| O | 0.749786 | 0.499857 | 0.590881 |
| O | 0.000000 | 0.749786 | 0.590881 |
| O | 0.499857 | 0.249929 | 0.590881 |
| O | 0.000000 | 0.249929 | 0.590881 |
| O | 0.499857 | 0.749786 | 0.590881 |
| O | 0.748801 | 0.751678 | 0.636414 |
| O | 0.251550 | 0.751644 | 0.636421 |
| O | 0.751509 | 0.251688 | 0.636421 |
| O | 0.248844 | 0.251727 | 0.636433 |
| O | 0.000328 | 0.749709 | 0.681369 |
| O | 0.500345 | 0.249707 | 0.681375 |
| O | 0.746633 | 0.503381 | 0.681704 |
| O | 0.246631 | 0.003365 | 0.681707 |
| O | 0.754024 | 0.003403 | 0.681719 |

```
O  0.254022  0.503416  0.681726
O  0.500421  0.756470  0.682542
O  0.000409  0.256472  0.682547
O  0.754963  0.755607  0.727483
O  0.254995  0.255617  0.727488
O  0.745229  0.255682  0.727490
O  0.245246  0.755678  0.727491
O  0.729073  0.009053  0.769620
O  0.229077  0.509067  0.769622
O  0.770705  0.509068  0.769642
O  0.270712  0.009066  0.769644
O  0.999920  0.240885  0.770769
O  0.499918  0.740873  0.770775
O  0.999906  0.781930  0.774678
O  0.499908  0.281952  0.774686
O  0.853816  0.252606  0.814409
O  0.353822  0.752553  0.814414
O  0.145990  0.252651  0.814415
O  0.645980  0.752595  0.814418
```

Structure 22. ZrO$_2$-terminated SrZrO$_3$ slab at $\Theta$=0.00 CO$_2$ coverage

```
_cell_length_a  8.39465800
_cell_length_b  8.39465800
_cell_length_c  46.17061600
_cell_angle_alpha  90.00000000
_cell_angle_beta  90.00000000
_cell_angle_gamma  90.00000000
_symmetry_space_group_name_H-M  'P 1'
loop_
 _atom_site_type_symbol
 _atom_site_fract_x
 _atom_site_fract_y
 _atom_site_fract_z
 Sr 0.000000 0.000000 0.545455
 Sr 0.996835 0.003177 0.637213
 Sr 0.999904 0.998202 0.729607
 Sr 0.000000 0.500000 0.545455
 Sr 0.998572 0.478496 0.636594
 Sr 0.000568 0.494508 0.730883
 Sr 0.500000 0.000000 0.545455
 Sr 0.499258 0.985873 0.634994
 Sr 0.500883 0.000840 0.729546
 Sr 0.500000 0.500000 0.545455
 Sr 0.496580 0.509113 0.636671
 Sr 0.498849 0.508141 0.727159
 Sr 0.000000 0.000000 0.454545
 Sr 0.996835 0.003177 0.362787
 Sr 0.999904 0.998202 0.270393
 Sr 0.000000 0.500000 0.454545
 Sr 0.998572 0.478496 0.363406
 Sr 0.000568 0.494508 0.269117
 Sr 0.500000 0.000000 0.454545
 Sr 0.499258 0.985873 0.365006
 Sr 0.500883 0.000840 0.270454
 Sr 0.500000 0.500000 0.454545
 Sr 0.496580 0.509113 0.363329
 Sr 0.498849 0.508141 0.272841
 Zr 0.250000 0.250000 0.500000
 Zr 0.250000 0.250000 0.590909
 Zr 0.252478 0.248342 0.681117
 Zr 0.259130 0.248813 0.768838
 Zr 0.250000 0.750000 0.500000
 Zr 0.250000 0.750000 0.590909
 Zr 0.245925 0.749325 0.680877
 Zr 0.243454 0.748269 0.768587
 Zr 0.750000 0.250000 0.500000
```

```
Zr  0.750000  0.250000  0.590909
Zr  0.746070  0.248561  0.681122
Zr  0.741419  0.249971  0.768928
Zr  0.750000  0.750000  0.500000
Zr  0.750000  0.750000  0.590909
Zr  0.752308  0.748950  0.680910
Zr  0.756132  0.746982  0.768632
Zr  0.250000  0.250000  0.409091
Zr  0.252478  0.248342  0.318883
Zr  0.259130  0.248813  0.231162
Zr  0.250000  0.750000  0.409091
Zr  0.245925  0.749325  0.319123
Zr  0.243454  0.748269  0.231413
Zr  0.750000  0.250000  0.409091
Zr  0.746070  0.248561  0.318878
Zr  0.741419  0.249971  0.231072
Zr  0.750000  0.750000  0.409091
Zr  0.752308  0.748950  0.319090
Zr  0.756132  0.746982  0.231368
O  0.250000  0.000000  0.500000
O  0.250000  0.000000  0.590909
O  0.202209  0.000346  0.682164
O  0.252710  0.997875  0.765913
O  0.250000  0.500000  0.500000
O  0.250000  0.500000  0.590909
O  0.297416  0.500441  0.680661
O  0.248105  0.497897  0.768158
O  0.750000  0.000000  0.500000
O  0.750000  0.000000  0.590909
O  0.796605  0.000344  0.683678
O  0.748276  0.997935  0.764966
O  0.750000  0.500000  0.500000
O  0.750000  0.500000  0.590909
O  0.702410  0.500239  0.679044
O  0.753647  0.497818  0.769983
O  0.250000  0.250000  0.545455
O  0.220512  0.253328  0.636625
O  0.302226  0.253479  0.726370
O  0.250000  0.750000  0.545455
O  0.283816  0.757499  0.636557
O  0.200066  0.739851  0.726176
O  0.750000  0.250000  0.545455
O  0.782175  0.246139  0.636667
O  0.698788  0.262058  0.726468
O  0.750000  0.750000  0.545455
O  0.718946  0.764895  0.636563
```

```
O  0.797721  0.731188  0.726228
O  0.000000  0.250000  0.500000
O  0.000000  0.250000  0.590909
O  0.000052  0.295188  0.687341
O  0.000520  0.245133  0.764503
O  0.000000  0.750000  0.500000
O  0.000000  0.750000  0.590909
O  0.999711  0.708946  0.672833
O  0.000126  0.754592  0.777342
O  0.500000  0.250000  0.500000
O  0.500000  0.250000  0.590909
O  0.500060  0.202343  0.674136
O  0.500330  0.243518  0.778914
O  0.500000  0.750000  0.500000
O  0.500000  0.750000  0.590909
O  0.499877  0.788832  0.688350
O  0.500104  0.747509  0.764011
O  0.250000  0.000000  0.409091
O  0.202209  0.000346  0.317836
O  0.252710  0.997875  0.234087
O  0.250000  0.500000  0.409091
O  0.297416  0.500441  0.319339
O  0.248105  0.497897  0.231842
O  0.750000  0.000000  0.409091
O  0.796605  0.000344  0.316322
O  0.748276  0.997935  0.235034
O  0.750000  0.500000  0.409091
O  0.702410  0.500239  0.320956
O  0.753647  0.497818  0.230017
O  0.250000  0.250000  0.454545
O  0.220512  0.253328  0.363375
O  0.302226  0.253479  0.273630
O  0.250000  0.750000  0.454545
O  0.283816  0.757499  0.363443
O  0.200066  0.739851  0.273824
O  0.750000  0.250000  0.454545
O  0.782175  0.246139  0.363333
O  0.698788  0.262058  0.273532
O  0.750000  0.750000  0.454545
O  0.718946  0.764895  0.363437
O  0.797721  0.731188  0.273772
O  0.000000  0.250000  0.409091
O  0.000052  0.295188  0.312659
O  0.000520  0.245133  0.235497
O  0.000000  0.750000  0.409091
O  0.999711  0.708946  0.327167
```

O  0.000126  0.754592  0.222658
O  0.500000  0.250000  0.409091
O  0.500060  0.202343  0.325864
O  0.500330  0.243518  0.221086
O  0.500000  0.750000  0.409091
O  0.499877  0.788832  0.311650
O  0.500104  0.747509  0.235989

Structure 23. ZrO$_2$-terminated SrZrO$_3$ slab at $\Theta$=0.25 CO$_2$ coverage

_cell_length_a  8.39465800
_cell_length_b  8.39465800
_cell_length_c  46.17061600
_cell_angle_alpha  90.00000000
_cell_angle_beta  90.00000000
_cell_angle_gamma  90.00000000
_symmetry_space_group_name_H-M  'P 1'
loop_
 _atom_site_type_symbol
 _atom_site_fract_x
 _atom_site_fract_y
 _atom_site_fract_z
 Sr  0.500013  0.997772  0.269879
 Sr  0.999923  0.491935  0.273026
 Sr  0.500149  0.503863  0.271053
 Sr  0.000171  0.004715  0.274042
 Sr  0.500562  0.497294  0.363317
 Sr  0.000562  0.989169  0.363249
 Sr  0.500106  0.016207  0.364393
 Sr  0.000148  0.519149  0.363962
 Sr  0.500000  0.500000  0.454545
 Sr  0.500000  0.000000  0.454545
 Sr  0.000000  0.500000  0.454545
 Sr  0.000000  0.000000  0.454545
 Sr  0.500000  0.500000  0.545455
 Sr  0.500000  0.000000  0.545455
 Sr  0.000000  0.500000  0.545455
 Sr  0.000000  0.000000  0.545455
 Sr  0.000148  0.519149  0.636038
 Sr  0.500106  0.016207  0.635607
 Sr  0.000562  0.989169  0.636751
 Sr  0.500562  0.497294  0.636683
 Sr  0.000171  0.004715  0.725958
 Sr  0.500149  0.503863  0.728947
 Sr  0.999923  0.491935  0.726974
 Sr  0.500013  0.997772  0.730121
 Zr  0.235843  0.751092  0.230321
 Zr  0.764292  0.751207  0.230319
 Zr  0.757671  0.250457  0.231380
 Zr  0.242445  0.250563  0.231386
 Zr  0.249447  0.251011  0.319229
 Zr  0.750896  0.250999  0.319232
 Zr  0.746086  0.751400  0.318394
 Zr  0.254266  0.751354  0.318396
 Zr  0.250000  0.250000  0.409091

Zr 0.250000 0.750000 0.409091
Zr 0.750000 0.250000 0.409091
Zr 0.750000 0.750000 0.409091
Zr 0.250000 0.250000 0.500000
Zr 0.250000 0.750000 0.500000
Zr 0.750000 0.250000 0.500000
Zr 0.750000 0.750000 0.500000
Zr 0.250000 0.250000 0.590909
Zr 0.250000 0.750000 0.590909
Zr 0.750000 0.250000 0.590909
Zr 0.750000 0.750000 0.590909
Zr 0.254266 0.751354 0.681604
Zr 0.746086 0.751400 0.681606
Zr 0.750896 0.250999 0.680768
Zr 0.249447 0.251011 0.680771
Zr 0.242445 0.250563 0.768614
Zr 0.757671 0.250457 0.768620
Zr 0.764292 0.751207 0.769681
Zr 0.235843 0.751092 0.769679
C 0.500075 0.743314 0.197621
C 0.500075 0.743314 0.802379
O 0.637894 0.745813 0.186131
O 0.362256 0.746182 0.186132
O 0.500071 0.738954 0.227151
O 0.000039 0.245381 0.222599
O 0.744536 0.999515 0.232998
O 0.255388 0.999494 0.233154
O 0.241583 0.502151 0.234086
O 0.758428 0.502165 0.234217
O 0.000062 0.756799 0.238822
O 0.500027 0.256299 0.235570
O 0.297749 0.754010 0.273354
O 0.799579 0.248253 0.273936
O 0.702530 0.754950 0.273354
O 0.200568 0.249088 0.273943
O 0.000098 0.705962 0.313122
O 0.500084 0.203314 0.312641
O 0.701170 0.499286 0.318076
O 0.199710 0.999724 0.318981
O 0.299141 0.499288 0.318289
O 0.800611 0.999757 0.319173
O 0.000075 0.299236 0.325210
O 0.500109 0.797009 0.325797
O 0.221906 0.743969 0.363358
O 0.718676 0.246379 0.363473
O 0.777741 0.742773 0.363358

```
O  0.280893  0.245162  0.363477
O  0.250000  0.000000  0.409091
O  0.750000  0.500000  0.409091
O  0.250000  0.500000  0.409091
O  0.750000  0.000000  0.409091
O  0.500000  0.250000  0.409091
O  0.500000  0.750000  0.409091
O  0.000000  0.750000  0.409091
O  0.000000  0.250000  0.409091
O  0.750000  0.250000  0.454545
O  0.750000  0.750000  0.454545
O  0.250000  0.750000  0.454545
O  0.250000  0.250000  0.454545
O  0.250000  0.500000  0.500000
O  0.250000  0.000000  0.500000
O  0.750000  0.500000  0.500000
O  0.750000  0.000000  0.500000
O  0.500000  0.250000  0.500000
O  0.500000  0.750000  0.500000
O  0.000000  0.250000  0.500000
O  0.000000  0.750000  0.500000
O  0.250000  0.250000  0.545455
O  0.250000  0.750000  0.545455
O  0.750000  0.250000  0.545455
O  0.750000  0.750000  0.545455
O  0.250000  0.500000  0.590909
O  0.250000  0.000000  0.590909
O  0.750000  0.500000  0.590909
O  0.750000  0.000000  0.590909
O  0.500000  0.250000  0.590909
O  0.500000  0.750000  0.590909
O  0.000000  0.250000  0.590909
O  0.000000  0.750000  0.590909
O  0.280893  0.245162  0.636523
O  0.777741  0.742773  0.636642
O  0.718676  0.246379  0.636527
O  0.221906  0.743969  0.636642
O  0.500109  0.797009  0.674203
O  0.000075  0.299236  0.674790
O  0.800611  0.999757  0.680827
O  0.299141  0.499288  0.681711
O  0.199710  0.999724  0.681019
O  0.701170  0.499286  0.681924
O  0.500084  0.203314  0.687359
O  0.000098  0.705962  0.686878
O  0.200568  0.249088  0.726057
```

O  0.702530  0.754950  0.726646
O  0.799579  0.248253  0.726064
O  0.297749  0.754010  0.726646
O  0.500027  0.256299  0.764430
O  0.000062  0.756799  0.761178
O  0.758428  0.502165  0.765783
O  0.241583  0.502151  0.765914
O  0.255388  0.999494  0.766846
O  0.744536  0.999515  0.767002
O  0.000039  0.245381  0.777401
O  0.500071  0.738954  0.772849
O  0.362256  0.746182  0.813868
O  0.637894  0.745813  0.813869

Structure 24. ZrO$_2$-terminated SrZrO$_3$ slab at $\Theta$=0.50 CO$_2$ coverage

_cell_length_a  8.39465800
_cell_length_b  8.39465800
_cell_length_c  46.17061600
_cell_angle_alpha  90.00000000
_cell_angle_beta  90.00000000
_cell_angle_gamma  90.00000000
_symmetry_space_group_name_H-M  'P 1'
loop_
 _atom_site_type_symbol
 _atom_site_fract_x
 _atom_site_fract_y
 _atom_site_fract_z
 Sr 0.999773 0.489837 0.272729
 Sr 0.499822 0.988843 0.272822
 Sr 0.000588 0.012225 0.273430
 Sr 0.500650 0.510800 0.273694
 Sr 0.002493 0.995304 0.363493
 Sr 0.502450 0.493797 0.363604
 Sr 0.000419 0.518320 0.364009
 Sr 0.500316 0.016712 0.364319
 Sr 0.000000 0.000000 0.454545
 Sr 0.000000 0.500000 0.454545
 Sr 0.500000 0.000000 0.454545
 Sr 0.500000 0.500000 0.454545
 Sr 0.000000 0.000000 0.545455
 Sr 0.000000 0.500000 0.545455
 Sr 0.500000 0.000000 0.545455
 Sr 0.500000 0.500000 0.545455
 Sr 0.500316 0.016712 0.635681
 Sr 0.000419 0.518320 0.635991
 Sr 0.502450 0.493797 0.636396
 Sr 0.002493 0.995304 0.636507
 Sr 0.500650 0.510800 0.726306
 Sr 0.000588 0.012225 0.726570
 Sr 0.499822 0.988843 0.727178
 Sr 0.999773 0.489837 0.727271
 Zr 0.736470 0.251980 0.230478
 Zr 0.263705 0.252305 0.230488
 Zr 0.236462 0.751803 0.230489
 Zr 0.763694 0.752082 0.230491
 Zr 0.749089 0.751899 0.318700
 Zr 0.252085 0.751736 0.318705
 Zr 0.249145 0.251716 0.318706
 Zr 0.751986 0.251529 0.318713
 Zr 0.750000 0.750000 0.409091

| | | | |
|---|---|---|---|
| Zr | 0.750000 | 0.250000 | 0.409091 |
| Zr | 0.250000 | 0.750000 | 0.409091 |
| Zr | 0.250000 | 0.250000 | 0.409091 |
| Zr | 0.750000 | 0.750000 | 0.500000 |
| Zr | 0.750000 | 0.250000 | 0.500000 |
| Zr | 0.250000 | 0.750000 | 0.500000 |
| Zr | 0.250000 | 0.250000 | 0.500000 |
| Zr | 0.750000 | 0.750000 | 0.590909 |
| Zr | 0.750000 | 0.250000 | 0.590909 |
| Zr | 0.250000 | 0.750000 | 0.590909 |
| Zr | 0.250000 | 0.250000 | 0.590909 |
| Zr | 0.751986 | 0.251529 | 0.681287 |
| Zr | 0.249145 | 0.251716 | 0.681294 |
| Zr | 0.252085 | 0.751736 | 0.681295 |
| Zr | 0.749089 | 0.751899 | 0.681300 |
| Zr | 0.763694 | 0.752082 | 0.769509 |
| Zr | 0.236462 | 0.751803 | 0.769511 |
| Zr | 0.263705 | 0.252305 | 0.769512 |
| Zr | 0.736470 | 0.251980 | 0.769522 |
| C | 0.000131 | 0.241767 | 0.198208 |
| C | 0.500109 | 0.743552 | 0.198219 |
| C | 0.500109 | 0.743552 | 0.801781 |
| C | 0.000131 | 0.241767 | 0.801792 |
| O | 0.137926 | 0.247146 | 0.186834 |
| O | 0.862344 | 0.247937 | 0.186837 |
| O | 0.637890 | 0.749119 | 0.186841 |
| O | 0.362312 | 0.749922 | 0.186852 |
| O | 0.000081 | 0.231231 | 0.227786 |
| O | 0.500076 | 0.732322 | 0.227788 |
| O | 0.237166 | 0.500739 | 0.233806 |
| O | 0.762444 | 0.500717 | 0.234133 |
| O | 0.737588 | 0.000729 | 0.234224 |
| O | 0.262060 | 0.000758 | 0.234546 |
| O | 0.500069 | 0.262322 | 0.238290 |
| O | 0.000060 | 0.762447 | 0.238333 |
| O | 0.796835 | 0.251411 | 0.273643 |
| O | 0.296707 | 0.748781 | 0.273651 |
| O | 0.203578 | 0.253523 | 0.273660 |
| O | 0.703700 | 0.750781 | 0.273661 |
| O | 0.500234 | 0.203324 | 0.313268 |
| O | 0.000273 | 0.702241 | 0.313568 |
| O | 0.199404 | 0.999758 | 0.318241 |
| O | 0.699482 | 0.499756 | 0.318695 |
| O | 0.801709 | 0.999821 | 0.318810 |
| O | 0.301571 | 0.499825 | 0.319297 |
| O | 0.500279 | 0.800799 | 0.324620 |

```
O  0.000242  0.299785  0.324850
O  0.721479  0.245343  0.363435
O  0.222169  0.747609  0.363441
O  0.276784  0.241061  0.363441
O  0.776353  0.743392  0.363445
O  0.750000  0.500000  0.409091
O  0.250000  0.000000  0.409091
O  0.750000  0.000000  0.409091
O  0.250000  0.500000  0.409091
O  0.000000  0.750000  0.409091
O  0.000000  0.250000  0.409091
O  0.500000  0.250000  0.409091
O  0.500000  0.750000  0.409091
O  0.250000  0.750000  0.454545
O  0.250000  0.250000  0.454545
O  0.750000  0.250000  0.454545
O  0.750000  0.750000  0.454545
O  0.750000  0.000000  0.500000
O  0.750000  0.500000  0.500000
O  0.250000  0.000000  0.500000
O  0.250000  0.500000  0.500000
O  0.000000  0.750000  0.500000
O  0.000000  0.250000  0.500000
O  0.500000  0.750000  0.500000
O  0.500000  0.250000  0.500000
O  0.750000  0.750000  0.545455
O  0.750000  0.250000  0.545455
O  0.250000  0.750000  0.545455
O  0.250000  0.250000  0.545455
O  0.750000  0.000000  0.590909
O  0.750000  0.500000  0.590909
O  0.250000  0.000000  0.590909
O  0.250000  0.500000  0.590909
O  0.000000  0.750000  0.590909
O  0.000000  0.250000  0.590909
O  0.500000  0.750000  0.590909
O  0.500000  0.250000  0.590909
O  0.776353  0.743392  0.636555
O  0.276784  0.241061  0.636559
O  0.222169  0.747609  0.636559
O  0.721479  0.245343  0.636565
O  0.000242  0.299785  0.675150
O  0.500279  0.800799  0.675380
O  0.301571  0.499825  0.680703
O  0.801709  0.999821  0.681190
O  0.699482  0.499756  0.681305
```

```
O  0.199404  0.999758  0.681759
O  0.000273  0.702241  0.686432
O  0.500234  0.203324  0.686732
O  0.703700  0.750781  0.726339
O  0.203578  0.253523  0.726340
O  0.296707  0.748781  0.726349
O  0.796835  0.251411  0.726357
O  0.000060  0.762447  0.761667
O  0.500069  0.262322  0.761710
O  0.262060  0.000758  0.765454
O  0.737588  0.000729  0.765776
O  0.762444  0.500717  0.765867
O  0.237166  0.500739  0.766194
O  0.500076  0.732322  0.772212
O  0.000081  0.231231  0.772214
O  0.362312  0.749922  0.813148
O  0.637890  0.749119  0.813159
O  0.862344  0.247937  0.813163
O  0.137926  0.247146  0.813166
```

Structure 25. $HfO_2$-terminated $SrHfO_3$ slab at $\Theta$=0.00 $CO_2$ coverage

_cell_length_a  8.28657000
_cell_length_b  8.28657000
_cell_length_c  45.57613800
_cell_angle_alpha  90.00000000
_cell_angle_beta  90.00000000
_cell_angle_gamma  90.00000000
_symmetry_space_group_name_H-M  'P 1'
loop_
 _atom_site_type_symbol
 _atom_site_fract_x
 _atom_site_fract_y
 _atom_site_fract_z
 Sr  0.500000  0.500000  0.545455
 Sr  0.498902  0.503017  0.636558
 Sr  0.498566  0.503783  0.728855
 Sr  0.500000  0.000000  0.545455
 Sr  0.499398  0.988516  0.635898
 Sr  0.498690  0.001926  0.729953
 Sr  0.000000  0.500000  0.545455
 Sr  0.999430  0.486833  0.636292
 Sr  0.998581  0.499448  0.730273
 Sr  0.000000  0.000000  0.545455
 Sr  0.998940  0.001655  0.636742
 Sr  0.998548  0.000332  0.729618
 Sr  0.500000  0.500000  0.454545
 Sr  0.498902  0.503017  0.363442
 Sr  0.498566  0.503783  0.271145
 Sr  0.500000  0.000000  0.454545
 Sr  0.499398  0.988516  0.364102
 Sr  0.498690  0.001926  0.270047
 Sr  0.000000  0.500000  0.454545
 Sr  0.999430  0.486833  0.363708
 Sr  0.998581  0.499448  0.269727
 Sr  0.000000  0.000000  0.454545
 Sr  0.998940  0.001655  0.363258
 Sr  0.998548  0.000332  0.270382
 Hf  0.250000  0.250000  0.500000
 Hf  0.250000  0.750000  0.500000
 Hf  0.750000  0.250000  0.500000
 Hf  0.750000  0.750000  0.500000
 Hf  0.750000  0.750000  0.590909
 Hf  0.750948  0.749333  0.681176
 Hf  0.753465  0.749326  0.769321
 Hf  0.750000  0.250000  0.590909
 Hf  0.746788  0.249323  0.681224

| | | | |
|---|---|---|---|
| Hf | 0.742885 | 0.250422 | 0.769378 |
| Hf | 0.250000 | 0.750000 | 0.590909 |
| Hf | 0.246780 | 0.749413 | 0.681167 |
| Hf | 0.243493 | 0.749485 | 0.769298 |
| Hf | 0.250000 | 0.250000 | 0.590909 |
| Hf | 0.251009 | 0.249224 | 0.681216 |
| Hf | 0.254159 | 0.250263 | 0.769356 |
| Hf | 0.750000 | 0.750000 | 0.409091 |
| Hf | 0.750948 | 0.749333 | 0.318824 |
| Hf | 0.753465 | 0.749326 | 0.230679 |
| Hf | 0.750000 | 0.250000 | 0.409091 |
| Hf | 0.746788 | 0.249323 | 0.318776 |
| Hf | 0.742885 | 0.250422 | 0.230622 |
| Hf | 0.250000 | 0.750000 | 0.409091 |
| Hf | 0.246780 | 0.749413 | 0.318833 |
| Hf | 0.243493 | 0.749485 | 0.230702 |
| Hf | 0.250000 | 0.250000 | 0.409091 |
| Hf | 0.251009 | 0.249224 | 0.318784 |
| Hf | 0.254159 | 0.250263 | 0.230644 |
| O | 0.250000 | 0.000000 | 0.500000 |
| O | 0.250000 | 0.500000 | 0.500000 |
| O | 0.750000 | 0.000000 | 0.500000 |
| O | 0.750000 | 0.500000 | 0.500000 |
| O | 0.000000 | 0.250000 | 0.500000 |
| O | 0.000000 | 0.750000 | 0.500000 |
| O | 0.500000 | 0.250000 | 0.500000 |
| O | 0.500000 | 0.750000 | 0.500000 |
| O | 0.500000 | 0.750000 | 0.590909 |
| O | 0.498857 | 0.790068 | 0.686572 |
| O | 0.498481 | 0.748592 | 0.766040 |
| O | 0.500000 | 0.250000 | 0.590909 |
| O | 0.498968 | 0.206110 | 0.676332 |
| O | 0.498506 | 0.249641 | 0.776771 |
| O | 0.000000 | 0.750000 | 0.590909 |
| O | 0.998901 | 0.707901 | 0.675973 |
| O | 0.998497 | 0.753065 | 0.776386 |
| O | 0.000000 | 0.250000 | 0.590909 |
| O | 0.998905 | 0.291550 | 0.686281 |
| O | 0.998527 | 0.247771 | 0.766120 |
| O | 0.750000 | 0.750000 | 0.545455 |
| O | 0.728232 | 0.756774 | 0.636499 |
| O | 0.785254 | 0.743174 | 0.726697 |
| O | 0.750000 | 0.250000 | 0.545455 |
| O | 0.770597 | 0.251625 | 0.636527 |
| O | 0.711087 | 0.253544 | 0.726738 |
| O | 0.250000 | 0.750000 | 0.545455 |

O  0.270852  0.755792  0.636500
O  0.211731  0.744368  0.726672
O  0.250000  0.250000  0.545455
O  0.228540  0.252477  0.636510
O  0.286287  0.252358  0.726722
O  0.750000  0.500000  0.590909
O  0.704766  0.500319  0.680849
O  0.750602  0.499604  0.769804
O  0.750000  0.000000  0.590909
O  0.792540  0.000326  0.682248
O  0.746574  0.999573  0.768206
O  0.250000  0.500000  0.590909
O  0.292817  0.500316  0.680998
O  0.246425  0.499593  0.769562
O  0.250000  0.000000  0.590909
O  0.204966  0.000322  0.682023
O  0.250459  0.999579  0.768321
O  0.500000  0.750000  0.409091
O  0.498857  0.790068  0.313428
O  0.498481  0.748592  0.233960
O  0.500000  0.250000  0.409091
O  0.498968  0.206110  0.323668
O  0.498506  0.249641  0.223229
O  0.000000  0.750000  0.409091
O  0.998901  0.707901  0.324027
O  0.998497  0.753065  0.223614
O  0.000000  0.250000  0.409091
O  0.998905  0.291550  0.313719
O  0.998527  0.247771  0.233880
O  0.750000  0.750000  0.454545
O  0.728232  0.756774  0.363501
O  0.785254  0.743174  0.273303
O  0.750000  0.250000  0.454545
O  0.770597  0.251625  0.363473
O  0.711087  0.253544  0.273262
O  0.250000  0.750000  0.454545
O  0.270852  0.755792  0.363500
O  0.211731  0.744368  0.273328
O  0.250000  0.250000  0.454545
O  0.228540  0.252477  0.363490
O  0.286287  0.252358  0.273278
O  0.750000  0.500000  0.409091
O  0.704766  0.500319  0.319151
O  0.750602  0.499604  0.230196
O  0.750000  0.000000  0.409091
O  0.792540  0.000326  0.317752

O  0.746574  0.999573  0.231794
O  0.250000  0.500000  0.409091
O  0.292817  0.500316  0.319002
O  0.246425  0.499593  0.230438
O  0.250000  0.000000  0.409091
O  0.204966  0.000322  0.317977
O  0.250459  0.999579  0.231679

Structure 26. HfO$_2$-terminated SrHfO$_3$ slab at $\Theta$=0.25 CO$_2$ coverage

```
_cell_length_a  8.28657000
_cell_length_b  8.28657000
_cell_length_c  45.57613800
_cell_angle_alpha  90.00000000
_cell_angle_beta  90.00000000
_cell_angle_gamma  90.00000000
_symmetry_space_group_name_H-M  'P 1'
loop_
 _atom_site_type_symbol
 _atom_site_fract_x
 _atom_site_fract_y
 _atom_site_fract_z
  Sr  0.500000  0.500000  0.545455
  Sr  0.499269  0.500105  0.636391
  Sr  0.499089  0.499834  0.728847
  Sr  0.500000  0.000000  0.545455
  Sr  0.499644  0.009411  0.635711
  Sr  0.499138  0.996016  0.729532
  Sr  0.000000  0.500000  0.545455
  Sr  0.999633  0.514356  0.636384
  Sr  0.999162  0.492535  0.727926
  Sr  0.000000  0.000000  0.545455
  Sr  0.999282  0.995506  0.636554
  Sr  0.999090  0.005949  0.727588
  Sr  0.500000  0.500000  0.454545
  Sr  0.499269  0.500105  0.363609
  Sr  0.499089  0.499834  0.271153
  Sr  0.500000  0.000000  0.454545
  Sr  0.499644  0.009411  0.364289
  Sr  0.499138  0.996016  0.270468
  Sr  0.000000  0.500000  0.454545
  Sr  0.999633  0.514356  0.363616
  Sr  0.999162  0.492535  0.272074
  Sr  0.000000  0.000000  0.454545
  Sr  0.999282  0.995506  0.363446
  Sr  0.999090  0.005949  0.272412
  Hf  0.250000  0.750000  0.500000
  Hf  0.250000  0.250000  0.500000
  Hf  0.750000  0.750000  0.500000
  Hf  0.750000  0.250000  0.500000
  Hf  0.750000  0.250000  0.590909
  Hf  0.749909  0.250625  0.680932
  Hf  0.753995  0.250090  0.769112
  Hf  0.750000  0.750000  0.590909
  Hf  0.746461  0.751028  0.681885
```

```
Hf  0.762433  0.749917  0.770625
Hf  0.250000  0.250000  0.590909
Hf  0.248655  0.250588  0.680932
Hf  0.244056  0.250014  0.769111
Hf  0.250000  0.750000  0.590909
Hf  0.252123  0.751080  0.681885
Hf  0.235606  0.749974  0.770620
Hf  0.750000  0.250000  0.409091
Hf  0.749909  0.250625  0.319068
Hf  0.753995  0.250090  0.230888
Hf  0.750000  0.750000  0.409091
Hf  0.746461  0.751028  0.318114
Hf  0.762433  0.749917  0.229375
Hf  0.250000  0.250000  0.409091
Hf  0.248655  0.250588  0.319068
Hf  0.244056  0.250014  0.230889
Hf  0.250000  0.750000  0.409091
Hf  0.252123  0.751080  0.318115
Hf  0.235606  0.749974  0.229380
C   0.498961  0.749908  0.803368
C   0.498961  0.749908  0.196632
O   0.498999  0.747126  0.773328
O   0.638719  0.751367  0.814875
O   0.359163  0.751108  0.814863
O   0.498999  0.747126  0.226672
O   0.638719  0.751367  0.185125
O   0.359163  0.751108  0.185137
O   0.250000  0.000000  0.500000
O   0.250000  0.500000  0.500000
O   0.750000  0.000000  0.500000
O   0.750000  0.500000  0.500000
O   0.000000  0.750000  0.500000
O   0.000000  0.250000  0.500000
O   0.500000  0.750000  0.500000
O   0.500000  0.250000  0.500000
O   0.500000  0.250000  0.590909
O   0.499285  0.208373  0.686261
O   0.499015  0.252297  0.766070
O   0.500000  0.750000  0.590909
O   0.499332  0.795304  0.676756
O   0.000000  0.250000  0.590909
O   0.999314  0.294830  0.676738
O   0.999010  0.246615  0.775638
O   0.000000  0.750000  0.590909
O   0.999300  0.708646  0.686124
O   0.999014  0.753220  0.764242
```

O  0.750000  0.250000  0.545455
O  0.729223  0.244204  0.636411
O  0.785150  0.253149  0.726378
O  0.750000  0.750000  0.545455
O  0.768379  0.747610  0.636648
O  0.711204  0.749182  0.727200
O  0.250000  0.250000  0.545455
O  0.270244  0.244907  0.636416
O  0.213035  0.252600  0.726375
O  0.250000  0.750000  0.545455
O  0.231060  0.747033  0.636643
O  0.287077  0.749716  0.727201
O  0.750000  0.500000  0.590909
O  0.704972  0.499764  0.681134
O  0.752848  0.501118  0.768258
O  0.750000  0.000000  0.590909
O  0.794528  0.000072  0.681783
O  0.747057  0.999268  0.767677
O  0.250000  0.500000  0.590909
O  0.293503  0.499767  0.681243
O  0.245089  0.501121  0.768167
O  0.250000  0.000000  0.590909
O  0.203963  0.000080  0.681673
O  0.251002  0.999274  0.767749
O  0.500000  0.250000  0.409091
O  0.499285  0.208373  0.313739
O  0.499015  0.252297  0.233930
O  0.500000  0.750000  0.409091
O  0.499332  0.795304  0.323244
O  0.000000  0.250000  0.409091
O  0.999314  0.294830  0.323262
O  0.999010  0.246615  0.224362
O  0.000000  0.750000  0.409091
O  0.999300  0.708646  0.313876
O  0.999014  0.753220  0.235758
O  0.750000  0.250000  0.454545
O  0.729223  0.244204  0.363589
O  0.785150  0.253149  0.273622
O  0.750000  0.750000  0.454545
O  0.768379  0.747610  0.363352
O  0.711204  0.749182  0.272800
O  0.250000  0.250000  0.454545
O  0.270244  0.244907  0.363584
O  0.213035  0.252600  0.273625
O  0.250000  0.750000  0.454545
O  0.231060  0.747033  0.363357

```
O  0.287077  0.749716  0.272799
O  0.750000  0.500000  0.409091
O  0.704972  0.499764  0.318866
O  0.752848  0.501118  0.231742
O  0.750000  0.000000  0.409091
O  0.794528  0.000072  0.318217
O  0.747057  0.999268  0.232323
O  0.250000  0.500000  0.409091
O  0.293503  0.499767  0.318757
O  0.245089  0.501121  0.231833
O  0.250000  0.000000  0.409091
O  0.203963  0.000080  0.318327
O  0.251002  0.999274  0.232251
```

Structure 27. HfO$_2$-terminated SrHfO$_3$ slab at $\Theta$=0.50 CO$_2$ coverage

```
_cell_length_a  8.28657000
_cell_length_b  8.28657000
_cell_length_c  45.57613800
_cell_angle_alpha  90.00000000
_cell_angle_beta  90.00000000
_cell_angle_gamma  90.00000000
_symmetry_space_group_name_H-M  'P 1'
loop_
_atom_site_type_symbol
_atom_site_fract_x
_atom_site_fract_y
_atom_site_fract_z
 Sr  0.000309  0.490353  0.272457
 Sr  0.500298  0.990045  0.272489
 Sr  0.000403  0.007825  0.272998
 Sr  0.500411  0.507342  0.273089
 Sr  0.000446  0.998174  0.363762
 Sr  0.500448  0.497675  0.363785
 Sr  0.000166  0.511858  0.364045
 Sr  0.500177  0.011274  0.364168
 Sr  0.500000  0.500000  0.454545
 Sr  0.500000  0.000000  0.454545
 Sr  0.000000  0.500000  0.454545
 Sr  0.000000  0.000000  0.454545
 Sr  0.500000  0.500000  0.545455
 Sr  0.500000  0.000000  0.545455
 Sr  0.000000  0.500000  0.545455
 Sr  0.000000  0.000000  0.545455
 Sr  0.500177  0.011274  0.635832
 Sr  0.000166  0.511858  0.635955
 Sr  0.500448  0.497675  0.636215
 Sr  0.000446  0.998174  0.636238
 Sr  0.500411  0.507342  0.726911
 Sr  0.000403  0.007825  0.727002
 Sr  0.500298  0.990045  0.727511
 Sr  0.000309  0.490353  0.727543
 Hf  0.263563  0.250464  0.229769
 Hf  0.237174  0.750312  0.229770
 Hf  0.763548  0.750361  0.229770
 Hf  0.737153  0.250426  0.229770
 Hf  0.251716  0.750858  0.318520
 Hf  0.749029  0.750903  0.318520
 Hf  0.249038  0.250811  0.318522
 Hf  0.751710  0.250776  0.318523
 Hf  0.250000  0.250000  0.409091
```

```
Hf  0.250000  0.750000  0.409091
Hf  0.750000  0.250000  0.409091
Hf  0.750000  0.750000  0.409091
Hf  0.750000  0.750000  0.500000
Hf  0.750000  0.250000  0.500000
Hf  0.250000  0.750000  0.500000
Hf  0.250000  0.250000  0.500000
Hf  0.250000  0.250000  0.590909
Hf  0.250000  0.750000  0.590909
Hf  0.750000  0.250000  0.590909
Hf  0.750000  0.750000  0.590909
Hf  0.751710  0.250776  0.681477
Hf  0.249038  0.250811  0.681478
Hf  0.749029  0.750903  0.681480
Hf  0.251716  0.750858  0.681480
Hf  0.737153  0.250426  0.770230
Hf  0.763548  0.750361  0.770230
Hf  0.237174  0.750312  0.770230
Hf  0.263563  0.250464  0.770231
C  0.000350  0.245357  0.197380
C  0.500362  0.745615  0.197381
C  0.500362  0.745615  0.802619
C  0.000350  0.245357  0.802620
O  0.140152  0.248586  0.186005
O  0.640150  0.748848  0.186005
O  0.860557  0.248651  0.186005
O  0.360578  0.748904  0.186006
O  0.500361  0.739480  0.227547
O  0.000355  0.239268  0.227548
O  0.242155  0.500237  0.232837
O  0.758533  0.500245  0.232874
O  0.742230  0.000248  0.232968
O  0.258464  0.000245  0.233004
O  0.000361  0.758519  0.235853
O  0.500358  0.258476  0.235855
O  0.788262  0.251072  0.273291
O  0.212476  0.251319  0.273291
O  0.288232  0.750056  0.273292
O  0.712510  0.750325  0.273294
O  0.500372  0.208030  0.313968
O  0.000370  0.707598  0.314061
O  0.205063  0.999865  0.318549
O  0.795845  0.999872  0.318621
O  0.705165  0.499876  0.318773
O  0.295742  0.499871  0.318834
O  0.500356  0.795704  0.323101
```

```
O  0.000359  0.295467  0.323163
O  0.731049  0.245634  0.363507
O  0.231197  0.746673  0.363508
O  0.269184  0.245184  0.363509
O  0.769041  0.746259  0.363511
O  0.500000  0.750000  0.409091
O  0.000000  0.250000  0.409091
O  0.500000  0.250000  0.409091
O  0.000000  0.750000  0.409091
O  0.250000  0.500000  0.409091
O  0.250000  0.000000  0.409091
O  0.750000  0.000000  0.409091
O  0.750000  0.500000  0.409091
O  0.750000  0.250000  0.454545
O  0.750000  0.750000  0.454545
O  0.250000  0.750000  0.454545
O  0.250000  0.250000  0.454545
O  0.750000  0.000000  0.500000
O  0.750000  0.500000  0.500000
O  0.250000  0.000000  0.500000
O  0.000000  0.750000  0.500000
O  0.250000  0.500000  0.500000
O  0.000000  0.250000  0.500000
O  0.500000  0.750000  0.500000
O  0.500000  0.250000  0.500000
O  0.250000  0.250000  0.545455
O  0.250000  0.750000  0.545455
O  0.750000  0.250000  0.545455
O  0.750000  0.750000  0.545455
O  0.500000  0.250000  0.590909
O  0.500000  0.750000  0.590909
O  0.000000  0.250000  0.590909
O  0.000000  0.750000  0.590909
O  0.250000  0.500000  0.590909
O  0.250000  0.000000  0.590909
O  0.750000  0.500000  0.590909
O  0.750000  0.000000  0.590909
O  0.769041  0.746259  0.636489
O  0.269184  0.245184  0.636491
O  0.231197  0.746673  0.636492
O  0.731049  0.245634  0.636493
O  0.000359  0.295467  0.676837
O  0.500356  0.795704  0.676899
O  0.295742  0.499871  0.681166
O  0.705165  0.499876  0.681227
O  0.795845  0.999872  0.681379
```

```
O  0.205063  0.999865  0.681451
O  0.000370  0.707598  0.685939
O  0.500372  0.208030  0.686032
O  0.712510  0.750325  0.726706
O  0.288232  0.750056  0.726708
O  0.212476  0.251319  0.726709
O  0.788262  0.251072  0.726709
O  0.500358  0.258476  0.764145
O  0.000361  0.758519  0.764147
O  0.258464  0.000245  0.766996
O  0.742230  0.000248  0.767032
O  0.758533  0.500245  0.767126
O  0.242155  0.500237  0.767163
O  0.000355  0.239268  0.772452
O  0.500361  0.739480  0.772453
O  0.360578  0.748904  0.813994
O  0.860557  0.248651  0.813995
O  0.640150  0.748848  0.813995
O  0.140152  0.248586  0.813995
```

Structure 28. TiO$_2$-terminated BaTiO$_3$ slab at $\Theta$=0.00 CO$_2$ coverage

_cell_length_a  8.07335600
_cell_length_b  8.07335600
_cell_length_c  44.40346100
_cell_angle_alpha  90.00000000
_cell_angle_beta  90.00000000
_cell_angle_gamma  90.00000000
_symmetry_space_group_name_H-M  'P 1'
loop_
_atom_site_type_symbol
_atom_site_fract_x
_atom_site_fract_y
_atom_site_fract_z
Ba 0.000000 0.000000 0.545455
Ba 0.997165 0.998236 0.636368
Ba 0.994941 0.001359 0.729666
Ba 0.000000 0.500000 0.545455
Ba 0.997167 0.498239 0.636368
Ba 0.994960 0.501368 0.729662
Ba 0.500000 0.000000 0.545455
Ba 0.497175 0.998235 0.636368
Ba 0.494956 0.001368 0.729669
Ba 0.500000 0.500000 0.545455
Ba 0.497165 0.498241 0.636368
Ba 0.494938 0.501358 0.729665
Ba 0.000000 0.000000 0.454545
Ba 0.997165 0.998236 0.363632
Ba 0.994941 0.001359 0.270334
Ba 0.000000 0.500000 0.454545
Ba 0.997167 0.498239 0.363632
Ba 0.994960 0.501368 0.270338
Ba 0.500000 0.000000 0.454545
Ba 0.497175 0.998235 0.363632
Ba 0.494956 0.001368 0.270331
Ba 0.500000 0.500000 0.454545
Ba 0.497165 0.498241 0.363632
Ba 0.494938 0.501358 0.270335
Ti 0.250000 0.250000 0.500000
Ti 0.250000 0.250000 0.590909
Ti 0.236917 0.239160 0.681460
Ti 0.236942 0.262199 0.770330
Ti 0.250000 0.750000 0.500000
Ti 0.250000 0.750000 0.590909
Ti 0.236914 0.739173 0.681461
Ti 0.236951 0.762208 0.770329
Ti 0.750000 0.250000 0.500000

```
Ti  0.750000  0.250000  0.590909
Ti  0.736911  0.239170  0.681461
Ti  0.736971  0.262223  0.770329
Ti  0.750000  0.750000  0.500000
Ti  0.750000  0.750000  0.590909
Ti  0.736910  0.739166  0.681462
Ti  0.736948  0.762189  0.770329
Ti  0.250000  0.250000  0.409091
Ti  0.236917  0.239160  0.318540
Ti  0.236942  0.262199  0.229670
Ti  0.250000  0.750000  0.409091
Ti  0.236914  0.739173  0.318539
Ti  0.236951  0.762208  0.229671
Ti  0.750000  0.250000  0.409091
Ti  0.736911  0.239170  0.318539
Ti  0.736971  0.262223  0.229671
Ti  0.750000  0.750000  0.409091
Ti  0.736910  0.739166  0.318538
Ti  0.736948  0.762189  0.229672
O   0.250000  0.000000  0.500000
O   0.250000  0.000000  0.590909
O   0.252691  0.009025  0.681804
O   0.255722  0.986999  0.773334
O   0.250000  0.500000  0.500000
O   0.250000  0.500000  0.590909
O   0.252705  0.509035  0.681805
O   0.255739  0.486985  0.773333
O   0.750000  0.000000  0.500000
O   0.750000  0.000000  0.590909
O   0.752704  0.009033  0.681805
O   0.755745  0.986973  0.773332
O   0.750000  0.500000  0.500000
O   0.750000  0.500000  0.590909
O   0.752686  0.509027  0.681803
O   0.755739  0.487013  0.773332
O   0.250000  0.250000  0.545455
O   0.250817  0.251785  0.636291
O   0.251713  0.250809  0.727561
O   0.250000  0.750000  0.545455
O   0.250814  0.751789  0.636291
O   0.251712  0.750822  0.727559
O   0.750000  0.250000  0.545455
O   0.750810  0.251786  0.636291
O   0.751732  0.250828  0.727559
O   0.750000  0.750000  0.545455
O   0.750811  0.751788  0.636291
```

| | | | |
|---|---|---|---|
| O | 0.751722 | 0.750806 | 0.727558 |
| O | 0.000000 | 0.250000 | 0.500000 |
| O | 0.000000 | 0.250000 | 0.590909 |
| O | 0.007464 | 0.254363 | 0.681844 |
| O | 0.012342 | 0.243241 | 0.773136 |
| O | 0.000000 | 0.750000 | 0.500000 |
| O | 0.000000 | 0.750000 | 0.590909 |
| O | 0.007456 | 0.754351 | 0.681845 |
| O | 0.012359 | 0.743217 | 0.773135 |
| O | 0.500000 | 0.250000 | 0.500000 |
| O | 0.500000 | 0.250000 | 0.590909 |
| O | 0.507455 | 0.254349 | 0.681846 |
| O | 0.512379 | 0.243234 | 0.773135 |
| O | 0.500000 | 0.750000 | 0.500000 |
| O | 0.500000 | 0.750000 | 0.590909 |
| O | 0.507459 | 0.754367 | 0.681846 |
| O | 0.512346 | 0.743233 | 0.773132 |
| O | 0.250000 | 0.000000 | 0.409091 |
| O | 0.252691 | 0.009025 | 0.318196 |
| O | 0.255722 | 0.986999 | 0.226666 |
| O | 0.250000 | 0.500000 | 0.409091 |
| O | 0.252705 | 0.509035 | 0.318195 |
| O | 0.255739 | 0.486985 | 0.226667 |
| O | 0.750000 | 0.000000 | 0.409091 |
| O | 0.752704 | 0.009033 | 0.318195 |
| O | 0.755745 | 0.986973 | 0.226668 |
| O | 0.750000 | 0.500000 | 0.409091 |
| O | 0.752686 | 0.509027 | 0.318197 |
| O | 0.755739 | 0.487013 | 0.226668 |
| O | 0.250000 | 0.250000 | 0.454545 |
| O | 0.250817 | 0.251785 | 0.363709 |
| O | 0.251713 | 0.250809 | 0.272439 |
| O | 0.250000 | 0.750000 | 0.454545 |
| O | 0.250814 | 0.751789 | 0.363709 |
| O | 0.251712 | 0.750822 | 0.272441 |
| O | 0.750000 | 0.250000 | 0.454545 |
| O | 0.750810 | 0.251786 | 0.363709 |
| O | 0.751732 | 0.250828 | 0.272441 |
| O | 0.750000 | 0.750000 | 0.454545 |
| O | 0.750811 | 0.751788 | 0.363709 |
| O | 0.751722 | 0.750806 | 0.272442 |
| O | 0.000000 | 0.250000 | 0.409091 |
| O | 0.007464 | 0.254363 | 0.318156 |
| O | 0.012342 | 0.243241 | 0.226864 |
| O | 0.000000 | 0.750000 | 0.409091 |
| O | 0.007456 | 0.754351 | 0.318155 |

O  0.012359  0.743217  0.226865
O  0.500000  0.250000  0.409091
O  0.507455  0.254349  0.318154
O  0.512379  0.243234  0.226865
O  0.500000  0.750000  0.409091
O  0.507459  0.754367  0.318154
O  0.512346  0.743233  0.226868

Structure 29. TiO$_2$-terminated BaTiO$_3$ slab at $\Theta$=0.25 CO$_2$ coverage

```
_cell_length_a   8.07335600
_cell_length_b   8.07335600
_cell_length_c   44.40346100
_cell_angle_alpha   90.00000000
_cell_angle_beta   90.00000000
_cell_angle_gamma   90.00000000
_symmetry_space_group_name_H-M   'P 1'
loop_
_atom_site_type_symbol
_atom_site_fract_x
_atom_site_fract_y
_atom_site_fract_z
 Ba  0.497163  0.002704  0.729169
 Ba  0.495664  0.503401  0.728809
 Ba  0.997825  0.007906  0.727807
 Ba  0.995788  0.502658  0.728496
 Ba  0.500000  0.000000  0.545455
 Ba  0.497863  0.000195  0.636286
 Ba  0.500000  0.500000  0.545455
 Ba  0.497864  0.504184  0.636187
 Ba  0.000000  0.000000  0.545455
 Ba  0.997656  0.000438  0.636223
 Ba  0.000000  0.500000  0.545455
 Ba  0.997474  0.503860  0.636212
 Ba  0.497163  0.002704  0.270831
 Ba  0.495664  0.503401  0.271191
 Ba  0.997825  0.007906  0.272193
 Ba  0.995788  0.502658  0.271504
 Ba  0.500000  0.000000  0.454545
 Ba  0.497863  0.000195  0.363714
 Ba  0.500000  0.500000  0.454545
 Ba  0.497864  0.504184  0.363813
 Ba  0.000000  0.000000  0.454545
 Ba  0.997656  0.000438  0.363777
 Ba  0.000000  0.500000  0.454545
 Ba  0.997474  0.503860  0.363788
 Ti  0.771373  0.767132  0.773054
 Ti  0.737204  0.264103  0.769702
 Ti  0.230107  0.766175  0.772882
 Ti  0.235908  0.264215  0.769700
 Ti  0.750000  0.750000  0.500000
 Ti  0.750000  0.750000  0.590909
 Ti  0.738426  0.761620  0.682537
 Ti  0.750000  0.250000  0.500000
 Ti  0.750000  0.250000  0.590909
```

```
Ti  0.737157  0.261797  0.681015
Ti  0.250000  0.750000  0.500000
Ti  0.250000  0.750000  0.590909
Ti  0.238194  0.761489  0.682361
Ti  0.250000  0.250000  0.500000
Ti  0.250000  0.250000  0.590909
Ti  0.238869  0.262136  0.680941
Ti  0.771373  0.767132  0.226946
Ti  0.737204  0.264103  0.230298
Ti  0.230107  0.766175  0.227118
Ti  0.235908  0.264215  0.230300
Ti  0.750000  0.750000  0.409091
Ti  0.738426  0.761620  0.317463
Ti  0.750000  0.250000  0.409091
Ti  0.737157  0.261797  0.318985
Ti  0.250000  0.750000  0.409091
Ti  0.238194  0.761489  0.317639
Ti  0.250000  0.250000  0.409091
Ti  0.238869  0.262136  0.319059
C  0.500920  0.739174  0.801986
C  0.500920  0.739174  0.198014
O  0.759070  0.987592  0.772517
O  0.753099  0.486061  0.771521
O  0.247239  0.986697  0.772403
O  0.251750  0.486413  0.771554
O  0.754424  0.748247  0.728091
O  0.751028  0.245607  0.726968
O  0.247805  0.747439  0.728123
O  0.253035  0.245806  0.727026
O  0.501544  0.743750  0.771511
O  0.511372  0.241136  0.772699
O  0.000917  0.741893  0.774783
O  0.010568  0.243604  0.772247
O  0.357728  0.738631  0.813904
O  0.644044  0.739571  0.813996
O  0.750000  0.000000  0.500000
O  0.750000  0.000000  0.590909
O  0.752598  0.991945  0.681815
O  0.750000  0.500000  0.500000
O  0.750000  0.500000  0.590909
O  0.752562  0.491931  0.681774
O  0.250000  0.000000  0.500000
O  0.250000  0.000000  0.590909
O  0.253501  0.991639  0.681876
O  0.250000  0.500000  0.500000
O  0.250000  0.500000  0.590909
```

O   0.253230   0.492158   0.681733
O   0.750000   0.750000   0.545455
O   0.750616   0.748715   0.636241
O   0.750000   0.250000   0.545455
O   0.751092   0.248601   0.636294
O   0.250000   0.750000   0.545455
O   0.251485   0.748726   0.636241
O   0.250000   0.250000   0.545455
O   0.251349   0.248605   0.636289
O   0.500000   0.750000   0.500000
O   0.500000   0.750000   0.590909
O   0.507712   0.746432   0.682439
O   0.500000   0.250000   0.500000
O   0.500000   0.250000   0.590909
O   0.507005   0.246649   0.681630
O   0.000000   0.750000   0.500000
O   0.000000   0.750000   0.590909
O   0.007657   0.746734   0.681820
O   0.000000   0.250000   0.500000
O   0.000000   0.250000   0.590909
O   0.007939   0.246977   0.681652
O   0.759070   0.987592   0.227483
O   0.753099   0.486061   0.228479
O   0.247239   0.986697   0.227597
O   0.251750   0.486413   0.228446
O   0.754424   0.748247   0.271909
O   0.751028   0.245607   0.273032
O   0.247805   0.747439   0.271877
O   0.253035   0.245806   0.272974
O   0.501544   0.743750   0.228489
O   0.511372   0.241136   0.227301
O   0.000917   0.741893   0.225217
O   0.010568   0.243604   0.227753
O   0.357728   0.738631   0.186096
O   0.644044   0.739571   0.186004
O   0.750000   0.000000   0.409091
O   0.752598   0.991945   0.318185
O   0.750000   0.500000   0.409091
O   0.752562   0.491931   0.318226
O   0.250000   0.000000   0.409091
O   0.253501   0.991639   0.318124
O   0.250000   0.500000   0.409091
O   0.253230   0.492158   0.318267
O   0.750000   0.750000   0.454545
O   0.750616   0.748715   0.363759
O   0.750000   0.250000   0.454545
O   0.750000   0.250000   0.454545

```
O  0.751092  0.248601  0.363706
O  0.250000  0.750000  0.454545
O  0.251485  0.748726  0.363759
O  0.250000  0.250000  0.454545
O  0.251349  0.248605  0.363711
O  0.500000  0.750000  0.409091
O  0.507712  0.746432  0.317561
O  0.500000  0.250000  0.409091
O  0.507005  0.246649  0.318370
O  0.000000  0.750000  0.409091
O  0.007657  0.746734  0.318180
O  0.000000  0.250000  0.409091
O  0.007939  0.246977  0.318348
```

Structure 30. TiO$_2$-terminated BaTiO$_3$ slab at $\Theta$=0.50 CO$_2$ coverage

_cell_length_a  8.07335600
_cell_length_b  8.07335600
_cell_length_c  44.40346100
_cell_angle_alpha  90.00000000
_cell_angle_beta  90.00000000
_cell_angle_gamma  90.00000000
_symmetry_space_group_name_H-M  'P 1'
loop_
_atom_site_type_symbol
_atom_site_fract_x
_atom_site_fract_y
_atom_site_fract_z
Ba  0.497665  0.498815  0.271799
Ba  0.997520  0.998831  0.271802
Ba  0.997719  0.492763  0.272825
Ba  0.497760  0.992749  0.272829
Ba  0.497776  0.498482  0.363814
Ba  0.997785  0.998473  0.363815
Ba  0.497896  0.997971  0.363908
Ba  0.997905  0.497983  0.363909
Ba  0.000000  0.000000  0.454545
Ba  0.000000  0.500000  0.454545
Ba  0.500000  0.000000  0.454545
Ba  0.500000  0.500000  0.454545
Ba  0.000000  0.000000  0.545455
Ba  0.000000  0.500000  0.545455
Ba  0.500000  0.000000  0.545455
Ba  0.500000  0.500000  0.545455
Ba  0.997905  0.497983  0.636091
Ba  0.497896  0.997971  0.636092
Ba  0.997785  0.998473  0.636185
Ba  0.497776  0.498482  0.636186
Ba  0.497760  0.992749  0.727171
Ba  0.997719  0.492763  0.727175
Ba  0.997520  0.998831  0.728198
Ba  0.497665  0.498815  0.728201
Ti  0.769669  0.731147  0.227489
Ti  0.269720  0.231377  0.227518
Ti  0.729111  0.231142  0.227546
Ti  0.229062  0.731413  0.227574
Ti  0.239804  0.238894  0.318017
Ti  0.739816  0.738832  0.318017
Ti  0.237896  0.739196  0.318066
Ti  0.737910  0.239147  0.318071
Ti  0.250000  0.250000  0.409091

Ti 0.250000 0.750000 0.409091
Ti 0.750000 0.250000 0.409091
Ti 0.750000 0.750000 0.409091
Ti 0.250000 0.250000 0.500000
Ti 0.250000 0.750000 0.500000
Ti 0.750000 0.250000 0.500000
Ti 0.750000 0.750000 0.500000
Ti 0.250000 0.250000 0.590909
Ti 0.250000 0.750000 0.590909
Ti 0.750000 0.250000 0.590909
Ti 0.750000 0.750000 0.590909
Ti 0.737910 0.239147 0.681929
Ti 0.237896 0.739196 0.681934
Ti 0.239804 0.238894 0.681983
Ti 0.739816 0.738832 0.681983
Ti 0.229062 0.731413 0.772426
Ti 0.729111 0.231142 0.772453
Ti 0.269720 0.231377 0.772482
Ti 0.769669 0.731147 0.772511
C 0.499345 0.759053 0.198630
C 0.999470 0.259034 0.198631
C 0.999470 0.259034 0.801369
C 0.499345 0.759053 0.801370
O 0.642493 0.758456 0.186729
O 0.142697 0.258933 0.186758
O 0.856254 0.258422 0.186759
O 0.356051 0.758987 0.186787
O 0.499393 0.259546 0.225602
O 0.999356 0.759579 0.225610
O 0.754633 0.512841 0.228922
O 0.743782 0.012706 0.228942
O 0.254525 0.013053 0.228968
O 0.243869 0.512964 0.228982
O 0.499512 0.755529 0.229197
O 0.999456 0.255554 0.229198
O 0.746789 0.253783 0.272314
O 0.246774 0.754031 0.272336
O 0.754823 0.753660 0.272401
O 0.254866 0.253914 0.272416
O 0.007594 0.253507 0.317780
O 0.507614 0.753483 0.317783
O 0.252955 0.508597 0.318183
O 0.752952 0.008517 0.318183
O 0.752659 0.508008 0.318220
O 0.252653 0.008098 0.318221
O 0.506804 0.253660 0.318291

```
O  0.006767  0.753673  0.318292
O  0.250861  0.251587  0.363687
O  0.750860  0.751528  0.363687
O  0.751233  0.251572  0.363701
O  0.251226  0.751627  0.363705
O  0.250000  0.500000  0.409091
O  0.750000  0.000000  0.409091
O  0.250000  0.000000  0.409091
O  0.750000  0.500000  0.409091
O  0.000000  0.250000  0.409091
O  0.000000  0.750000  0.409091
O  0.500000  0.750000  0.409091
O  0.500000  0.250000  0.409091
O  0.750000  0.250000  0.454545
O  0.750000  0.750000  0.454545
O  0.250000  0.750000  0.454545
O  0.250000  0.250000  0.454545
O  0.250000  0.000000  0.500000
O  0.250000  0.500000  0.500000
O  0.750000  0.000000  0.500000
O  0.750000  0.500000  0.500000
O  0.000000  0.250000  0.500000
O  0.000000  0.750000  0.500000
O  0.500000  0.250000  0.500000
O  0.500000  0.750000  0.500000
O  0.250000  0.250000  0.545455
O  0.250000  0.750000  0.545455
O  0.750000  0.250000  0.545455
O  0.750000  0.750000  0.545455
O  0.250000  0.000000  0.590909
O  0.250000  0.500000  0.590909
O  0.750000  0.000000  0.590909
O  0.750000  0.500000  0.590909
O  0.000000  0.250000  0.590909
O  0.000000  0.750000  0.590909
O  0.500000  0.250000  0.590909
O  0.500000  0.750000  0.590909
O  0.251226  0.751627  0.636294
O  0.751233  0.251572  0.636299
O  0.250861  0.251587  0.636313
O  0.750860  0.751528  0.636313
O  0.006767  0.753673  0.681708
O  0.506804  0.253660  0.681709
O  0.252653  0.008098  0.681779
O  0.752659  0.508008  0.681780
O  0.752952  0.008517  0.681817
```

```
O  0.252955  0.508597  0.681817
O  0.507614  0.753483  0.682217
O  0.007594  0.253507  0.682220
O  0.254866  0.253914  0.727584
O  0.754823  0.753660  0.727599
O  0.246774  0.754031  0.727664
O  0.746789  0.253783  0.727686
O  0.999456  0.255554  0.770802
O  0.499512  0.755529  0.770803
O  0.243869  0.512964  0.771018
O  0.254525  0.013053  0.771032
O  0.743782  0.012706  0.771058
O  0.754633  0.512841  0.771078
O  0.999356  0.759579  0.774390
O  0.499393  0.259546  0.774398
O  0.356051  0.758987  0.813213
O  0.856254  0.258422  0.813241
O  0.142697  0.258933  0.813242
O  0.642493  0.758456  0.813271
```

Structure 31. ZrO$_2$-terminated BaZrO$_3$ slab at $\Theta$=0.00 CO$_2$ coverage

_cell_length_a  8.51149400
_cell_length_b  8.51149400
_cell_length_c  46.81321300
_cell_angle_alpha  90.00000000
_cell_angle_beta  90.00000000
_cell_angle_gamma  90.00000000
_symmetry_space_group_name_H-M  'P 1'
loop_
 _atom_site_type_symbol
 _atom_site_fract_x
 _atom_site_fract_y
 _atom_site_fract_z
 Ba 0.000000 0.000000 0.545455
 Ba 0.999663 0.000202 0.636864
 Ba 0.998984 0.000414 0.729915
 Ba 0.000000 0.500000 0.545455
 Ba 0.999626 0.500243 0.636882
 Ba 0.998973 0.500429 0.729914
 Ba 0.500000 0.000000 0.545455
 Ba 0.499629 0.000186 0.636870
 Ba 0.498952 0.000403 0.729912
 Ba 0.500000 0.500000 0.545455
 Ba 0.499671 0.500141 0.636869
 Ba 0.498984 0.500414 0.729906
 Ba 0.000000 0.000000 0.454545
 Ba 0.999663 0.000202 0.363136
 Ba 0.998984 0.000414 0.270085
 Ba 0.000000 0.500000 0.454545
 Ba 0.999626 0.500243 0.363118
 Ba 0.998973 0.500429 0.270086
 Ba 0.500000 0.000000 0.454545
 Ba 0.499629 0.000186 0.363130
 Ba 0.498952 0.000403 0.270088
 Ba 0.500000 0.500000 0.454545
 Ba 0.499671 0.500141 0.363131
 Ba 0.498984 0.500414 0.270094
 Zr 0.250000 0.250000 0.500000
 Zr 0.250000 0.250000 0.590909
 Zr 0.249369 0.250299 0.681444
 Zr 0.248961 0.250433 0.770418
 Zr 0.250000 0.750000 0.500000
 Zr 0.250000 0.750000 0.590909
 Zr 0.249353 0.750319 0.681445
 Zr 0.248938 0.750409 0.770424
 Zr 0.750000 0.250000 0.500000

Zr 0.750000 0.250000 0.590909
Zr 0.749359 0.250281 0.681444
Zr 0.748946 0.250408 0.770417
Zr 0.750000 0.750000 0.500000
Zr 0.750000 0.750000 0.590909
Zr 0.749344 0.750298 0.681445
Zr 0.748918 0.750408 0.770423
Zr 0.250000 0.250000 0.409091
Zr 0.249369 0.250299 0.318556
Zr 0.248961 0.250433 0.229582
Zr 0.250000 0.750000 0.409091
Zr 0.249353 0.750319 0.318555
Zr 0.248938 0.750409 0.229576
Zr 0.750000 0.250000 0.409091
Zr 0.749359 0.250281 0.318556
Zr 0.748946 0.250408 0.229583
Zr 0.750000 0.750000 0.409091
Zr 0.749344 0.750298 0.318555
Zr 0.748918 0.750408 0.229577
O 0.250000 0.000000 0.500000
O 0.250000 0.000000 0.590909
O 0.245272 0.000308 0.681701
O 0.249391 0.000446 0.771309
O 0.250000 0.500000 0.500000
O 0.250000 0.500000 0.590909
O 0.253321 0.500310 0.681674
O 0.248391 0.500443 0.771322
O 0.750000 0.000000 0.500000
O 0.750000 0.000000 0.590909
O 0.753320 0.000288 0.681667
O 0.748419 0.000422 0.771309
O 0.750000 0.500000 0.500000
O 0.750000 0.500000 0.590909
O 0.745252 0.500292 0.681703
O 0.749405 0.500416 0.771318
O 0.250000 0.250000 0.545455
O 0.249679 0.250093 0.636497
O 0.249159 0.250432 0.727639
O 0.250000 0.750000 0.545455
O 0.249557 0.750245 0.636498
O 0.249131 0.750345 0.727644
O 0.750000 0.250000 0.545455
O 0.749576 0.250233 0.636497
O 0.749079 0.250339 0.727638
O 0.750000 0.750000 0.545455
O 0.749663 0.750064 0.636498

```
O  0.749024  0.750369  0.727642
O  0.000000  0.250000  0.500000
O  0.000000  0.250000  0.590909
O  0.999337  0.254333  0.681691
O  0.998941  0.249953  0.771339
O  0.000000  0.750000  0.500000
O  0.000000  0.750000  0.590909
O  0.999320  0.746306  0.681706
O  0.998914  0.750939  0.771360
O  0.500000  0.250000  0.500000
O  0.500000  0.250000  0.590909
O  0.499337  0.246261  0.681701
O  0.498940  0.250943  0.771359
O  0.500000  0.750000  0.500000
O  0.500000  0.750000  0.590909
O  0.499320  0.754376  0.681690
O  0.498913  0.749954  0.771374
O  0.250000  0.000000  0.409091
O  0.245272  0.000308  0.318299
O  0.249391  0.000446  0.228691
O  0.250000  0.500000  0.409091
O  0.253321  0.500310  0.318326
O  0.248391  0.500443  0.228678
O  0.750000  0.000000  0.409091
O  0.753320  0.000288  0.318333
O  0.748419  0.000422  0.228691
O  0.750000  0.500000  0.409091
O  0.745252  0.500292  0.318297
O  0.749405  0.500416  0.228682
O  0.250000  0.250000  0.454545
O  0.249679  0.250093  0.363503
O  0.249159  0.250432  0.272361
O  0.250000  0.750000  0.454545
O  0.249557  0.750245  0.363502
O  0.249131  0.750345  0.272356
O  0.750000  0.250000  0.454545
O  0.749576  0.250233  0.363503
O  0.749079  0.250339  0.272362
O  0.750000  0.750000  0.454545
O  0.749663  0.750064  0.363502
O  0.749024  0.750369  0.272358
O  0.000000  0.250000  0.409091
O  0.999337  0.254333  0.318309
O  0.998941  0.249953  0.228661
O  0.000000  0.750000  0.409091
O  0.999320  0.746306  0.318294
```

```
O  0.998914  0.750939  0.228640
O  0.500000  0.250000  0.409091
O  0.499337  0.246261  0.318299
O  0.498940  0.250943  0.228641
O  0.500000  0.750000  0.409091
O  0.499320  0.754376  0.318310
O  0.498913  0.749954  0.228626
```

Structure 32. ZrO$_2$-terminated BaZrO$_3$ slab at $\Theta$=0.25 CO$_2$ coverage

```
_cell_length_a  8.51149400
_cell_length_b  8.51149400
_cell_length_c  46.81321300
_cell_angle_alpha  90.00000000
_cell_angle_beta  90.00000000
_cell_angle_gamma  90.00000000
_symmetry_space_group_name_H-M  'P 1'
loop_
_atom_site_type_symbol
_atom_site_fract_x
_atom_site_fract_y
_atom_site_fract_z
 Ba  0.499669  0.500156  0.270880
 Ba  0.499662  0.000838  0.270937
 Ba  0.999667  0.495692  0.271865
 Ba  0.999669  0.005441  0.271896
 Ba  0.499866  0.502364  0.363300
 Ba  0.499865  0.998257  0.363316
 Ba  0.999866  0.502414  0.363320
 Ba  0.999868  0.998066  0.363352
 Ba  0.000000  0.000000  0.454545
 Ba  0.000000  0.500000  0.454545
 Ba  0.500000  0.000000  0.454545
 Ba  0.500000  0.500000  0.454545
 Ba  0.000000  0.000000  0.545455
 Ba  0.000000  0.500000  0.545455
 Ba  0.500000  0.000000  0.545455
 Ba  0.500000  0.500000  0.545455
 Ba  0.999868  0.998066  0.636648
 Ba  0.999866  0.502414  0.636680
 Ba  0.499865  0.998257  0.636684
 Ba  0.499866  0.502364  0.636700
 Ba  0.999669  0.005441  0.728104
 Ba  0.999667  0.495692  0.728135
 Ba  0.499662  0.000838  0.729063
 Ba  0.499669  0.500156  0.729120
 Zr  0.765626  0.750552  0.227934
 Zr  0.233644  0.750541  0.227935
 Zr  0.249096  0.250484  0.230141
 Zr  0.750212  0.250475  0.230142
 Zr  0.250510  0.750394  0.317802
 Zr  0.749002  0.750397  0.317802
 Zr  0.250229  0.250393  0.318969
 Zr  0.749284  0.250391  0.318969
 Zr  0.250000  0.250000  0.409091
```

```
Zr 0.250000 0.750000 0.409091
Zr 0.750000 0.250000 0.409091
Zr 0.750000 0.750000 0.409091
Zr 0.250000 0.250000 0.500000
Zr 0.250000 0.750000 0.500000
Zr 0.750000 0.250000 0.500000
Zr 0.750000 0.750000 0.500000
Zr 0.250000 0.250000 0.590909
Zr 0.250000 0.750000 0.590909
Zr 0.750000 0.250000 0.590909
Zr 0.750000 0.750000 0.590909
Zr 0.250229 0.250393 0.681031
Zr 0.749284 0.250391 0.681031
Zr 0.250510 0.750394 0.682198
Zr 0.749002 0.750397 0.682198
Zr 0.750212 0.250475 0.769858
Zr 0.249096 0.250484 0.769859
Zr 0.233644 0.750541 0.772065
Zr 0.765626 0.750552 0.772066
C 0.499617 0.752565 0.197733
C 0.499617 0.752565 0.802267
O 0.634582 0.753110 0.186076
O 0.364641 0.753150 0.186079
O 0.499624 0.751049 0.226973
O 0.999644 0.250503 0.229170
O 0.499646 0.251064 0.229465
O 0.999631 0.751460 0.229587
O 0.247596 0.501730 0.230566
O 0.751661 0.501733 0.230582
O 0.751145 0.000144 0.230782
O 0.248123 0.000146 0.230792
O 0.262095 0.750239 0.271949
O 0.737295 0.750315 0.271950
O 0.247852 0.251154 0.272947
O 0.751564 0.251081 0.272948
O 0.999751 0.746362 0.316811
O 0.499751 0.246064 0.318046
O 0.999752 0.254467 0.318343
O 0.245097 0.000516 0.318457
O 0.754352 0.000517 0.318470
O 0.746042 0.500268 0.318543
O 0.253411 0.500268 0.318556
O 0.499752 0.754622 0.319188
O 0.752625 0.750246 0.363404
O 0.247114 0.750289 0.363405
O 0.250571 0.250075 0.363659
```

```
O  0.749169  0.250122  0.363659
O  0.250000  0.000000  0.409091
O  0.250000  0.500000  0.409091
O  0.750000  0.000000  0.409091
O  0.750000  0.500000  0.409091
O  0.000000  0.250000  0.409091
O  0.000000  0.750000  0.409091
O  0.500000  0.250000  0.409091
O  0.500000  0.750000  0.409091
O  0.250000  0.250000  0.454545
O  0.750000  0.250000  0.454545
O  0.250000  0.750000  0.454545
O  0.750000  0.750000  0.454545
O  0.250000  0.000000  0.500000
O  0.250000  0.500000  0.500000
O  0.750000  0.000000  0.500000
O  0.750000  0.500000  0.500000
O  0.000000  0.250000  0.500000
O  0.000000  0.750000  0.500000
O  0.500000  0.250000  0.500000
O  0.500000  0.750000  0.500000
O  0.250000  0.250000  0.545455
O  0.250000  0.750000  0.545455
O  0.750000  0.250000  0.545455
O  0.750000  0.750000  0.545455
O  0.250000  0.000000  0.590909
O  0.250000  0.500000  0.590909
O  0.750000  0.000000  0.590909
O  0.750000  0.500000  0.590909
O  0.000000  0.250000  0.590909
O  0.000000  0.750000  0.590909
O  0.500000  0.250000  0.590909
O  0.500000  0.750000  0.590909
O  0.250571  0.250075  0.636341
O  0.749169  0.250122  0.636341
O  0.247114  0.750289  0.636595
O  0.752625  0.750246  0.636596
O  0.499752  0.754622  0.680812
O  0.253411  0.500268  0.681444
O  0.746042  0.500268  0.681457
O  0.754352  0.000517  0.681530
O  0.245097  0.000516  0.681543
O  0.999752  0.254467  0.681657
O  0.499751  0.246064  0.681954
O  0.999751  0.746362  0.683189
O  0.751564  0.251081  0.727052
```

O 0.247852 0.251154 0.727053
O 0.737295 0.750315 0.728050
O 0.262095 0.750239 0.728051
O 0.248123 0.000146 0.769208
O 0.751145 0.000144 0.769218
O 0.751661 0.501733 0.769418
O 0.247596 0.501730 0.769434
O 0.999631 0.751460 0.770413
O 0.499646 0.251064 0.770535
O 0.999644 0.250503 0.770830
O 0.499624 0.751049 0.773027
O 0.364641 0.753150 0.813921
O 0.634582 0.753110 0.813924

Structure 33. ZrO$_2$-terminated BaZrO$_3$ slab at Θ=0.50 CO$_2$ coverage

```
_cell_length_a  8.51149400
_cell_length_b  8.51149400
_cell_length_c  46.81321300
_cell_angle_alpha  90.00000000
_cell_angle_beta  90.00000000
_cell_angle_gamma  90.00000000
_symmetry_space_group_name_H-M  'P 1'
loop_
 _atom_site_type_symbol
 _atom_site_fract_x
 _atom_site_fract_y
 _atom_site_fract_z
 Ba 0.999642 0.003972 0.272546
 Ba 0.499642 0.503972 0.272546
 Ba 0.999640 0.494615 0.272636
 Ba 0.499641 0.994615 0.272636
 Ba 0.999832 0.000166 0.363531
 Ba 0.999830 0.500052 0.363531
 Ba 0.499829 0.000051 0.363531
 Ba 0.499831 0.500164 0.363531
 Ba 0.000000 0.000000 0.454545
 Ba 0.000000 0.500000 0.454545
 Ba 0.500000 0.000000 0.454545
 Ba 0.500000 0.500000 0.454545
 Ba 0.000000 0.000000 0.545455
 Ba 0.000000 0.500000 0.545455
 Ba 0.500000 0.000000 0.545455
 Ba 0.500000 0.500000 0.545455
 Ba 0.999832 0.000166 0.636469
 Ba 0.999830 0.500052 0.636469
 Ba 0.499829 0.000051 0.636469
 Ba 0.499831 0.500164 0.636469
 Ba 0.999640 0.494615 0.727364
 Ba 0.499641 0.994615 0.727364
 Ba 0.999642 0.003972 0.727454
 Ba 0.499642 0.503972 0.727454
 Zr 0.265291 0.247393 0.228575
 Zr 0.765288 0.747393 0.228575
 Zr 0.233976 0.747390 0.228576
 Zr 0.733978 0.247389 0.228576
 Zr 0.249456 0.249988 0.318278
 Zr 0.749456 0.749987 0.318278
 Zr 0.249931 0.749989 0.318279
 Zr 0.749932 0.249988 0.318279
 Zr 0.250000 0.250000 0.409091
```

```
Zr  0.250000  0.750000  0.409091
Zr  0.750000  0.250000  0.409091
Zr  0.750000  0.750000  0.409091
Zr  0.250000  0.250000  0.500000
Zr  0.250000  0.750000  0.500000
Zr  0.750000  0.250000  0.500000
Zr  0.750000  0.750000  0.500000
Zr  0.250000  0.250000  0.590909
Zr  0.250000  0.750000  0.590909
Zr  0.750000  0.250000  0.590909
Zr  0.750000  0.750000  0.590909
Zr  0.249931  0.749989  0.681721
Zr  0.749932  0.249988  0.681721
Zr  0.249456  0.249988  0.681722
Zr  0.749456  0.749987  0.681722
Zr  0.233976  0.747390  0.771424
Zr  0.733978  0.247389  0.771424
Zr  0.765288  0.747393  0.771425
Zr  0.265291  0.247393  0.771425
C  0.499622  0.751806  0.198637
C  0.999617  0.251800  0.198637
C  0.999617  0.251800  0.801363
C  0.499622  0.751806  0.801363
O  0.134613  0.251797  0.187087
O  0.634609  0.751804  0.187088
O  0.364625  0.751817  0.187088
O  0.864636  0.251810  0.187088
O  0.999626  0.251142  0.227981
O  0.499623  0.751143  0.227982
O  0.499632  0.252795  0.230874
O  0.999632  0.752795  0.230875
O  0.248568  0.501408  0.231903
O  0.748568  0.001407  0.231903
O  0.250709  0.001409  0.231907
O  0.750709  0.501408  0.231907
O  0.235514  0.250639  0.272589
O  0.735512  0.750638  0.272589
O  0.263818  0.750607  0.272590
O  0.763816  0.250607  0.272590
O  0.999675  0.746302  0.316732
O  0.499675  0.246301  0.316733
O  0.245639  0.000133  0.318635
O  0.745639  0.500133  0.318635
O  0.253568  0.500134  0.318643
O  0.753568  0.000133  0.318644
O  0.999677  0.254207  0.319330
```

```
O  0.499677  0.754208  0.319330
O  0.252891  0.250012  0.363571
O  0.752892  0.750010  0.363571
O  0.246776  0.750053  0.363572
O  0.746778  0.250053  0.363572
O  0.250000  0.500000  0.409091
O  0.750000  0.000000  0.409091
O  0.250000  0.000000  0.409091
O  0.750000  0.500000  0.409091
O  0.000000  0.250000  0.409091
O  0.000000  0.750000  0.409091
O  0.500000  0.750000  0.409091
O  0.500000  0.250000  0.409091
O  0.750000  0.250000  0.454545
O  0.750000  0.750000  0.454545
O  0.250000  0.750000  0.454545
O  0.250000  0.250000  0.454545
O  0.250000  0.000000  0.500000
O  0.250000  0.500000  0.500000
O  0.750000  0.000000  0.500000
O  0.750000  0.500000  0.500000
O  0.000000  0.250000  0.500000
O  0.000000  0.750000  0.500000
O  0.500000  0.250000  0.500000
O  0.500000  0.750000  0.500000
O  0.250000  0.250000  0.545455
O  0.250000  0.750000  0.545455
O  0.750000  0.250000  0.545455
O  0.750000  0.750000  0.545455
O  0.250000  0.000000  0.590909
O  0.250000  0.500000  0.590909
O  0.750000  0.000000  0.590909
O  0.750000  0.500000  0.590909
O  0.000000  0.250000  0.590909
O  0.000000  0.750000  0.590909
O  0.500000  0.250000  0.590909
O  0.500000  0.750000  0.590909
O  0.246776  0.750053  0.636428
O  0.746778  0.250053  0.636428
O  0.252891  0.250012  0.636429
O  0.752892  0.750010  0.636429
O  0.999677  0.254207  0.680670
O  0.499677  0.754208  0.680670
O  0.753568  0.000133  0.681356
O  0.253568  0.500134  0.681357
O  0.245639  0.000133  0.681365
```

O 0.745639 0.500133 0.681365
O 0.499675 0.246301 0.683267
O 0.999675 0.746302 0.683268
O 0.263818 0.750607 0.727410
O 0.763816 0.250607 0.727410
O 0.235514 0.250639 0.727411
O 0.735512 0.750638 0.727411
O 0.250709 0.001409 0.768093
O 0.750709 0.501408 0.768093
O 0.748568 0.001407 0.768097
O 0.248568 0.501408 0.768097
O 0.999632 0.752795 0.769125
O 0.499632 0.252795 0.769126
O 0.499623 0.751143 0.772018
O 0.999626 0.251142 0.772019
O 0.634609 0.751804 0.812912
O 0.364625 0.751817 0.812912
O 0.864636 0.251810 0.812912
O 0.134613 0.251797 0.812913

Structure 34. HfO$_2$-terminated BaHfO$_3$ slab at $\Theta$=0.00 CO$_2$ coverage

_cell_length_a  8.41088800
_cell_length_b  8.41088800
_cell_length_c  46.25988800
_cell_angle_alpha  90.00000000
_cell_angle_beta  90.00000000
_cell_angle_gamma  90.00000000
_symmetry_space_group_name_H-M  'P 1'
loop_
 _atom_site_type_symbol
 _atom_site_fract_x
 _atom_site_fract_y
 _atom_site_fract_z
 Ba 0.000000 0.000000 0.545455
 Ba 0.000716 0.000328 0.636749
 Ba 0.001812 0.000894 0.729555
 Ba 0.000000 0.500000 0.545455
 Ba 0.000699 0.500363 0.636737
 Ba 0.001812 0.500865 0.729553
 Ba 0.500000 0.000000 0.545455
 Ba 0.500765 0.000319 0.636752
 Ba 0.501739 0.000887 0.729556
 Ba 0.500000 0.500000 0.545455
 Ba 0.500762 0.500386 0.636741
 Ba 0.501828 0.500825 0.729554
 Ba 0.000000 0.000000 0.454545
 Ba 0.000716 0.000328 0.363251
 Ba 0.001812 0.000894 0.270445
 Ba 0.000000 0.500000 0.454545
 Ba 0.000699 0.500363 0.363263
 Ba 0.001812 0.500865 0.270447
 Ba 0.500000 0.000000 0.454545
 Ba 0.500765 0.000319 0.363248
 Ba 0.501739 0.000887 0.270444
 Ba 0.500000 0.500000 0.454545
 Ba 0.500762 0.500386 0.363259
 Ba 0.501828 0.500825 0.270446
 Hf 0.250000 0.250000 0.500000
 Hf 0.250000 0.250000 0.590909
 Hf 0.251230 0.250593 0.681395
 Hf 0.251854 0.250933 0.770191
 Hf 0.250000 0.750000 0.500000
 Hf 0.250000 0.750000 0.590909
 Hf 0.251243 0.750573 0.681400
 Hf 0.251869 0.750862 0.770190
 Hf 0.750000 0.250000 0.500000

Hf 0.750000 0.250000 0.590909
Hf 0.751234 0.250592 0.681401
Hf 0.751874 0.250852 0.770194
Hf 0.750000 0.750000 0.500000
Hf 0.750000 0.750000 0.590909
Hf 0.751236 0.750606 0.681407
Hf 0.751846 0.750925 0.770193
Hf 0.250000 0.250000 0.409091
Hf 0.251230 0.250593 0.318605
Hf 0.251854 0.250933 0.229809
Hf 0.250000 0.750000 0.409091
Hf 0.251243 0.750573 0.318600
Hf 0.251869 0.750862 0.229810
Hf 0.750000 0.250000 0.409091
Hf 0.751234 0.250592 0.318599
Hf 0.751874 0.250852 0.229806
Hf 0.750000 0.750000 0.409091
Hf 0.751236 0.750606 0.318593
Hf 0.751846 0.750925 0.229807
O 0.250000 0.000000 0.500000
O 0.250000 0.000000 0.590909
O 0.252034 0.000568 0.681729
O 0.251754 0.000896 0.771535
O 0.250000 0.500000 0.500000
O 0.250000 0.500000 0.590909
O 0.250369 0.500572 0.681676
O 0.251989 0.500893 0.771593
O 0.750000 0.000000 0.500000
O 0.750000 0.000000 0.590909
O 0.750362 0.000583 0.681671
O 0.751974 0.000897 0.771610
O 0.750000 0.500000 0.500000
O 0.750000 0.500000 0.590909
O 0.752014 0.500585 0.681738
O 0.751749 0.500893 0.771532
O 0.250000 0.250000 0.545455
O 0.250662 0.250217 0.636432
O 0.251476 0.250880 0.727460
O 0.250000 0.750000 0.545455
O 0.250643 0.750407 0.636433
O 0.251597 0.750564 0.727458
O 0.750000 0.250000 0.545455
O 0.750650 0.250431 0.636434
O 0.751593 0.250530 0.727463
O 0.750000 0.750000 0.545455
O 0.750682 0.750203 0.636435

```
O  0.751489  0.750958  0.727462
O  0.000000  0.250000  0.500000
O  0.000000  0.250000  0.590909
O  0.001203  0.249739  0.681696
O  0.001869  0.251010  0.771583
O  0.000000  0.750000  0.500000
O  0.000000  0.750000  0.590909
O  0.001211  0.751384  0.681714
O  0.001866  0.750774  0.771560
O  0.500000  0.250000  0.500000
O  0.500000  0.250000  0.590909
O  0.501208  0.251409  0.681706
O  0.501865  0.250785  0.771561
O  0.500000  0.750000  0.500000
O  0.500000  0.750000  0.590909
O  0.501214  0.749738  0.681693
O  0.501863  0.751007  0.771574
O  0.250000  0.000000  0.409091
O  0.252034  0.000568  0.318271
O  0.251754  0.000896  0.228465
O  0.250000  0.500000  0.409091
O  0.250369  0.500572  0.318324
O  0.251989  0.500893  0.228407
O  0.750000  0.000000  0.409091
O  0.750362  0.000583  0.318329
O  0.751974  0.000897  0.228390
O  0.750000  0.500000  0.409091
O  0.752014  0.500585  0.318262
O  0.751749  0.500893  0.228468
O  0.250000  0.250000  0.454545
O  0.250662  0.250217  0.363568
O  0.251476  0.250880  0.272540
O  0.250000  0.750000  0.454545
O  0.250643  0.750407  0.363567
O  0.251597  0.750564  0.272542
O  0.750000  0.250000  0.454545
O  0.750650  0.250431  0.363566
O  0.751593  0.250530  0.272537
O  0.750000  0.750000  0.454545
O  0.750682  0.750203  0.363565
O  0.751489  0.750958  0.272538
O  0.000000  0.250000  0.409091
O  0.001203  0.249739  0.318303
O  0.001869  0.251010  0.228417
O  0.000000  0.750000  0.409091
O  0.001211  0.751384  0.318286
```

```
O  0.001866  0.750774  0.228440
O  0.500000  0.250000  0.409091
O  0.501208  0.251409  0.318294
O  0.501865  0.250785  0.228439
O  0.500000  0.750000  0.409091
O  0.501214  0.749738  0.318307
O  0.501863  0.751007  0.228426
```

Structure 35. HfO$_2$-terminated BaHfO$_3$ slab at $\Theta$=0.25 CO$_2$ coverage

_cell_length_a  8.41088800
_cell_length_b  8.41088800
_cell_length_c  46.25988800
_cell_angle_alpha  90.00000000
_cell_angle_beta  90.00000000
_cell_angle_gamma  90.00000000
_symmetry_space_group_name_H-M  'P 1'
loop_
 _atom_site_type_symbol
 _atom_site_fract_x
 _atom_site_fract_y
 _atom_site_fract_z
 Ba  0.500890  0.500432  0.271100
 Ba  0.500888  0.001191  0.271129
 Ba  0.000891  0.497214  0.271883
 Ba  0.000890  0.004366  0.271895
 Ba  0.500348  0.502010  0.363383
 Ba  0.000345  0.501977  0.363386
 Ba  0.000346  0.998611  0.363387
 Ba  0.500346  0.998584  0.363387
 Ba  0.000000  0.000000  0.454545
 Ba  0.000000  0.500000  0.454545
 Ba  0.500000  0.000000  0.454545
 Ba  0.500000  0.500000  0.454545
 Ba  0.000000  0.000000  0.545455
 Ba  0.000000  0.500000  0.545455
 Ba  0.500000  0.000000  0.545455
 Ba  0.500000  0.500000  0.545455
 Ba  0.000346  0.998611  0.636613
 Ba  0.500346  0.998584  0.636613
 Ba  0.000345  0.501977  0.636614
 Ba  0.500348  0.502010  0.636617
 Ba  0.000890  0.004366  0.728105
 Ba  0.000891  0.497214  0.728117
 Ba  0.500888  0.001191  0.728871
 Ba  0.500890  0.500432  0.728900
 Hf  0.234747  0.750857  0.227838
 Hf  0.767263  0.750869  0.227849
 Hf  0.250438  0.250834  0.230321
 Hf  0.751388  0.250836  0.230324
 Hf  0.251088  0.750518  0.317886
 Hf  0.750145  0.750522  0.317888
 Hf  0.251035  0.250544  0.318940
 Hf  0.750155  0.250544  0.318940
 Hf  0.250000  0.250000  0.409091

Hf 0.250000 0.750000 0.409091
Hf 0.750000 0.250000 0.409091
Hf 0.750000 0.750000 0.409091
Hf 0.250000 0.250000 0.500000
Hf 0.250000 0.750000 0.500000
Hf 0.750000 0.250000 0.500000
Hf 0.750000 0.750000 0.500000
Hf 0.250000 0.250000 0.590909
Hf 0.250000 0.750000 0.590909
Hf 0.750000 0.250000 0.590909
Hf 0.750000 0.750000 0.590909
Hf 0.251035 0.250544 0.681060
Hf 0.750155 0.250544 0.681060
Hf 0.750145 0.750522 0.682112
Hf 0.251088 0.750518 0.682114
Hf 0.751388 0.250836 0.769676
Hf 0.250438 0.250834 0.769679
Hf 0.767263 0.750869 0.772151
Hf 0.234747 0.750857 0.772162
C 0.501097 0.752135 0.197667
C 0.501097 0.752135 0.802333
O 0.364241 0.752448 0.186034
O 0.637992 0.752439 0.186049
O 0.501033 0.751239 0.227339
O 0.001006 0.750765 0.227548
O 0.500931 0.250696 0.229114
O 0.000930 0.251007 0.229127
O 0.249569 0.501335 0.230048
O 0.752355 0.501349 0.230065
O 0.249227 0.000528 0.230162
O 0.752706 0.000548 0.230169
O 0.256487 0.750573 0.271991
O 0.745230 0.750612 0.271998
O 0.250229 0.250947 0.273082
O 0.751270 0.250925 0.273085
O 0.000611 0.750705 0.317465
O 0.500585 0.250704 0.318201
O 0.000588 0.250301 0.318313
O 0.500605 0.750287 0.318405
O 0.250686 0.000640 0.318429
O 0.750478 0.000643 0.318435
O 0.750901 0.500411 0.318460
O 0.250260 0.500411 0.318462
O 0.249433 0.750317 0.363440
O 0.751205 0.750309 0.363442
O 0.250564 0.250237 0.363700

```
O  0.750039  0.250251  0.363701
O  0.250000  0.000000  0.409091
O  0.250000  0.500000  0.409091
O  0.750000  0.000000  0.409091
O  0.750000  0.500000  0.409091
O  0.000000  0.250000  0.409091
O  0.000000  0.750000  0.409091
O  0.500000  0.250000  0.409091
O  0.500000  0.750000  0.409091
O  0.250000  0.250000  0.454545
O  0.750000  0.250000  0.454545
O  0.250000  0.750000  0.454545
O  0.750000  0.750000  0.454545
O  0.250000  0.000000  0.500000
O  0.250000  0.500000  0.500000
O  0.750000  0.000000  0.500000
O  0.750000  0.500000  0.500000
O  0.000000  0.250000  0.500000
O  0.000000  0.750000  0.500000
O  0.500000  0.250000  0.500000
O  0.500000  0.750000  0.500000
O  0.250000  0.250000  0.545455
O  0.250000  0.750000  0.545455
O  0.750000  0.250000  0.545455
O  0.750000  0.750000  0.545455
O  0.250000  0.000000  0.590909
O  0.250000  0.500000  0.590909
O  0.750000  0.000000  0.590909
O  0.750000  0.500000  0.590909
O  0.000000  0.250000  0.590909
O  0.000000  0.750000  0.590909
O  0.500000  0.250000  0.590909
O  0.500000  0.750000  0.590909
O  0.750039  0.250251  0.636299
O  0.250564  0.250237  0.636300
O  0.751205  0.750309  0.636558
O  0.249433  0.750317  0.636560
O  0.250260  0.500411  0.681538
O  0.750901  0.500411  0.681540
O  0.750478  0.000643  0.681565
O  0.250686  0.000640  0.681571
O  0.500605  0.750287  0.681595
O  0.000588  0.250301  0.681687
O  0.500585  0.250704  0.681799
O  0.000611  0.750705  0.682535
O  0.751270  0.250925  0.726915
```

O  0.250229  0.250947  0.726918
O  0.745230  0.750612  0.728002
O  0.256487  0.750573  0.728009
O  0.752706  0.000548  0.769831
O  0.249227  0.000528  0.769838
O  0.752355  0.501349  0.769935
O  0.249569  0.501335  0.769952
O  0.000930  0.251007  0.770873
O  0.500931  0.250696  0.770886
O  0.001006  0.750765  0.772452
O  0.501033  0.751239  0.772661
O  0.637992  0.752439  0.813951
O  0.364241  0.752448  0.813966

Structure 36. HfO$_2$-terminated BaHfO$_3$ slab at $\Theta$=0.50 CO$_2$ coverage

```
_cell_length_a  8.41088800
_cell_length_b  8.41088800
_cell_length_c  46.25988800
_cell_angle_alpha  90.00000000
_cell_angle_beta  90.00000000
_cell_angle_gamma  90.00000000
_symmetry_space_group_name_H-M  'P 1'
loop_
_atom_site_type_symbol
_atom_site_fract_x
_atom_site_fract_y
_atom_site_fract_z
Ba 0.000930 0.003889 0.272503
Ba 0.500932 0.503891 0.272503
Ba 0.500927 0.997266 0.272507
Ba 0.000925 0.497264 0.272508
Ba 0.000482 0.000336 0.363545
Ba 0.500482 0.500335 0.363546
Ba 0.000470 0.500177 0.363550
Ba 0.500469 0.000177 0.363550
Ba 0.000000 0.000000 0.454545
Ba 0.000000 0.500000 0.454545
Ba 0.500000 0.000000 0.454545
Ba 0.500000 0.500000 0.454545
Ba 0.000000 0.000000 0.545455
Ba 0.000000 0.500000 0.545455
Ba 0.500000 0.000000 0.545455
Ba 0.500000 0.500000 0.545455
Ba 0.000470 0.500177 0.636450
Ba 0.500469 0.000177 0.636450
Ba 0.500482 0.500335 0.636454
Ba 0.000482 0.000336 0.636455
Ba 0.000925 0.497264 0.727492
Ba 0.500927 0.997266 0.727493
Ba 0.000930 0.003889 0.727497
Ba 0.500932 0.503891 0.727497
Hf 0.735172 0.250586 0.228417
Hf 0.235168 0.750588 0.228418
Hf 0.266724 0.250599 0.228422
Hf 0.766722 0.750600 0.228422
Hf 0.750837 0.250447 0.318282
Hf 0.250836 0.750448 0.318283
Hf 0.250724 0.250451 0.318284
Hf 0.750723 0.750450 0.318284
Hf 0.250000 0.250000 0.409091
```

```
Hf 0.250000 0.750000 0.409091
Hf 0.750000 0.250000 0.409091
Hf 0.750000 0.750000 0.409091
Hf 0.250000 0.250000 0.500000
Hf 0.250000 0.750000 0.500000
Hf 0.750000 0.250000 0.500000
Hf 0.750000 0.750000 0.500000
Hf 0.250000 0.250000 0.590909
Hf 0.250000 0.750000 0.590909
Hf 0.750000 0.250000 0.590909
Hf 0.750000 0.750000 0.590909
Hf 0.250724 0.250451 0.681716
Hf 0.750723 0.750450 0.681716
Hf 0.250836 0.750448 0.681717
Hf 0.750837 0.250447 0.681718
Hf 0.266724 0.250599 0.771578
Hf 0.766722 0.750600 0.771578
Hf 0.235168 0.750588 0.771582
Hf 0.735172 0.250586 0.771583
C 0.000980 0.250947 0.198425
C 0.500977 0.750938 0.198426
C 0.500977 0.750938 0.801574
C 0.000980 0.250947 0.801575
O 0.864091 0.251008 0.186909
O 0.364080 0.750996 0.186910
O 0.637895 0.750968 0.186914
O 0.137906 0.250981 0.186914
O 0.000955 0.250773 0.228225
O 0.500951 0.750772 0.228226
O 0.000942 0.750387 0.228592
O 0.500944 0.250389 0.228593
O 0.751060 0.000617 0.231311
O 0.251061 0.500618 0.231315
O 0.250730 0.000616 0.231325
O 0.750730 0.500616 0.231326
O 0.757727 0.250486 0.272601
O 0.257713 0.750502 0.272602
O 0.244099 0.250573 0.272604
O 0.744108 0.750587 0.272604
O 0.500790 0.251213 0.317439
O 0.000789 0.751215 0.317443
O 0.500786 0.749729 0.318560
O 0.000786 0.249727 0.318561
O 0.751593 0.500449 0.318566
O 0.251592 0.000450 0.318570
O 0.250093 0.500450 0.318585
```

```
O  0.750094  0.000450  0.318587
O  0.749328  0.250286  0.363594
O  0.249337  0.750276  0.363595
O  0.251481  0.250208  0.363596
O  0.751470  0.750194  0.363596
O  0.250000  0.500000  0.409091
O  0.750000  0.000000  0.409091
O  0.250000  0.000000  0.409091
O  0.750000  0.500000  0.409091
O  0.000000  0.250000  0.409091
O  0.000000  0.750000  0.409091
O  0.500000  0.750000  0.409091
O  0.500000  0.250000  0.409091
O  0.750000  0.250000  0.454545
O  0.750000  0.750000  0.454545
O  0.250000  0.750000  0.454545
O  0.250000  0.250000  0.454545
O  0.250000  0.000000  0.500000
O  0.250000  0.500000  0.500000
O  0.750000  0.000000  0.500000
O  0.750000  0.500000  0.500000
O  0.000000  0.250000  0.500000
O  0.000000  0.750000  0.500000
O  0.500000  0.250000  0.500000
O  0.500000  0.750000  0.500000
O  0.250000  0.250000  0.545455
O  0.250000  0.750000  0.545455
O  0.750000  0.250000  0.545455
O  0.750000  0.750000  0.545455
O  0.250000  0.000000  0.590909
O  0.250000  0.500000  0.590909
O  0.750000  0.000000  0.590909
O  0.750000  0.500000  0.590909
O  0.000000  0.250000  0.590909
O  0.000000  0.750000  0.590909
O  0.500000  0.250000  0.590909
O  0.500000  0.750000  0.590909
O  0.251481  0.250208  0.636404
O  0.751470  0.750194  0.636404
O  0.249337  0.750276  0.636405
O  0.749328  0.250286  0.636406
O  0.750094  0.000450  0.681413
O  0.250093  0.500450  0.681415
O  0.251592  0.000450  0.681430
O  0.751593  0.500449  0.681434
O  0.000786  0.249727  0.681439
```

```
O  0.500786  0.749729  0.681440
O  0.000789  0.751215  0.682557
O  0.500790  0.251213  0.682561
O  0.244099  0.250573  0.727396
O  0.744108  0.750587  0.727396
O  0.257713  0.750502  0.727398
O  0.757727  0.250486  0.727399
O  0.750730  0.500616  0.768674
O  0.250730  0.000616  0.768675
O  0.251061  0.500618  0.768685
O  0.751060  0.000617  0.768689
O  0.500944  0.250389  0.771407
O  0.000942  0.750387  0.771408
O  0.500951  0.750772  0.771774
O  0.000955  0.250773  0.771775
O  0.637895  0.750968  0.813086
O  0.137906  0.250981  0.813086
O  0.364080  0.750996  0.813090
O  0.864091  0.251008  0.813091
```